\documentclass[prc,letterpaper,onecolumn,showpacs,showkeys,lengthcheck,
               floatfix,nofootinbib,preprintnumbers,superscriptaddress]{revtex4-1} 

\usepackage{mathtools}
\usepackage{amssymb,amsfonts}
\usepackage{bm}
\usepackage{slashed}

\usepackage[svgnames]{xcolor}
\usepackage[english]{babel}
\usepackage{blindtext}
\usepackage{microtype}
\usepackage{tikz}
\usepackage{dcolumn}
\usepackage{multirow}
\usepackage[]{units}

\usepackage{graphicx}
\usepackage[caption=false]{subfig}

\usepackage[colorlinks=true]{hyperref}
\usepackage{ifpdf}

\graphicspath{{.}{figures/}} 

\def\a{\alpha}
\def\b{\beta}
\def\d{\delta}
\def\g{\gamma}

\def\ve{\varepsilon}

\def\k{\kappa}
\def\l{\lambda}
\def\s{\sigma}
\def\o{\omega}

\def\G{\Gamma}
\def\L{\Lambda}
\def\O{\Omega}
\def\S{\Sigma}

\def\pl{\partial}
\def\hs{\hspace}

\def\ol{\overline}
\def\no{\nonumber}

\def\lf{\left}
\def\rg{\right}

\def\lra{\longrightarrow}

\newcommand{\ph}[1]{\phantom{#1}}

\newcommand{\sh}[1]{\slashed{#1}}

\font\bb=bbmss10 scaled 1200
\def\ident{\mbox{\bb 1}}

\def\calQ{\mathcal{Q}}
\def\calD{\mathcal{D}}

\begin{document}

\preprint{ADP-14-14/T872}

\title{Role of diquark correlations and the pion cloud in nucleon elastic form factors}

\author{Ian~C.~Clo\"et}
\affiliation{Physics Division, Argonne National Laboratory, Argonne, Illinois 60439, USA}

\author{Wolfgang~Bentz}
\affiliation{Department of Physics, School of Science, Tokai University,
             Hiratsuka-shi, Kanagawa 259-1292, Japan}

\author{Anthony~W.~Thomas}
\affiliation{CSSM and ARC Centre of Excellence for Particle Physics at the Terascale, \\ 
School of Chemistry and Physics, 
University of Adelaide, Adelaide SA 5005, Australia}

\begin{abstract}
Electromagnetic form factors of the nucleon in the space-like region are investigated 
within the framework of a covariant and confining Nambu--Jona-Lasinio model.
The bound state amplitude of the nucleon is obtained as the solution of a 
relativistic Faddeev equation, where diquark correlations appear naturally 
as a consequence of the strong coupling in the colour $\bar{3}$ $qq$ channel. 
Pion degrees of freedom are included as a perturbation to the ``quark-core'' 
contribution obtained using the Poincar\'e covariant Faddeev amplitude. 
While no model parameters are fit to form factor data, excellent agreement is 
obtained with the empirical nucleon form factors (including the magnetic moments 
and radii) where pion loop corrections play a critical role for 
$Q^2 \lesssim 1\,$GeV$^2$. Using charge symmetry, the nucleon form factors can be expressed as proton quark 
sector form factors. The latter are studied in detail, leading, for example, to 
the conclusion that the $d$-quark sector of the Dirac form factor is much softer 
than the $u$-quark sector, a consequence of the dominance of scalar diquark 
correlations in the proton wave function. On the other hand, for the proton quark sector 
Pauli form factors we find that the effect of the pion cloud and axialvector diquark correlations 
overcomes the effect of scalar diquark dominance, leading to a larger $d$-quark anomalous magnetic moment 
and a form factor in the $u$-quark sector that is slightly softer than in 
the $d$-quark sector.
\end{abstract}


\maketitle
\section{Introduction}
The electromagnetic form factors of a nucleon provide information on its internal 
momentum space distribution of charge and magnetization, thus furnishing a unique 
window into the quark and gluon substructure of the nucleon. Building a bridge 
between QCD and the observed nucleon properties is a key challenge for modern 
hadron physics and recent form factor measurements, for example, demonstrate 
that a robust understanding of nucleon properties founded in QCD is just beginning.

A key example of the impact of such measurements is provided by the polarization 
transfer experiments~\cite{Jones:1999rz,Gayou:2001qd,Gayou:2001qt,Puckett:2010ac,Puckett:2011xg}, 
which revealed that the ratio of the proton's electric to magnetic Sachs form factors, 
$\mu_p\,G_{Ep}(Q^2)/G_{Mp}(Q^2)$, is not constant but instead  decreases almost 
linearly with $Q^2$. These experiments dispelled decades of perceived wisdom 
which perpetuated the view that the nucleon contained similar distributions 
of charge and magnetization. Nucleon form factor data at large $Q^2$ can 
also be used to test the scaling behaviour predicted by perturbative QCD, 
which, for example, makes the prediction that $Q^2\,F_{2p}(Q^2)/F_{1p}(Q^2)$ should 
tend to a constant as $Q^2 \to \infty$~\cite{Brodsky:1973kr,Brodsky:1974vy}. 
However, recent data extending to $Q^2 \simeq 8\,$GeV$^2$~\cite{Puckett:2010ac,Puckett:2011xg}, find scaling behaviour much closer to $Q\,F_{2p}(Q^2)/F_{1p}(Q^2)$, 
which has been attributed to the quark component of the nucleon wave function 
possessing sizeable orbital angular momentum~\cite{Ralston:2003mt}. 
An interesting recent example, which demonstrates that there is much 
of a fundamental nature still to learn in hadron physics, involves the muonic 
hydrogen experiments~\cite{Pohl:2010zza,Antognini:1900ns} that found a proton charge radius 
some 4\% smaller than that measured in elastic electron scattering or 
electronic hydrogen, representing a 7$\s$ discrepancy. As yet there is no 
accepted resolution to this puzzle~\cite{Cloet:2010qa,Miller:2011yw,Carroll:2011rv}. 

It is clear, therefore, that a quantitative theoretical understanding of 
nucleon form factors in terms of the fundamental degrees of freedom of QCD, namely 
the quarks and gluons, remains an important goal. This task is particularly 
challenging because nucleon form factors parameterize the amplitude for a nucleon 
to interact through a current and remain a nucleon, for arbitrary space-like 
momentum transfer. Therefore, long distance non-perturbative effects associated 
with quark binding and confinement must play an important role at all $Q^2$, while,
because of asymptotic freedom at short distances, perturbative QCD must also be 
relevant at large momentum transfer. This scenario is somewhat in contrast to 
that found with the structure functions measured in deep inelastic scattering, 
which can be factorized into short distance Wilson coefficients, calculable 
in perturbative QCD, and the long distance parton distribution functions 
(PDFs) which encode non-perturbative information on the structure of the bound 
state. A consequence of factorization is that once the PDFs are known at a 
scale $Q_0^2 \gg \L_{\text{QCD}}^2$, the $Q^2$ evolution of the PDFs, on 
the Bjorken $x$ domain relevant to hadron structure, is governed by 
the DGLAP evolution equations~\cite{Gribov:1972rt,Altarelli:1977zs,Dokshitzer:1977sg}. 
An analogous factorization is not possible for the 
nucleon electromagnetic form factors.

Here we investigate the nucleon electromagnetic form factors using the 
Nambu--Jona-Lasinio (NJL) model~\cite{Nambu:1961tp,Nambu:1961fr,Vogl:1991qt,Hatsuda:1994pi,Klevansky:1992qe}, which is a Poincar\'e covariant quantum field theory with many of the same 
low-energy properties as QCD. For example, it encapsulates 
the key emergent phenomena of dynamical 
chiral symmetry breaking and confinement.\footnote{Standard implementations of the 
NJL model are not confining. This can be seen in results for hadron propagators 
which develop imaginary pieces in particular kinematical domains, indicating 
that the hadron can decay into quarks. In the version of the NJL model used here 
quark confinement is introduced via a particular regularization prescription 
which eliminates these unphysical thresholds. This regularization procedure 
is discussed in Sect.~\ref{sec:NJL}.} This model also has the same flavour 
symmetries as QCD and should therefore provide a robust chiral effective theory 
of QCD valid at low to intermediate energies. The NJL model is solved 
non-perturbatively, using the standard leading order truncation. Finally, in 
order to respect chiral symmetry effectively, we also include pion 
degrees of freedom in a perturbative manner. This proves 
essential~\cite{Thomas:1981vc,Theberge:1982xs,Thomas:1982kv} for a 
good description of the nucleon form factors below $Q^2 \sim 1\,$GeV$^2$. 

The outline of the paper is as follows: Sect.~\ref{sec:NJL} gives an introduction 
to the NJL model, encompassing the gap equation, the Bethe-Salpeter equation 
and the relativistic Faddeev equation. In Sect.~\ref{sec:nucleon_current} we explain 
how to calculate the matrix elements of the quark electromagnetic current which 
give the nucleon electromagnetic form factors. A key ingredient is the dressed 
quark-photon vertex; the interaction of a virtual photon with a non-pointlike 
constituent, or dressed quark, which is detailed in Sect.~\ref{sec:QPV}. 
Pion loop effects at the constituent quark level are also discussed and results for 
dressed quark form factors are presented. Because the nucleon emerges as a 
quark--diquark bound state, a critical step in determining the nucleon form factors is 
to determine the electromagnetic current for the relevant diquarks. This is 
discussed in Sect.~\ref{sec:diquark_results} for scalar and axialvector diquarks, 
together with form factor results for the pion and rho mesons, which are 
the $\bar{q}q$ analogs of these diquarks. The electromagnetic current of the 
nucleon is determined in Sect.~\ref{sec:nucleon_results}, where the role of 
pion loop effects is discussed in detail. Careful attention is paid to the flavour 
decomposition of the nucleon form factors and the interpretation of their 
$Q^2$ dependence in terms of the interplay between the roles of diquark correlations 
and pionic effects within the nucleon. Comparisons with experiment are presented 
and inferences drawn regarding features of the data and connections to 
the quark structure within the nucleon. 
Conclusions are presented in Sect.~\ref{sec:conclusion}.

\section{Nambu--Jona-Lasinio Model \label{sec:NJL}}
The Nambu--Jona-Lasinio (NJL) model, while originally a theory of elementary nucleons~\cite{Nambu:1961tp,Nambu:1961fr}, is now interpreted as a QCD motivated chiral effective quark theory characterized by a 4-fermion contact interaction between the quarks~\cite{Vogl:1991qt,Hatsuda:1994pi,Klevansky:1992qe}. A salient feature of the model is that it is a Poincar\'e covariant quantum field theory where interactions dynamically break chiral symmetry, giving rise to dynamically generated dressed quark masses, a pion that is a $\bar{q}q$ bound state with the properties of a pseudo-Goldstone boson and a large mass splitting between low lying chiral partners. The NJL model has a long history of success in the study of meson properties~\cite{Vogl:1991qt,Klevansky:1992qe} and more recently as a tool to investigate baryons as 3-quark bound states using the relativistic Faddeev equation~\cite{Ishii:1993np,Ishii:1993rt,Ishii:1995bu}. Recent examples include the study of nucleon parton distribution functions (PDFs)~\cite{Mineo:2003vc,Cloet:2005rt,Cloet:2005pp,Cloet:2006bq,Cloet:2007em}, quark fragmentation functions~\cite{Ito:2009zc,Matevosyan:2010hh} and transverse momentum dependent  PDFs~\cite{Matevosyan:2011vj,Matevosyan:2012ga}. Finally, we mention that the NJL model has been used to study the self-consistent modification of the structure of the nucleon in-medium and its role in the binding of atomic nuclei~\cite{Bentz:2001vc}.

The $SU(2)$ flavour NJL Lagrangian relevant to this study, in the $\bar{q}q$ interaction channel, reads\footnote{The complete $SU(2)$ flavour NJL interaction Lagrangian can in principle also contain the chiral singlet terms
\begin{multline*}
\tfrac{1}{2}\,G_\eta\lf[ \lf(\bar{\psi}\,\vec{\tau}\,\psi\rg)^2 - \lf(\bar{\psi}\,\g_5\,\psi\rg)^2\rg] 
- \tfrac{1}{2}\,G_f\lf(\bar{\psi}\,\g^\mu\g_5\,\psi\rg)^2 \\
- \tfrac{1}{2}\,G_T\lf[\lf(\bar{\psi}\,i\sigma^{\mu\nu}\psi\rg)^2 - \lf(\bar{\psi}\,i\sigma^{\mu\nu}\vec{\tau}\,\psi\rg)^2\rg].
\end{multline*}
The complete Lagrangian explicitly breaks $U_A(1)$ symmetry unless $G_\eta = G_\pi$ and $G_T = 0$. These are the conditions imposed on the NJL  Lagrangian by chiral symmetry if the chiral group is enlarged to three flavours, where the $U_A(1)$ symmetry is usually broken by introducing a 6-fermion interaction~\cite{Vogl:1991qt,Klevansky:1992qe}.}
\begin{align}
\mathcal{L} &= \bar{\psi}\lf(i\sh{\pl}-\hat{m}\rg)\psi \no \\
&\hs{5mm}
+ \frac{1}{2}\,G_\pi\lf[\lf(\bar{\psi}\psi\rg)^2 - \lf(\bar{\psi}\,\g_5\vec{\tau}\,\psi\rg)^2\rg]
- \frac{1}{2}\,G_\o\lf(\bar{\psi}\,\g^\mu\,\psi\rg)^2 \no \\
&\hs{5mm}
- \frac{1}{2}\,G_\rho\lf[\lf(\bar{\psi}\,\g^\mu\vec{\tau}\,\psi\rg)^2 + \lf(\bar{\psi}\,\g^\mu\g_5\vec{\tau}\,\psi\rg)^2\rg],
\label{eq:njllagrangian}
\end{align}
where $\hat{m} \equiv \text{diag}[m_u,\,m_d]$ is the current quark mass matrix and the 4-fermion coupling constants in each chiral channel are labelled by $G_\pi$, $G_\o$ and $G_\rho$. Throughout this paper we take $m_u = m_d = m$. The interaction Lagrangian can be Fierz symmetrized, with the consequence that after a redefinition of the 4-fermion couplings one need only consider direct terms in the elementary interaction~\cite{Ishii:1995bu}. The elementary quark--antiquark interaction kernel is then given by
\begin{align}
\mathcal{K}_{\a\b,\g\d} &= \sideset{}{_\Omega} \sum K_\Omega\,\Omega_{\a\b}\, \bar{\Omega}_{\g\d}\no \\
&\hs{-8mm}
= 2i\,G_\pi \lf[\lf(\ident\rg)_{\a\b}\lf(\ident\rg)_{\g\d} - \lf(\g_5\tau_i\rg)_{\a\b}\lf(\g_5\tau_i\rg)_{\g\d}\rg] \no \\
&\hs{-8mm}
-2i\,G_\rho \lf[\lf(\g_\mu\tau_i\rg)_{\a\b}\lf(\g^\mu\tau_i\rg)_{\g\d} + \lf(\g_\mu\g_5\tau_i\rg)_{\a\b}\lf(\g^\mu\g_5\tau_i\rg)_{\g\d}\rg] \no \\
&\hs{-8mm}
-2i\,G_\o\lf(\g_\mu\rg)_{\a\b}\lf(\g^\mu\rg)_{\g\d},
\label{eq:qbarqkernel}
\end{align}
where the indices label Dirac, colour and isospin.

\begin{figure}[tbp]
\centering\includegraphics[width=\columnwidth,clip=true,angle=0]{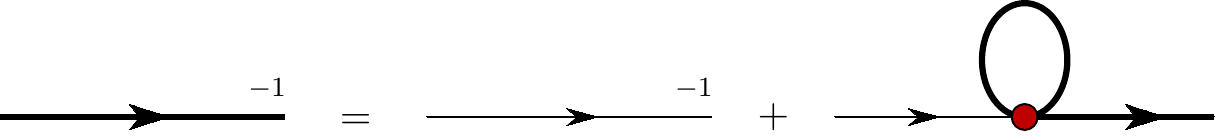}
\caption{(Colour online) The NJL gap equation in the Hartree-Fock approximation, where the thin line represents the elementary quark propagator, $S_0^{-1}(k) = \sh{k} - m + i\ve$, and the shaded circle the $\bar{q}q$ interaction kernel given in Eq.~\eqref{eq:qbarqkernel}. 
Higher order terms, attributed to meson loops, for example, are not included in the gap equation kernel.}
\label{fig:gapequation}
\end{figure}

The building blocks of mesons and baryons in the NJL model are the quark propagators. The NJL dressed quark propagator is obtained by solving the gap equation, which at the level of approximation used here is illustrated in Fig.~\ref{fig:gapequation} and reads\footnote{In principle there is an infinite tower of higher order terms that can appear in the NJL gap equation kernel, with meson loops an important example. However, in keeping with the standard treatment, these higher order terms are not included. We will however include a single pion loop as a  perturbative correction to the quark-photon vertex. This is discussed in Sect.~\ref{sec:QPV}.}
\begin{align}
S^{-1}(k) = S_0^{-1}(k) - \sum_\O K_\O\, \O \int \frac{d^4\ell}{(2\pi)^4} \ \text{Tr}\lf[\bar{\O}\,S(\ell)\rg],
\label{eq:gapequation}
\end{align}
where $S_0^{-1}(k) = \sh{k} - m + i\ve$ is the bare quark propagator and the trace is over Dirac, colour and isospin indices. The only piece of the $\bar{q}q$ interaction kernel given in Eq.~\eqref{eq:qbarqkernel} that contributes to the gap equation expressed in Eq.~\eqref{eq:gapequation} is the isoscalar-scalar interaction $2i\,G_\pi\lf(\ident\rg)_{\a\b}\lf(\ident\rg)_{\g\d}$. This yields a solution of the form
\begin{align}
S(k) = \frac{1}{\sh{k} - M + i\ve}.
\label{eq:quarkpropagator}
\end{align}
The interaction kernel in the gap equation of Fig.~\ref{fig:gapequation} is local and therefore the dressed quark mass, $M$, is a constant and satisfies
\begin{align}
M = m + 12\,i\,G_\pi \int \frac{d^4\ell}{(2\pi)^4}\,\mathrm{Tr}_D\lf[S(\ell)\rg],
\label{eq:gap}
\end{align}
where the remaining trace is over Dirac indices. For sufficiently strong coupling, $G_\pi > G_{\text{critial}}$, Eq.~\eqref{eq:gap} supports a non-trivial solution with $M > m$, which survives even in the chiral limit ($m=0$).\footnote{In the proper-time regularization scheme defined in Eq.~\eqref{eq:propertime} the critical coupling in the chiral limit has the value: $G_{\text{critial}} = \frac{\pi^2}{3}\lf(\L_{UV}^2 - \L_{IR}^2\rg)^{-1}$.} This solution is a consequence of dynamical chiral symmetry breaking (DCSB) in the Nambu-Goldstone mode and it is readily demonstrated, by calculating the total energy~\cite{Buballa:2003qv}, that this phase corresponds to the ground state of the vacuum.

The NJL model is a non-renormalizable quantum field theory, therefore a regularization prescription must be specified to fully define the model. We choose the proper-time regularization  scheme~\cite{Ebert:1996vx,Hellstern:1997nv,Bentz:2001vc}, which is introduced formally via the relation
\begin{align}
\frac{1}{X^n} &= \frac{1}{(n-1)!}\int_0^\infty d\tau \, \tau^{n-1}\, e^{-\tau\,X}, \no \\
&\hs{15mm}
\lra \frac{1}{(n-1)!}\int_{1/\Lambda^2_{UV}}^{1/\Lambda^2_{IR}} d\tau \, \tau^{n-1}\, e^{-\tau\,X},
\label{eq:propertime}
\end{align}
where $X$ represents a product of propagators that have been combined using Feynman parametrization. Only the ultraviolet cutoff, $\Lambda_{UV}$, is needed to render the theory finite, however, for bound states of quarks we also include the infrared cutoff, $\Lambda_{IR}$. This has the effect of eliminating unphysical thresholds for the decay of hadrons into free quarks and therefore simulates aspects of quark confinement in QCD. 

\begin{figure}[t]
\centering\includegraphics[width=\columnwidth,clip=true,angle=0]{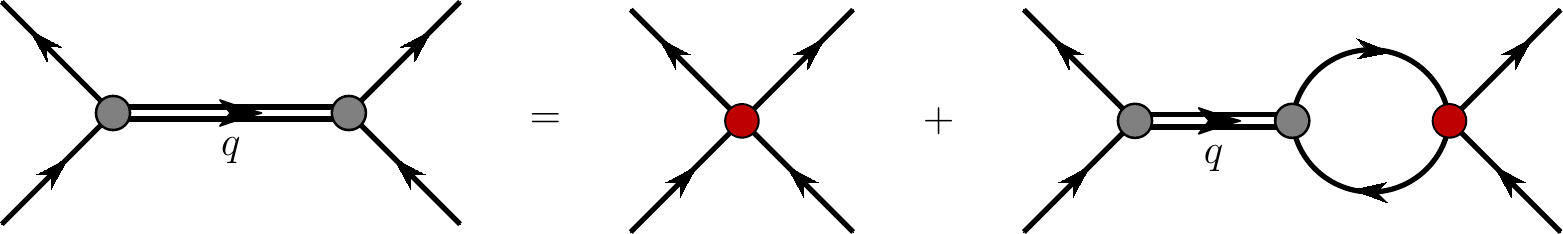}
\caption{(Colour online) NJL Bethe-Salpeter equation for the quark--antiquark $t$-matrix, represented as the double line with the vertices. The single line corresponds to the dressed quark propagator and the BSE $\bar{q}q$ interaction kernel, consistent with the gap equation kernel used in Eq.~\eqref{eq:gap}, is given by Eq.~\eqref{eq:qbarqkernel}.}
\label{fig:bethesalpeter}
\end{figure}

Mesons in the NJL model are quark--antiquark bound states whose properties are determined by first solving the Bethe-Salpeter equation (BSE). The kernels of the gap and BSEs are intimately related, as exemplified by the vector and axialvector Ward--Takahashi identities, which relate the quark propagator to inhomogeneous Bethe-Salpeter vertices~\cite{Maris:2003vk}. The NJL BSE, consistent with the gap equation of Fig.~\ref{fig:gapequation}, is illustrated in Fig.~\ref{fig:bethesalpeter} and reads
\begin{align}
\mathcal{T}(q) = \mathcal{K} +
\int \frac{d^4k}{(2\pi)^4}\, \mathcal{K}\, S(k+q)\, S(k)\,\mathcal{T}(q),
\label{eq:mesonbse}
\end{align}
where $q$ is the total momentum of the two-body system, $\mathcal{T}$ is the two-body $t$-matrix and $\mathcal{K}$ is the $\bar{q}q$ interaction kernel given in Eq.~\eqref{eq:qbarqkernel}. Dirac, colour and isospin indices have been suppressed in Eq.~\eqref{eq:mesonbse}. Solutions to the BSE in the $\bar{q}q$ channels with quantum numbers that correspond to those of the pion,\footnote{The NJL Lagrangian of Eq.~\eqref{eq:njllagrangian} implies that the $G_\rho\lf(\bar{\psi}\,\g^\mu\g_5\vec{\tau}\,\psi\rg)^2$ $\bar{q}q$ interaction should also contribute in the pionic channel, giving rise to $\pi$--$a_1$ mixing. However, since the $a_1$ meson is much heavier than the pion, the amount of mixing is small and we therefore ignore $\pi$--$a_1$ mixing in this work.} rho and omega have the form
\begin{align}
\label{eq:piontmatrix}
\mathcal{T}_\pi(q)_{\a\b,\g\d} &= \lf(\g_5\tau_i\rg)_{\a\b}\hs{1.5mm}\, \tau_\pi(q) \hs{3mm}\lf(\g_5\tau_i\rg)_{\g\d}, \\
\mathcal{T}_\rho(q)_{\a\b,\g\d} &= \lf(\g_\mu\tau_i\rg)_{\a\b}\ \, \tau^{\mu\nu}_\rho(q)\ \lf(\g_\nu\tau_i\rg)_{\g\d},  \\
\label{eq:omegatmatrix}
\mathcal{T}_\o(q)_{\a\b,\g\d}    &= \lf(\g_\mu\rg)_{\a\b}\hs{4mm}\, \tau^{\mu\nu}_\o(q)\, \, \lf(\g_\nu\rg)_{\g\d},
\end{align}
where $\tau_i$ are the Pauli matrices and
\begin{align}
\label{eq:tpion}
\tau_\pi(q) &= \frac{-2i\,G_\pi}{1 + 2\,G_\pi\,\Pi_{PP}(q^2)}, \\
\label{eq:trho}
\tau^{\mu\nu}_\rho(q) 
&= \frac{-2i\,G_\rho}{1+2\,G_\rho\,\Pi_{VV}(q^2)} 
\lf[g^{\mu\nu} + 2\,G_\rho\,\Pi_{VV}(q^2)\,\frac{q^\mu q^\nu}{q^2}\rg], \\
\label{eq:tomega}
\tau^{\mu\nu}_\o(q) 
&= \frac{-2i\,G_\o}{1+2\,G_\o\,\Pi_{VV}(q^2)} 
\lf[g^{\mu\nu} + 2\,G_\o\,\Pi_{VV}(q^2)\,\frac{q^\mu q^\nu}{q^2}\rg].
\end{align}
The functions $\tau_\pi(q)$, $\tau^{\mu\nu}_\rho(q)$ and $\tau^{\mu\nu}_\o(q)$ are the reduced $t$-matrices, which are interpreted as propagators for the pion, rho and omega mesons. The bubble diagrams in Eqs.~\eqref{eq:tpion}--\eqref{eq:tomega} have the form
\begin{align}
\label{eq:bubble_PP}
&\Pi_{PP}\lf(q^2\rg)\delta_{ij} = 3i \int \frac{d^4k}{(2\pi)^4}\ \mathrm{Tr}\lf[\g_5\,\tau_i\,S(k)\,\g_5\,\tau_j\,S(k+q)\rg], \\
\label{eq:bubble_VV}
&\Pi_{VV}(q^2)\lf(g^{\mu\nu} - \frac{q^\mu q^\nu}{q^2}\rg)\delta_{ij} \no \\
&\hs{11mm}
= 3i \int \frac{d^4k}{(2\pi)^4}\ \mathrm{Tr}\lf[\g^\mu\tau_i\,S(k)\,\g^\nu\tau_j\,S(k+q)\rg],
\end{align}
where the traces are over Dirac and isospin indices. Meson masses are then defined by the pole in the corresponding two-body $t$-matrix.

In a covariant formulation a two-body $t$-matrix, near a bound state pole of mass $m_i$, behaves as
\begin{align}
\mathcal{T}(q) \to \frac{\G_i(q)\,\ol{\G}_i(q)}{q^2 - m_i^2},
\end{align}
where $\G_i(q)$ is the normalized homogeneous Bethe-Salpeter vertex function for the bound state. Expanding the $t$-matrices in Eqs.~\eqref{eq:piontmatrix}--\eqref{eq:omegatmatrix} about the pole masses gives
\begin{align}
\hs*{-1.5mm}\G^i_\pi &= \sqrt{Z_\pi}\,\g_5\,\tau_i, ~~
\G^{\mu,i}_\rho = \sqrt{Z_\rho}\,\g^\mu\,\tau_i, ~~
\G^\mu_\o = \sqrt{Z_\o}\,\g^\mu,
\end{align}
where $i$ is an isospin index and the normalization factors are given by
\begin{align}
\label{eq:Zpi}
Z_\pi^{-1} &= -\frac{\pl}{\pl q^2}\,\Pi_{PP}(q^2)\Big\rvert_{q^2 = m_\pi^2}, \\[0.1ex]
Z_{\rho,\o}^{-1} &= -\frac{\pl}{\pl q^2}\,\Pi_{VV}(q^2)\Big\rvert_{q^2 = m_{\rho,\o}^2}.
\end{align}
These residues are interpreted as the effective meson-quark-quark coupling constants. Homogeneous Bethe-Salpeter vertex functions are an essential ingredient in, for example, triangle diagrams that determine the meson form factors.

Baryons in the NJL model are naturally described as bound states of three dressed quarks. The properties of these bound states are determined by the relativistic Faddeev equation whose solution gives the Poincar\'e covariant Faddeev amplitude. To construct the interaction kernel of the Faddeev equation we require the elementary quark-quark interaction kernel. Using Fierz transformations to rewrite Eq.~\eqref{eq:njllagrangian} as a sum of $qq$ interactions, keeping only the isoscalar--scalar ($0^+,T=0$) and isovector--axialvector ($1^+,T=1$) two-body channels, the NJL interaction Lagrangian takes the form
\begin{multline}
\mathcal{L}_{I,qq} = G_s \Bigl[\bar{\psi}\,\g_5\, C\,\tau_2\,\beta_A\, \bar{\psi}^T\Bigr]
                              \Bigl[\psi^T\,C^{-1}\g_5\,\tau_2\,\beta_A\, \psi\Bigr] \\
+ G_a \Bigl[\bar{\psi}\,\g_\mu\,C\,\tau_i\tau_2\,\beta_A\, \bar{\psi}^T\Bigr]
                              \Bigl[\psi^T\,C^{-1}\g^{\mu}\,\tau_2\tau_i\, \beta_A\, \psi\Bigr],
\label{eq:qqlagrangian}
\end{multline}
where $C = i\g_2\g_0$ is the charge conjugation matrix and the couplings $G_s$ and $G_a$ give the strength of the scalar and axialvector $qq$ interactions. Because only colour $\bar{3}$ $qq$ states can couple to a third quark to form a colourless three-quark state, we must have $\beta_A = \sqrt{\tfrac{3}{2}}\,\lambda_A~(A=2,5,7)$~\cite{Ishii:1995bu}. The Lagrangian of Eq.~\eqref{eq:qqlagrangian} gives the following elementary $qq$ interaction kernel
\begin{align}
\mathcal{K}_{\a\b,\g\d} &= 4i\,G_s\lf(\g_5\,C\,\tau_2\,\b_A\rg)_{\a\b}\lf(C^{-1}\,\g_5\,\tau_2\,\b_{A}\rg)_{\g\d} \no \\
&\hs{-5mm}
+ 4i\,G_a\lf(\g_\mu\,C\,\tau_i\tau_2\,\b_A\rg)_{\a\b}\lf(C^{-1}\,\g^\mu\,\tau_2\tau_i\,\b_{A}\rg)_{\g\d}.
\label{eq:qqkernel}
\end{align}
This kernel has been truncated to support only scalar and axialvector diquark correlations because the pseudoscalar and vector diquark components of the nucleon must predominantly be in $\ell = 1$ states and are therefore suppressed. Pseudoscalar and vector diquarks are also usually found to be considerably heavier than their scalar and axialvector counterparts~\cite{Roberts:2011cf}.

Using Eq.~\eqref{eq:qqkernel} as the interaction kernel in the Faddeev equation allows us to first sum all two-body $qq$ interactions to form the scalar and axialvector diquark $t$-matrices. Diquark correlations in the nucleon are therefore a natural consequence of the strong coupling in the colour $\bar{3}$ quark-quark interaction channel. The BSE in the $qq$ channel for our NJL model reads
\begin{align}
\mathcal{T}(q) = \mathcal{K} +
\frac{1}{2}\int \frac{d^4k}{(2\pi)^4}\, \mathcal{K}\,S(k+q)\,S(-k)\,\mathcal{T}(q),
\label{eq:diquarkbse}
\end{align}
where $\mathcal{K}$ is given in Eq.~\eqref{eq:qqkernel} and there is a symmetry factor of $\tfrac{1}{2}$ relative to the $\bar{q}q$ BSE of Eq.~\eqref{eq:mesonbse}. The solutions to the BSE in the scalar and axialvector diquark channels are
\begin{align}
\mathcal{T}_s(q)_{\a\b,\g\d} &= \lf(\g_5\,C\,\tau_2\,\beta_A\rg)_{\a\b}\tau_s(q)\lf(C^{-1}\g_5\,\tau_2\,\beta_A\rg)_{\g\d}, \\
\mathcal{T}_a(q)_{\a\b,\g\d} &= \lf(\g_\mu\,C\,\tau_i\tau_2\,\beta_A\rg)_{\a\b} \tau^{\mu\nu}_a(q) \lf(C^{-1}\g_\nu\,\tau_2\tau_i\,\beta_A\rg)_{\g\d}, 
\end{align}
where
\begin{align}
\label{eq:tscalar}
\tau_s(q) &= \frac{-4i\,G_s}{1 + 2\,G_s\,\Pi_{PP}(q^2)}, \\
\label{eq:taxial}
\tau^{\mu\nu}_a(q) 
&= \frac{-4i\,G_a}{1+2\,G_a\,\Pi_{VV}(q^2)} 
\lf[g^{\mu\nu} + 2\,G_a\,\Pi_{VV}(q^2)\,\frac{q^\mu q^\nu}{q^2}\rg].
\end{align}
The scalar and axialvector diquark masses are defined as the poles\footnote{In QCD these poles should not exist, since diquarks, as coloured objects, are not part of the physical spectrum. Nevertheless, diquark states play a very important role in many phenomenological studies, for example in the spin and flavor dependence of nucleon PDFs~\cite{Close:1988br,Schreiber:1991qx}. They have also been observed in lattice QCD studies~\cite{Hess:1998sd} as well as model studies of QCD, for example, in the rainbow-ladder truncation of the Dyson-Schwinger equations (DSEs). In the DSE approach diagrams beyond the rainbow-ladder truncation have been shown to remove the pole in the diquark $t$-matrix~\cite{Bender:1996bb}.} in Eqs.~\eqref{eq:tscalar} and \eqref{eq:taxial}, respectively, and the homogeneous Bethe-Salpeter vertices read
\begin{align}
\G_s &= \sqrt{Z_s}\,\g_5\,C\,\tau_2\,\beta_A,~ &
\G^{\mu,i}_a &= \sqrt{Z_a}\,\g^\mu\,C\,\tau_i\,\tau_2\,\beta_A,
\label{eq:homogeneousBSvertex}
\end{align}
where $i$ is an isospin index. The pole residues are given by
\begin{align}
\label{eq:Zs}
Z_s^{-1} &= -\frac{1}{2}\,\frac{\pl}{\pl q^2}\,\Pi_{PP}(q^2)\Big\rvert_{q^2 = M_s^2}, \\[0.2ex]
Z_a^{-1} &= -\frac{1}{2}\,\frac{\pl}{\pl q^2}\,\Pi_{VV}(q^2)\Big\rvert_{q^2 = M_a^2},
\end{align}
where $M_s$ and $M_a$ are the scalar and axialvector diquark masses. These pole residues are interpreted as the effective diquark--quark-quark couplings.

\begin{figure}[t]
\centering\includegraphics[width=\columnwidth,clip=true,angle=0]{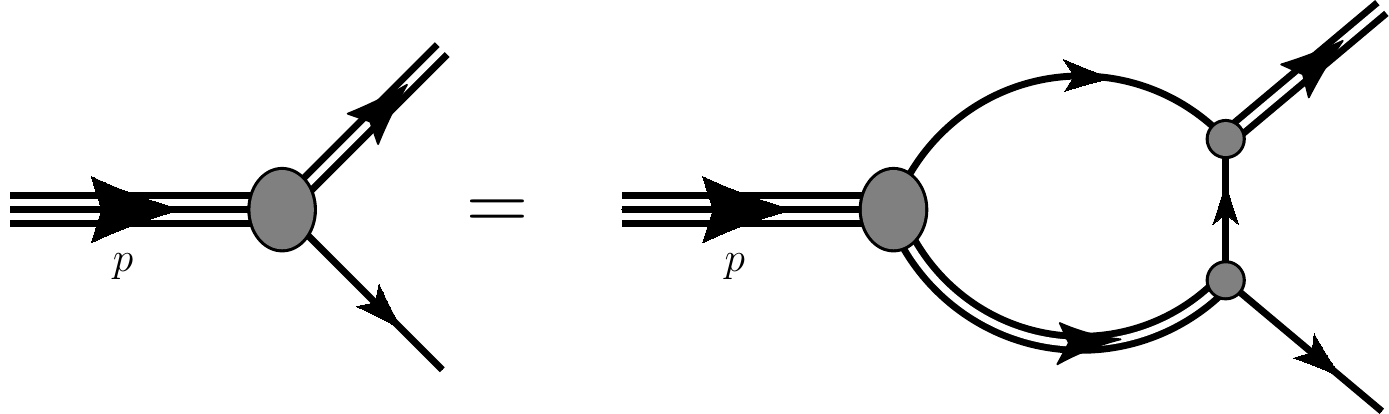}
\caption{Homogeneous Faddeev equation for the nucleon in the NJL model. The single line represents the quark propagator and the double line the diquark propagators. Both scalar and axialvector diquarks are included in these calculations.}
\label{fig:homofaddeev}
\end{figure}

The homogeneous Faddeev equation is illustrated in Fig.~\ref{fig:homofaddeev}, where 
diquark correlations have been made explicit.
The relativistic Faddeev equation in the NJL model has been solved numerically in 
Refs.~\cite{Buck:1992wz,Mineo:1999eq,Mineo:2002bg}, 
where the integrals were regularized using 
the Lepage--Brodsky and transverse momentum cutoff schemes. 
In the proper-time regularization 
scheme used here, solving the Faddeev equation is much more challenging and 
we therefore employ the static approximation to the quark exchange kernel. 
In this approximation the propagator of the exchanged quark becomes 
$S(k) \to -\tfrac{1}{M}$~\cite{Buck:1992wz}. 
The nucleon vertex function then takes the form
\begin{align}
&\G_N(p) = \sqrt{-Z_N}\ \G = \sqrt{-Z_N}\ \begin{bmatrix} \G_s(p) \\ \G_a^{\mu,i}(p) \end{bmatrix}  \no \\
&\hs{0mm}
= \sqrt{-Z_N}
\begin{bmatrix} \a_1 \\ \lf(\a_2\,\frac{p^{\mu}}{M_N}\,\g_5\ + \a_3\,\g^\mu\g_5\rg) \frac{\tau_i}{\sqrt{3}} \end{bmatrix} \chi(t)\,u(p),
\label{eq:faddeevvertex}
\end{align}
where $i$ is an isospin index and $\chi_N(t)$ is the nucleon isospinor:
\begin{align}
\chi\!\lf( \tfrac{1}{2}\rg) &= \begin{pmatrix} 1 \\ 0 \end{pmatrix},  &
\chi\!\lf(-\tfrac{1}{2}\rg) &= \begin{pmatrix} 0 \\ 1 \end{pmatrix}.
\end{align}
The first element in the column vector of Eq.~\eqref{eq:faddeevvertex} 
represents the piece of the nucleon vertex function consisting
of a quark and scalar diquark, while the second element 
represents the quark and axialvector diquark component. 
The nucleon mass is labelled by $M_N$ and the Dirac spinor 
is normalized such that $\bar{u}_N\,u_N =1$. 
$Z_N$ is the nucleon vertex function normalization and $\a_1,~\a_2,~\a_3$ 
are obtained by solving the Faddeev equation. 
After projection onto positive parity, spin one-half and
isospin one-half, the homogeneous Faddeev equation is given by~\cite{Ishii:1995bu}
\begin{align}
\G_N(p,s) = K(p)\, \G_N(p,s),
\label{eq:faddeev}
\end{align}
which in matrix form reads
\begin{align}
\begin{bmatrix} \G_s \\ \G_a^\mu \end{bmatrix}
= \frac{3}{M}
\begin{bmatrix} \Pi_{Ns}     &  \sqrt{3}\g_{\a}\g_5\,\Pi^{\a\b}_{Na}\\
\sqrt{3}\g_5\g^{\mu}\,\Pi_{Ns} & -\g_{\a}\g^{\mu}\,\Pi^{\a\b}_{Na}
\end{bmatrix}
\begin{bmatrix} \G_s \\ \G_{a,\b} \end{bmatrix}.
\end{align}
The quark-diquark bubble diagrams are defined as
\begin{align}
\label{eq:snucleon}
\Pi_{Ns}(p) &= \int \frac{d^4k}{(2\pi)^4}\, \tau_s(p-k)\,S(k), \\
\label{eq:anucleon}
\Pi^{\mu\nu}_{Na}(p) &= \int \frac{d^4k}{(2\pi)^4}\, \tau^{\mu\nu}_a(p-k)\,S(k).
\end{align}
The vertex normalization of Eq.~\eqref{eq:faddeevvertex} is given by
\begin{align}
Z_N = \lf[\ol{\G}\,\frac{\pl\,\Pi_N(p)}{\pl p^2}\,\G\rg]_{p^2=M_N^2}^{-1},
\end{align}
where
\begin{align}
\Pi_N(p) = \begin{bmatrix} \Pi_{Ns}(p)  &  0 \\ 0 & \Pi^{\a\b}_{Na}(p) \end{bmatrix}.
\label{eq:nucleon_bubble_matrix}
\end{align}

Regulating expressions such as those in Eqs.~\eqref{eq:snucleon} 
and \eqref{eq:anucleon} using
the proper-time scheme is tedious. Therefore, to render the Faddeev equation and 
form factor calculations tractable we make the pole approximation 
to the meson and diquark $t$-matrices, for example, 
Eqs.~\eqref{eq:tscalar} and \eqref{eq:taxial} become
\begin{align}
\label{eq:scalarpropagatorpoleform}
\tau_s(q) &\to  -\frac{i\,Z_s}{q^2 - M_s^2+i\,\ve}, \\
\label{eq:axialpropagatorpoleform}
\tau^{\mu\nu}_a(q) &\to  -\frac{i\,Z_a}{q^2 - M_a^2+i\,\ve}
               \lf(g^{\mu\nu} - \frac{q^{\mu}q^{\nu}}{M_a^2}\rg).
\end{align}
Similar expressions are obtained in the meson sector.

In summary, the model parameters consist of the two regularization scales $\L_{IR}$ and $\L_{UV}$, the dressed quark mass $M$,\footnote{Alternatively, one could specify a current quark mass, as one determines the other through the gap equation.} and the Lagrangian coupling constants $G_\pi$, $G_\rho$, $G_\omega$, $G_s$, $G_a$. The infrared regularization scale is associated with confinement and therefore should be of the order $\L_{\text{QCD}}$, and we  choose $\L_{IR} = 0.240\,$GeV and for the constituent quark mass take $M = 0.4\,$GeV. The physical pion mass ($m_\pi = 140\,$MeV) and decay constant ($f_\pi = 92\,$MeV) determine $\L_{UV}$ and $G_\pi$. The physical masses of the rho ($m_\rho = 770\,$MeV) and omega ($m_\o = 782\,$MeV) mesons constrain $G_\rho$ and $G_\omega$, respectively, while the physical nucleon ($M_N = 940\,$MeV) and $\Delta$ ($M_\Delta = 1232\,$MeV) baryon masses determine $G_s$ and $G_a$.\footnote{The relativistic Faddeev equation for the Delta baryon is discussed, for example, in Ref.~\cite{Ishii:1995bu}.}  Numerical values are given in Tab.~\ref{tab:parameters}.

Using the parameters given in Tab.~\ref{tab:parameters}, we obtain the following results for the residues of the two-body $t$-matrices: $Z_\pi = 17.9$, $Z_\rho = 6.96$, $Z_\o = 6.63$, $Z_s = 11.1$ and $Z_a = 6.73$. For the nucleon vertex function of Eq.~\eqref{eq:faddeevvertex} we find $Z_N = 28.1$ and $(\a_1,\,\a_2,\,\a_3) = (0.55,\,0.05,\,-0.40)$, where the scalar and axialvector diquark masses are $M_s = 0.768\,$GeV and $M_a = 0.903\,$GeV.

\begin{table}[tbp]
\addtolength{\tabcolsep}{5.0pt}
\addtolength{\extrarowheight}{2.2pt}
\begin{tabular}{cccccccc}
\hline\hline
$\L_{IR}$ & $\L_{UV}$ & $M$ & $G_\pi$ & $G_\rho$ & $G_\o$ & $G_s$  & $G_a$ \\
\hline
0.240    & 0.645    & 0.4 & 19.0   & 11.0    & 10.4   & 5.8   & 4.9    \\
\hline\hline
\end{tabular}
\caption{Model parameters constrained to reproduce the physical 
pion, rho and omega masses;
the pion decay constant; and the nucleon and delta baryon masses. 
The infrared regulator
and the dressed quark mass are assigned their values \textit{a priori}.
The regularization parameters
and dressed quark mass are in units of GeV, 
while the couplings are in units of GeV\,$^{-2}$.} 
\label{tab:parameters}
\end{table}

\section{Nucleon Electromagnetic Current \label{sec:nucleon_current}}
The electromagnetic current of an on-shell nucleon, expressed in terms of the Dirac and Pauli form factors, has the form
\begin{align}
&j^\mu_{\lambda' \, \lambda}(p',p) 
= \lf<p',\,\l'\lf|J^\mu_{\text{em}}\rg|p,\,\l\rg> \no \\
&
= u(p',\,\l')\lf[\g^\mu\,F_{1}(Q^2) + \frac{i\s^{\mu\nu}q_\nu}{2\,M_N}\,F_{2}(Q^2)\rg]u(p,\,\l),
\end{align}
where $q = p' - p$ is the 4-momentum transfer, $Q^2 \equiv -q^2$ and $\lambda$, $\lambda'$ represent
the initial and final nucleon helicity respectively. The nucleon's electric and magnetic Sachs form factors~\cite{Ernst:1960zza}, 
which diagonalize the Rosenbluth cross-section, are then given by
\begin{align}
G_E(Q^2) &= F_{1}(Q^2) - \frac{Q^2}{4\,M_N^2}\,F_{2}(Q^2), \\
G_M(Q^2) &= F_{1}(Q^2) + F_{2}(Q^2).
\end{align}

Hadron form factors can be decomposed into a sum over the quark charges multiplied by quark sector form factors, such that
\begin{align}
F_h(Q^2) = \sideset{}{_q}\sum \, e_q\, F_h^q(Q^2).
\label{eq:quark_sectorform_factors}
\end{align}
The quark sector form factors $F^q_h(Q^2)$ represent the contribution of the current quarks of flavour $q$ to the total hadron form factor $F_h(Q^2)$. The proton and neutron form factors expressed in terms of quark sector form factors read
\begin{align}
\label{eq:protonflavoursector}
F_{ip}(Q^2) &= e_u\,F^u_{ip}(Q^2) + e_d\,F^d_{ip}(Q^2) + \ldots \\
\label{eq:neutronflavoursector}
F_{in}(Q^2) &= e_u\,F^u_{in}(Q^2) + e_d\,F^d_{in}(Q^2) + \ldots
\end{align}
where $i = 1,\,2$. Note that in light of the experimental discovery that the strange quarks contribute very little to the nucleon electromagnetic form factors~~\cite{Young:2006jc,Aniol:2005zf,Armstrong:2005hs,Baunack:2009gy}, we will neglect their contribution to Eqs.~\eqref{eq:protonflavoursector} and \eqref{eq:neutronflavoursector}. Assuming equal $u$ and $d$ current quark masses and neglecting electroweak corrections, the $u$ and $d$ quark sector form factors of the nucleon must satisfy the charge symmetry constraints:
\begin{align}
F^d_{in}(Q^2) = F^u_{ip}(Q^2) \quad \text{and} \quad F^u_{in}(Q^2) = F^d_{ip}(Q^2).
\end{align}
Experimentally, if electroweak and heavy quark effects are small, the $u$ and $d$ quark sector form factors are given accurately by
\begin{align}
\label{eq:flavourseparation}
F^u_{ip} &= 2\,F_{ip} + F_{in},  &  F^d_{ip} &= F_{ip} + 2\,F_{in}.
\end{align}
Recent accurate data for the neutron form factors has enabled a precise determination of the quark sector proton form factors~\cite{Cates:2011pz}. We will discuss results for these quark sector form factors in Sect.~\ref{sec:nucleon_results}.

The slope of an electromagnetic form factor at $Q^2 = 0$ is a measure of either the squared rms charge or magnetic radius of a hadron. Unless stated otherwise all squared rms radii are defined by
\begin{align}
\lf<r^2\rg> = -\frac{6}{\eta}\ \frac{\pl\, f(Q^2)}{\pl Q^2}\bigg\lvert_{Q^2 = 0} ~~ \Big| ~~
\eta =\begin{cases}
      1     & \text{if~}f(0)=0,\\[0.4ex]
      f(0)  & \text{if~}f(0)\neq 0,
      \end{cases}
\label{eq:radii}
\end{align}
where $f(Q^2)$ is an arbitrary form factor. This definition reproduces the standard nucleon results for the charge and magnetic radii defined by the Sachs form factors:
\begin{align}
\label{eq:charge_radius}
\lf<r_E^2\rg> &= -6\ \ \frac{\pl\, G_E(Q^2)}{\pl Q^2}\Big\lvert_{Q^2 = 0}, \\[0.2ex]
\label{eq:magnetic_radius}
\lf<r_M^2\rg> &= -\frac{6}{G_M(0)}\ \frac{\pl\, G_M(Q^2)}{\pl Q^2}\Big\lvert_{Q^2 = 0},
\end{align}
but also generalizes to radii defined with respect to the Dirac and Pauli form factors and quark sector form factors. Hadronic radii, in units of fm, will be obtained from the result of Eq.~\eqref{eq:radii} using
\begin{align}
r \equiv \text{sign}\lf(\lf<r^2\rg>\rg)\ \sqrt{\lf|\lf<r^2\rg>\rg|}.
\end{align}

\begin{figure}[t]
\centering\includegraphics[width=\columnwidth,clip=true,angle=0]{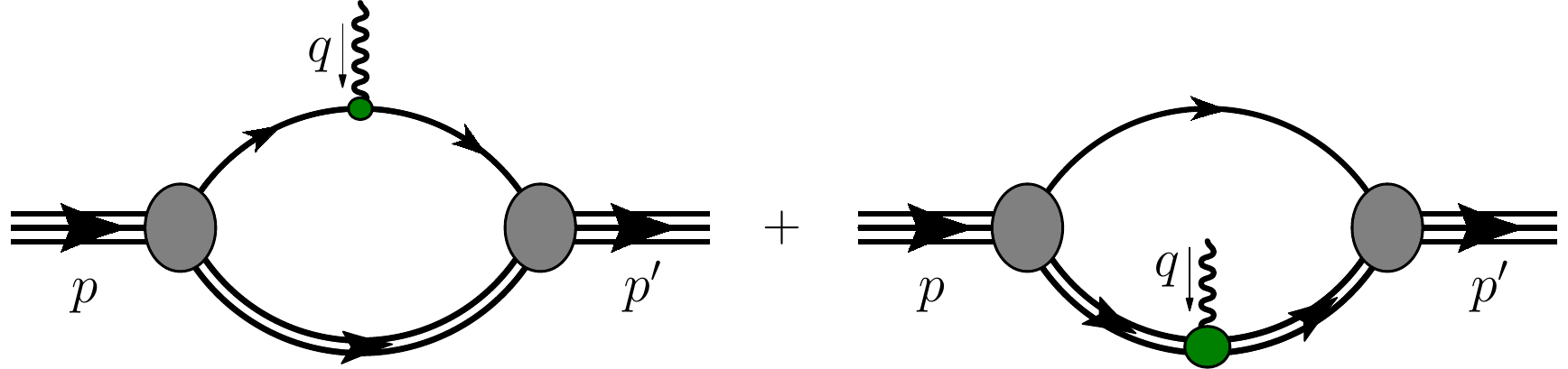}
\caption{(Colour online) Feynman diagrams representing the nucleon electromagnetic current. The diagram on the left is called the \textit{quark diagram} and the diagram on the right the \textit{diquark diagram}. In the diquark diagram the photon interacts with each quark inside the non-pointlike diquark.}
\label{fig:nucleon_em_current_feynman_diagrams}
\end{figure}

To calculate the nucleon electromagnetic current and therefore the Dirac and Pauli form factors, one must know the manner in which the nucleon described in Sect.~\ref{sec:NJL} couples to the photon, guaranteeing electromagnetic gauge invariance. The necessary Feynman diagrams are illustrated in Fig.~\ref{fig:nucleon_em_current_feynman_diagrams} and a proof of gauge invariance is given in App.~\ref{sec:gaugeinvariance}. We include both scalar and axialvector diquarks in our nucleon wave function and therefore the diagrams in Fig.~\ref{fig:nucleon_em_current_feynman_diagrams}  represent six distinct Feynman diagrams. The diagram on the left, referred to as the \textit{quark diagram}, represents the processes where the photon couples to a dressed quark with either a scalar or axialvector diquark as a spectator. The \textit{diquark diagram}, on the right in Fig.~\ref{fig:nucleon_em_current_feynman_diagrams}, represents four Feynman diagrams; the photon can couple to a scalar diquark, an axialvector diquark or cause a transition between these two diquark states. Importantly, in the \textit{diquark diagram} the photon couples to the quarks inside each diquark, thereby resolving internal diquark structure and resulting in, for example, diquarks with a finite size. The coupling of a photon to a dressed quark and to the diquarks is discussed in Sects.~\ref{sec:QPV} and \ref{sec:diquark_results}.

\section{Quark--Photon Vertex \label{sec:QPV}}
\begin{figure}[t]
\centering\includegraphics[width=\columnwidth,clip=true,angle=0]{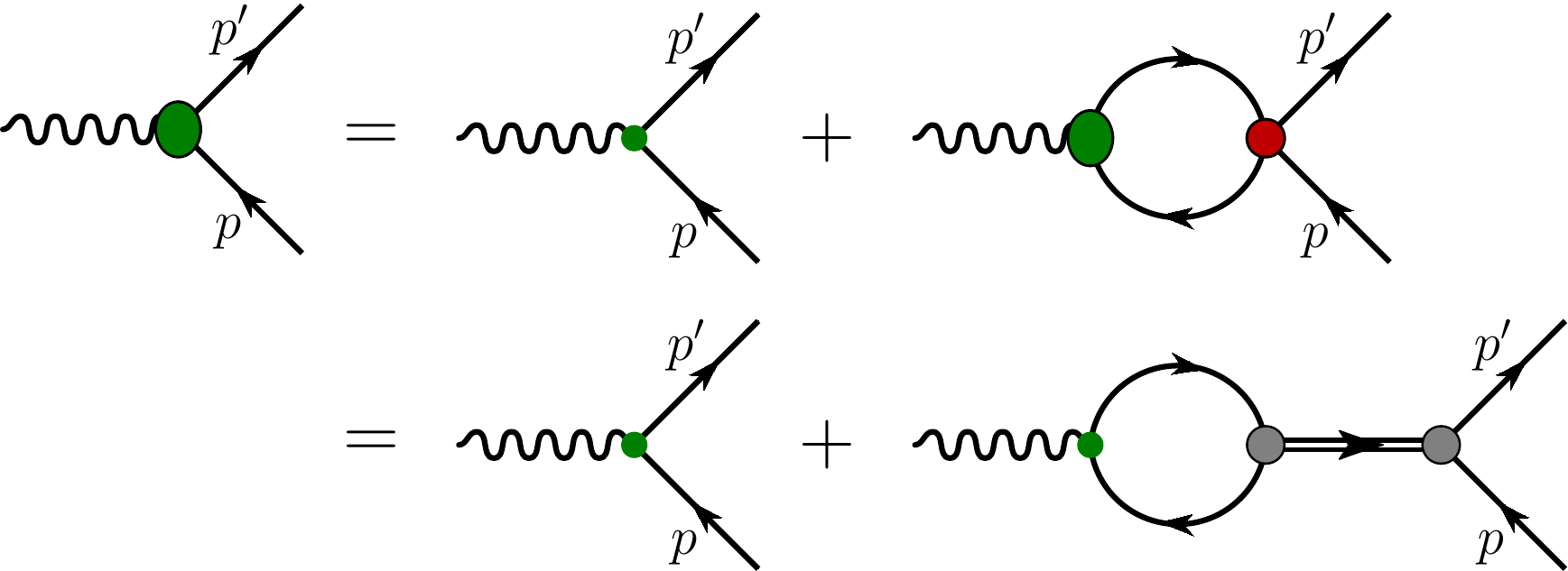}
\caption{(Colour online) Inhomogeneous Bethe-Salpeter equation whose solution gives the quark-photon vertex, represented as the large shaded oval. The small dot is the inhomogeneous driving term, while the shaded circle is the $\bar{q}q$ interaction kernel given in Eq.~\eqref{eq:qbarqkernel}. Only the $\rho$ and $\omega$ interaction channels contribute. This integral equation can equivalently be represented using the elementary quark-photon interaction and the $\rho$ and $\omega$ $t$-matrices, given in Eqs.~\eqref{eq:trho} and \eqref{eq:tomega}. This case is depicted by the second equality.}
\label{fig:quarkphotonvertex}
\end{figure}

The quark-photon vertex in the NJL model, and other field theoretic approaches, is given by the solution to an inhomogeneous BSE. The NJL model version of this equation, consistent with the truncation used in the gap and BSEs discussed Sect.~\ref{sec:NJL}, is represented diagrammatically in Fig.~\ref{fig:quarkphotonvertex}. The large oval represents the quark-photon vertex, $\,\L^\mu_{\g Q}(p',p)$, the 4-fermion interaction kernel is given in Eq.~\eqref{eq:qbarqkernel} and the elementary vertex, which gives the inhomogeneous driving term, has the form $\g^\mu\,\hat{Q}$ (where $\hat{Q}$ is the quark charge operator). The second equality in Fig.~\ref{fig:quarkphotonvertex} expresses this equation in an equivalent form using the $\bar{q}q$ $t$-matrices.

The quark charge operator is
\begin{align}
\hat{Q} &= \begin{pmatrix} e_u & 0 \\ 0 & e_d\end{pmatrix}
=\frac{1}{6} + \frac{\tau_3}{2},
\end{align}
where $e_u = \tfrac{2}{3}$ and $e_d = -\tfrac{1}{3}$ are the $u$ and $d$ quark charges. The quark-photon vertex therefore has both an isoscalar and isovector component, which may in general be expressed in the form
\begin{align}
\L^\mu_{\g Q}(p',p) 
&= \frac{1}{6}\,\L^\mu_{\o}(p',p) + \frac{\tau_3}{2}\,\L^\mu_{\rho}(p',p).
\label{eq:quark_photon_vertex_iso}
\end{align}
The quark-photon vertex, separated into flavour sectors defined by the dressed quarks, reads
\begin{align}
\L^\mu_{\g Q}(p',p) 
= \L^\mu_{U}(p',p)\ \frac{1+\tau_3}{2} + \L^\mu_{D}(p',p)\ \frac{1-\tau_3}{2}.
\label{eq:quark_photon_vertex}
\end{align}
In general each dressed quark component of the quark-photon vertex contains contributions from both the $u$ and $d$ current quarks. This will prove important when we consider quark sector form factors and associated charge symmetry constraints. Note that throughout this manuscript we use a capital $Q = (U,\,D)$ to indicate that an object is associated with dressed quarks and a lowercase $q = (u,\,d)$ to represent the current quarks of the NJL and QCD Lagrangians.

The quark-photon vertex has in general 12 Lorentz structures~\cite{Maris:1999bh}, 4 longitudinal and 8 pieces transverse to the photon momentum, where each Lorentz structure is accompanied by a scalar function of the three variables $q^2$, $p'^2$ and $p^2$.\footnote{The 12 Lorentz structures in the quark-photon vertex are not all independent, since, for example, the Ward--Takahashi identity and time reversal invariance place additional constraints.}  The standard NJL $\bar{q}q$ interaction kernel, as employed in Sect.~\ref{sec:NJL} for the gap and BSE equations, is momentum independent which implies that the quark-photon vertex can only depend on the momentum transfer $q = p'-p$, not $p'$ and $p$ separately. Therefore, in this work, the contributions to the  vertex functions of Eq.~\eqref{eq:quark_photon_vertex_iso} from the NJL BSE, take the form
\begin{align}
\label{eq:quark_photon_vertex_parts_omega}
\L^{(\text{bse})\mu}_{\o}(q)  &= \g^\mu + \lf(\!\g^\mu - \frac{q^\mu\sh{q}}{q^2}\rg)\!\hat{F}_{1\o}(q^2) + \frac{i\s^{\mu\nu}q_\nu}{2\,M} F_{2\o}(q^2), \\
\label{eq:quark_photon_vertex_parts_rho}
\L^{(\text{bse})\mu}_{\rho}(q) &= \g^\mu + \lf(\!\g^\mu - \frac{q^\mu\sh{q}}{q^2}\rg)\!\hat{F}_{1\rho}(q^2) + \frac{i\s^{\mu\nu}q_\nu}{2\,M} F_{2\rho}(q^2).
\end{align}
With the quark propagator of Eq.~\eqref{eq:quarkpropagator}, these results satisfy the Ward--Takahashi identity:
\begin{align}
q_\mu\,\L^\mu_{\g Q}(p',p) &= \hat{Q}\lf[S^{-1}(p') - S^{-1}(p)\rg],
\end{align}
demanded by $U(1)$ vector gauge invariance.

Current conservation at the hadron level implies that the $q^\mu\sh{q}/q^2$ term in Eqs.~\eqref{eq:quark_photon_vertex_parts_omega} and \eqref{eq:quark_photon_vertex_parts_rho} cannot contribute to hadron form factors. We therefore write our effective vertex as
\begin{align}
\L^{(\text{bse})\mu}_{i}(q)  &= \g^\mu\,F_{1i}(q^2) + \frac{i\s^{\mu\nu}q_\nu}{2\,M}\, F_{2i}(q^2),
\label{eq:onshellquarkvertex}
\end{align}
where $i = (\o,\,\rho)$ and $F_{1i}(q^2) = 1 + \hat{F}_{1i}(q^2)$. This vertex has the same form as the electromagnetic current for an on-shell spin-half fermion. For a pointlike quark $F_{1\o}(q^2) = 1 = F_{1\rho}(q^2)$ and $F_{2\o}(q^2) = 0 = F_{2\rho}(q^2)$. However, interactions in the NJL model not only dynamically generate a dressed quark mass but also generate non-trivial dressed quark form factors.

The inhomogeneous BSE for the quark-photon vertex, depicted in Fig.~\ref{fig:quarkphotonvertex}, has the form
\begin{align}
&\L^\mu_{\g Q}(p',p) = \g^\mu\lf(\frac{1}{6} + \frac{\tau_3}{2}\rg) + \sideset{}{_\O}\sum \,K_\O\ \O   \no \\
&\hs{6mm}
\times
\ i\!\int \frac{d^4k}{(2\pi)^4}\ \mathrm{Tr}\lf[\bar{\O}\,S(k+q)\,\L^\mu_{\g Q}(p',p)\,S(k)\rg],
\label{eq:quark_photon_bse}
\end{align}
where $\sum_\O \,K_\O\ \O_{\a\b}\,\bar{\O}_{\l\ve}$ represents the interaction kernel given in Eq.~\eqref{eq:qbarqkernel}. The Dirac and isospin structure of $\L^\mu_{\g Q}(p',p)$, given in Eqs.~\eqref{eq:quark_photon_vertex_iso}, \eqref{eq:quark_photon_vertex_parts_omega} and \eqref{eq:quark_photon_vertex_parts_rho}, implies that of the interaction channels in Eq.~\eqref{eq:qbarqkernel} only the isovector--vector, $-2i\,G_\rho\lf(\g_\mu\vec{\tau}\rg)_{\a\b}\lf(\g^\mu\vec{\tau}\rg)_{\g\d}$, and isoscalar--vector, $-2i\,G_\o\lf(\g_\mu\rg)_{\a\b}\lf(\g^\mu\rg)_{\g\d}$, pieces can contribute.

\begin{figure}[t]
\subfloat{\centering\includegraphics[width=\columnwidth,clip=true,angle=0]{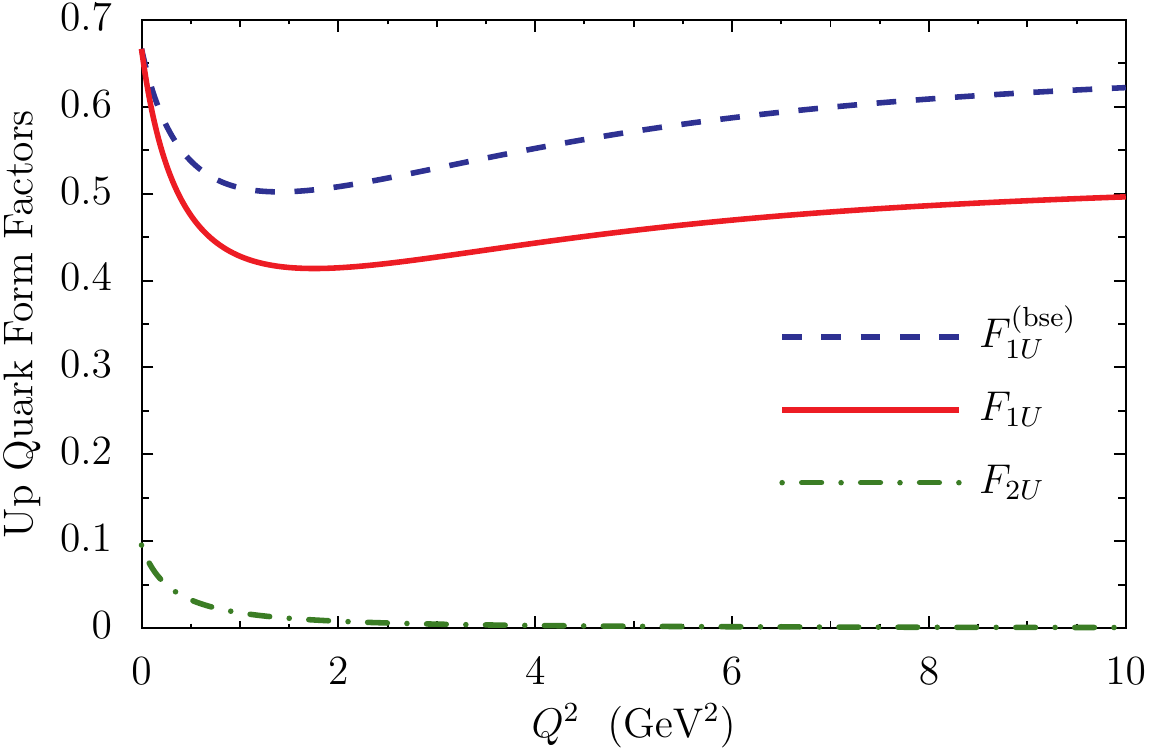}}\\
\subfloat{\centering\includegraphics[width=\columnwidth,clip=true,angle=0]{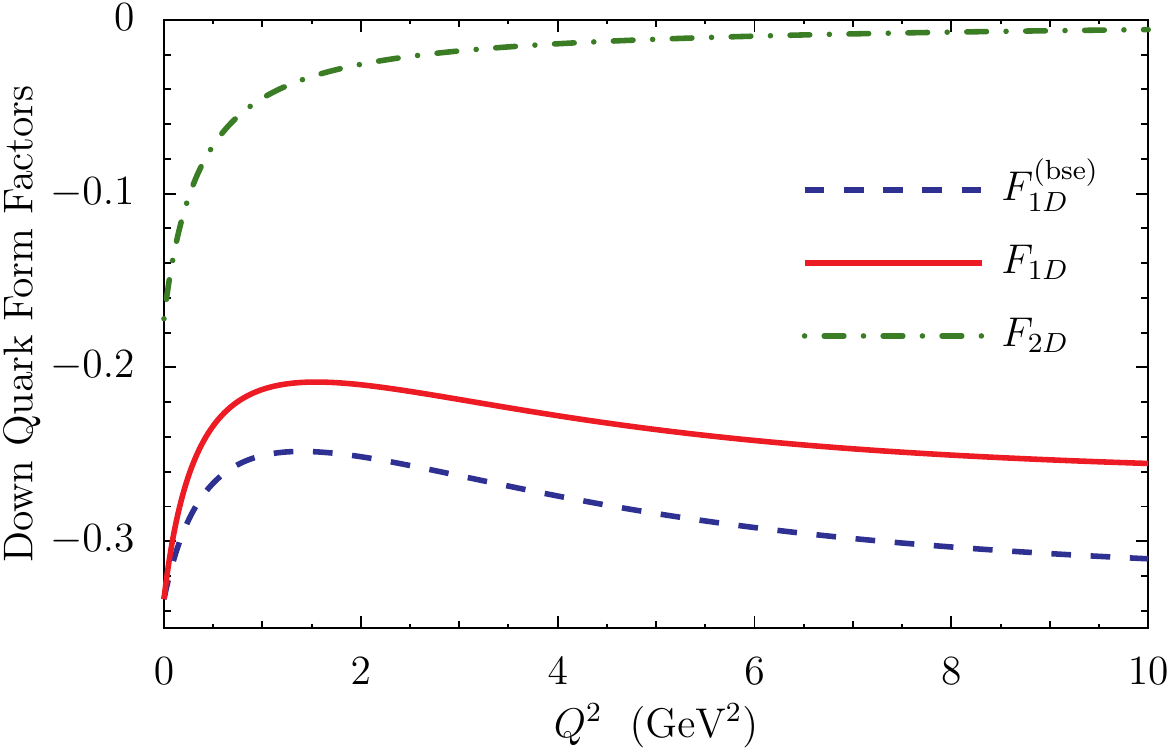}}
\caption{(Colour online) \textit{Upper panel:} Dressed up quark form factors: the dashed line is the Dirac form factor obtained from the BSE of Eq.~\eqref{eq:quark_photon_bse}; the solid and dash-dotted lines are respectively the Dirac and Pauli form factors generated by also including the pion loop corrections illustrated in Fig.~\ref{fig:quarkpionphotonvertex}. \textit{Lower panel:} Dressed down quark form factors: each curve represents an analogous form factor to those in the upper panel.}
\label{fig:constituentquarkformfactors}
\end{figure}

The dressed quark form factors obtained from the inhomogeneous BSE, associated with the electromagnetic current of Eq.~\eqref{eq:onshellquarkvertex}, are
\begin{align}
F_{1i}(q^2) = \frac{1}{1 + 2\,G_i\,\Pi_{VV}(q^2)}, \hs{12mm}
F_{2i}(q^2) = 0, 
\label{eq:bseformfactors}
\end{align}
where $i = \o,\,\rho$. Comparison with Eqs.~\eqref{eq:trho} and \eqref{eq:tomega} indicate that $F_{1\o}$ and $F_{1\rho}$ have a pole at $q^2 = m_\o^2$ and $m_\rho^2$, respectively. The NJL BSE kernel of Eq.~\eqref{eq:qbarqkernel} does not generate Pauli form factors for the dressed quarks because it does not
include the tensor--tensor 4-fermion interaction. The dressed up and down quark form factors given by the BSE therefore read
\begin{align}
\label{eq:bseconstituent}
F^{\text{(bse)}}_{1Q}(Q^2) &=  \frac{1}{6}\,F_{1\o}(Q^2) \pm \frac{1}{2}\,F_{1\rho}(Q^2),
\end{align}
where the plus sign is associated with a dressed up quark. The superscript (bse) indicates that these form factors are obtained solely from the BSE. 

Results for the dressed quark BSE form factors are illustrated as the dashed lines in Figs.~\ref{fig:constituentquarkformfactors}. A notable feature of these results is that they do not drop to zero as $Q^2 \to \infty$, but instead behave as
\begin{align}
F^{\text{(bse)}}_{1U}(Q^2) &\stackrel{Q^2 \to \infty}{=} e_u, & F^{\text{(bse)}}_{1D}(Q^2) &\stackrel{Q^2 \to \infty}{=} e_d,
\end{align}
signifying that at infinite $Q^2$ the photon interacts with a bare current quark. This result is consistent with QCD expectations based on asymptotic freedom. 

\begin{figure}[tbp]
\centering\includegraphics[width=0.6\columnwidth,clip=true,angle=0]{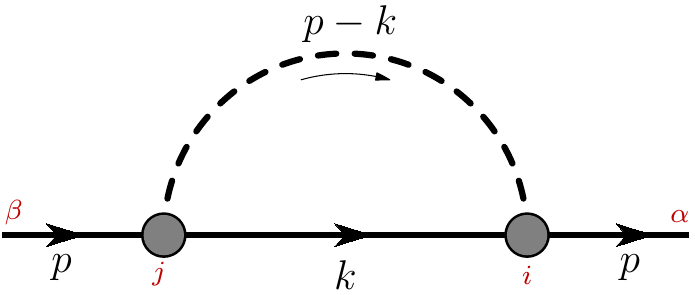}
\caption{(Colour online) Pion loop contribution to the dressed quark self-energy. The pion couples to the dressed quark via $\g_5\,\tau_i$ and the pion $t$-matrix is approximated by its pole form.}
\label{fig:quarkpionloop}
\end{figure}

Pion loop corrections to the quark-photon vertex will also be considered and treated as a perturbation to the dressed quark form factors obtained from the BSE, as given in Eqs.~\eqref{eq:bseformfactors} and \eqref{eq:bseconstituent}.
In this case the dressed quark propagator receives an additional self-energy correction which is illustrated in Fig.~\ref{fig:quarkpionloop}.\footnote{Chiral symmetry as expressed in the the NJL Lagrangian of Eq.~\eqref{eq:njllagrangian} demands that the sigma meson be included also, however since the sigma has charge zero and (in this work) $m_\pi/m_\s \simeq 0.18$, these additional correction are small and will not be included.} In addition to the pionic self-energies on the dressed quarks, pion exchange between quarks should also be included in the two-body kernels that enter the Bethe-Salpeter and Faddeev equations. However, it is straightforward to show that in the limit where the nucleon and $\Delta$ are mass degenerate, including only self-energy correction on the dressed quarks yields essentially the correct leading non-analytic behaviour of the electromagnetic form factors as a function of quark mass. Further, in form factor calculations diagrams with a photon coupling to an exchanged pion do not contribute because of the cancellation between $\pi^+$ and $\pi^-$ exchange. This self-energy is evaluated using a pole approximation, where the external quark is assumed on-mass-shell. The pion loop therefore shifts the dressed quark mass by a constant, giving a quark propagator of the form
\begin{align}
\tilde{S}(k) = Z\,S(k), \qquad Z = 1 + \frac{\pl\, \S(p)}{\pl \sh{p}}\bigg\rvert_{\sh{p}=M},
\end{align}
where $S(k)$ is the usual Feynman propagator for a dressed quark of mass\footnote{When including pion loops on the dressed quarks we renormalize the $G_\pi$ coupling in the NJL Lagrangian to keep the dressed quark mass fixed.} $M$ and the self-energy reads\footnote{The pion--quark-quark vertex in Fig.~\ref{fig:quarkpionloop} can be read directly from the pion $t$-matrix, given by Eq.~\eqref{eq:piontmatrix}, and takes the form $\g_5\,\tau_i$. A pseudovector component to the vertex would be generated through $\pi$--$a_1$ mixing in the BSE kernel, however the strength of this vertex is suppressed by $m_\pi/m_{a_1} \sim 0.1$ relative to the dominant pseudoscalar component. Therefore, we do not include a $\pi$--$a_1$ mixing in the pion--quark-quark vertex.}
\begin{align}
\S(p) = -\int \frac{d^4k}{(2\pi)^2}\ \g_5\,\tau_i\,S(p-k)\,\g_5\,\tau_i\,\tau_\pi(k).
\label{eq:quarkselfenergypion}
\end{align}
When evaluating $\S(p)$ the reduced pion $t$-matrix is approximated by its pole form, that is
\begin{align}
\tau_\pi(k) \to \frac{i\,Z_\pi}{k^2 - m_\pi^2 + i\ve}.
\end{align}
The quark wave function renormalization factor, $Z$, represents the probability to strike a dressed quark without the pion cloud and is essential to maintain charge conservation. Using the parameters in Tab.~\ref{tab:parameters} gives $Z = 0.80$.

\begin{figure}[t]
\centering\includegraphics[width=\columnwidth,clip=true,angle=0]{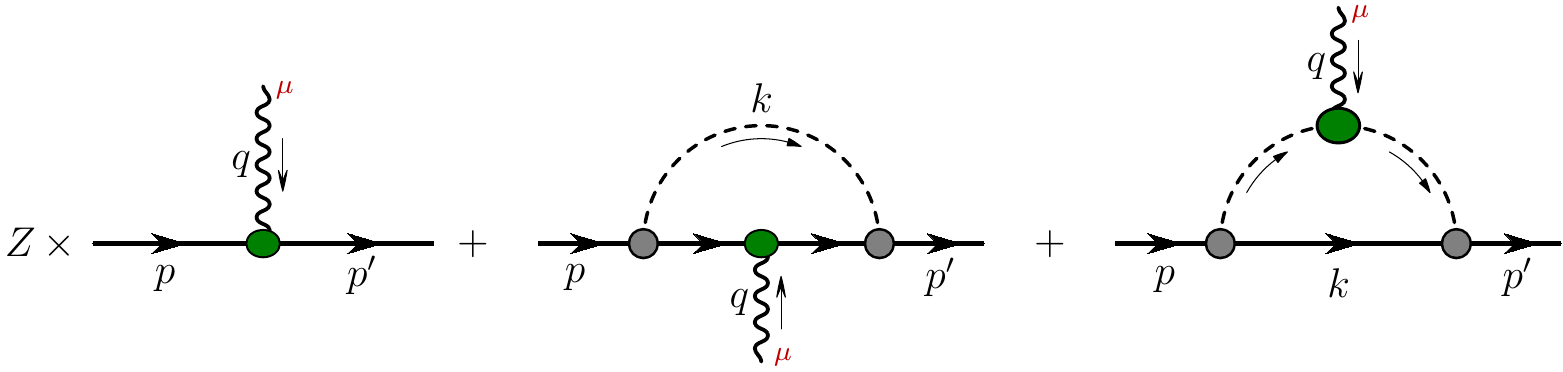}
\caption{(Colour online) Pion loop contributions to the quark-photon vertex. The quark wave function renormalization factor $Z$ represents the probability of striking a dressed quark without a pion cloud. In the first two diagrams the photon couples to the dressed quark with a vertex of the general form given by Eq.~\eqref{eq:quark_photon_vertex_iso}; and defined by Eqs.~\eqref{eq:onshellquarkvertex} and \eqref{eq:bseformfactors}. The shaded oval in the third diagram represents the quark--pion vertex, which we approximate by its pole form. It is therefore given by $\lf(\ell'+\ell\rg)F_\pi(Q^2)$,  where $F_\pi(Q^2)$ is the usual pion form factor (see Eq.~\eqref{eq:pionformfactor} and associated discussion).}
\label{fig:quarkpionphotonvertex}
\end{figure}

\begin{figure}[tbp]
\centering\includegraphics[width=\columnwidth,clip=true,angle=0]{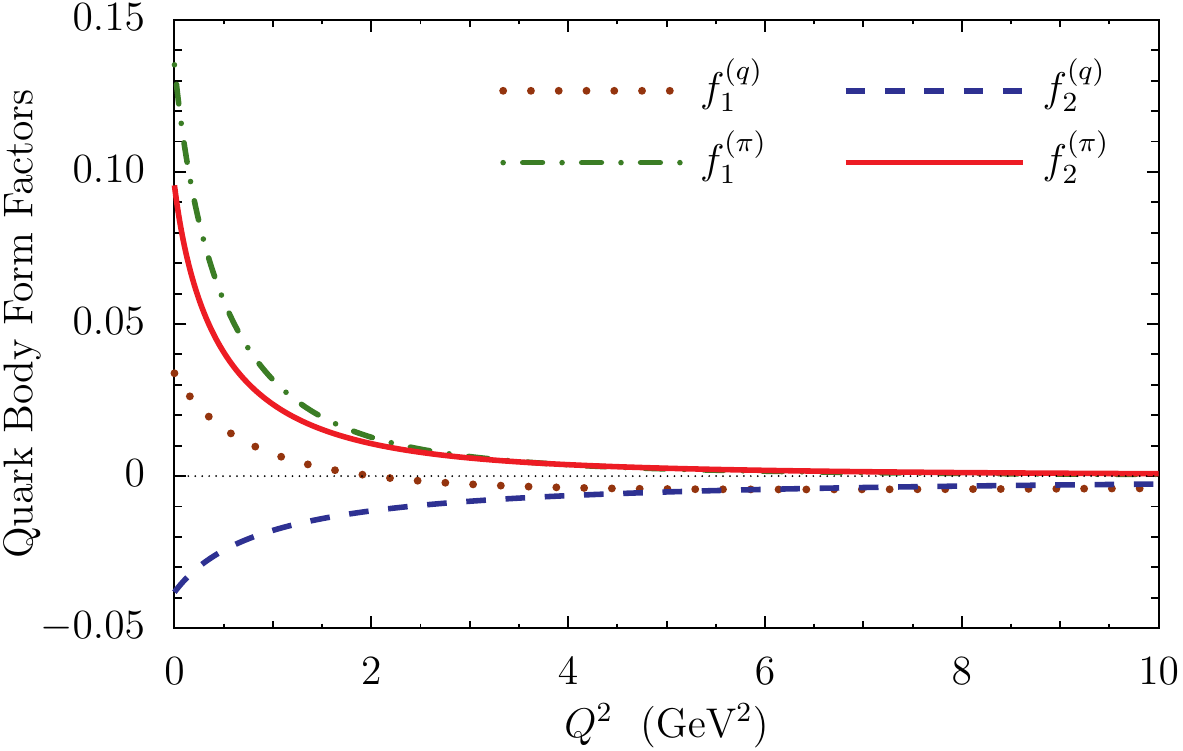}
\caption{(Colour online) Dressed quark body form factors associated with the pion loop corrections, where $f_{1}^{(q)}$ and $f_{2}^{(q)}$ originate from the second diagram in Fig.~\ref{fig:quarkpionphotonvertex} and $f_{1}^{(\pi)}$ and $f_{2}^{(\pi)}$ from the third diagram.}
\label{fig:quarkbodyformfactor}
\end{figure}

The quark electromagnetic current, including pion loops, is illustrated in Fig.~\ref{fig:quarkpionphotonvertex}. Evaluating this current between on-shell constituent quarks, gives for the dressed quark sector currents of Eq.~\eqref{eq:quark_photon_vertex}:
\begin{align}
\L_Q^\mu(p',p) &= \g^\mu\,F_{1Q}(Q^2) + \frac{i\s^{\mu\nu}q_\nu}{2\,M}\,F_{2Q}(Q^2),
\label{eq:dressedquarkcurrent}
\end{align}
where $Q =(U,\,D)$. The dressed quark form factors read
\begin{align}
\label{eq:f1U_quarkpion}
&F_{1U} = Z\lf[\tfrac{1}{6}\,F_{1\o}+\tfrac{1}{2}\,F_{1\rho}\rg]
+ \lf[F_{1\o}-F_{1\rho}\rg]f_{1}^{(q)} + F_{1\rho}\, f_{1}^{(\pi)}, \\
&F_{1D} =  Z\lf[\tfrac{1}{6}\,F_{1\o}-\tfrac{1}{2}\,F_{1\rho}\rg]
+ \lf[F_{1\o}+F_{1\rho}\rg]f_{1}^{(q)} - F_{1\rho}\, f_{1}^{(\pi)}, \\
\label{eq:f2U_quarkpion}
&F_{2U} =  \lf[F_{1\o}-F_{1\rho}\rg] f_{2}^{(q)} + F_{1\rho}\, f_{2}^{(\pi)}, \\
\label{eq:f2D_quarkpion}
&F_{2D} =  \lf[F_{1\o}+F_{1\rho}\rg] f_{2}^{(q)} - F_{1\rho}\, f_{2}^{(\pi)},
\end{align}
where the $Q^2$ dependence of each form factor has been omitted. The body form factors, $f_{1}^{(q)}$ and $f_{2}^{(q)}$, originate from the second diagram in Fig.~\ref{fig:quarkpionphotonvertex}, while $f_{1}^{(\pi)}$ and $f_{2}^{(\pi)}$ are the body form factors from the third diagram, which also contain the pion body form factor (see discussion associated with  Eq.~\eqref{eq:pionformfactor}). These body form factors are illustrated in Fig.~\ref{fig:quarkbodyformfactor}. When evaluating the pion loop diagrams in Figs.~\ref{fig:quarkpionloop} and \ref{fig:quarkpionphotonvertex} we use the 
proper-time regularization scheme, however, in this case the pions should not be confined and we therefore set $\L_{IR} =0\,$GeV. This procedure guarantees that the leading-order non-analytic behaviour of the hadron form factors as a function of the pion mass is retained.

\begin{figure}[t]
\subfloat{\centering\includegraphics[width=\columnwidth,clip=true,angle=0]{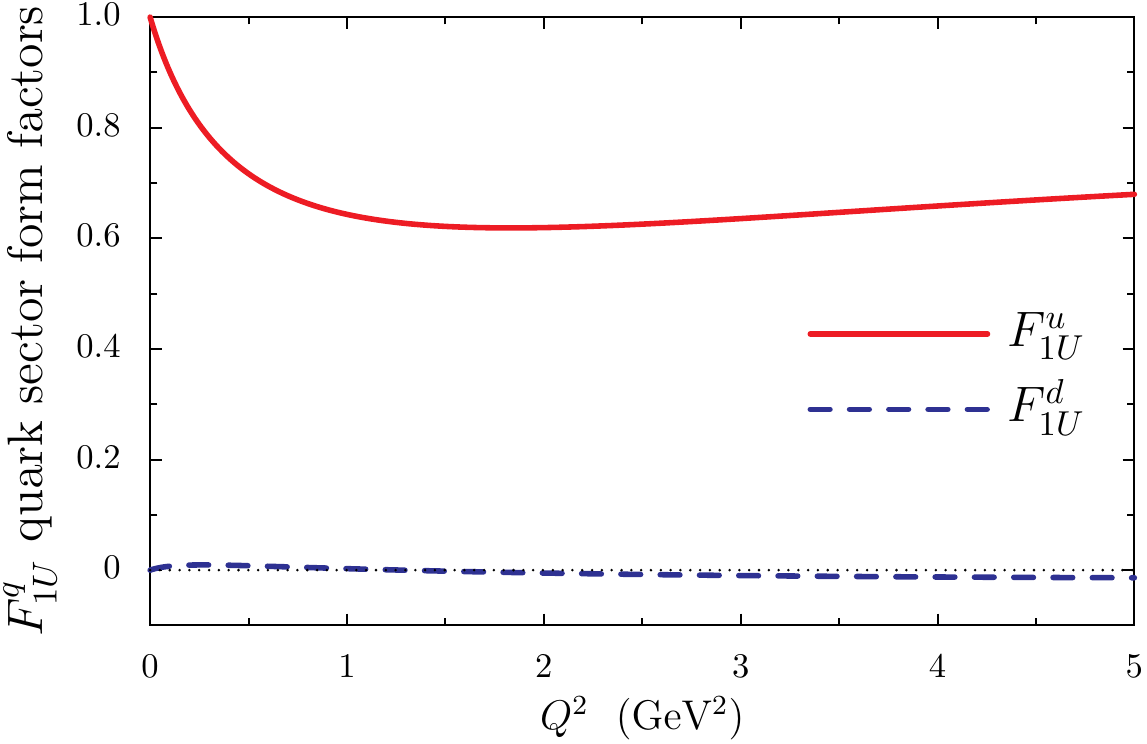}}\\
\subfloat{\centering\includegraphics[width=\columnwidth,clip=true,angle=0]{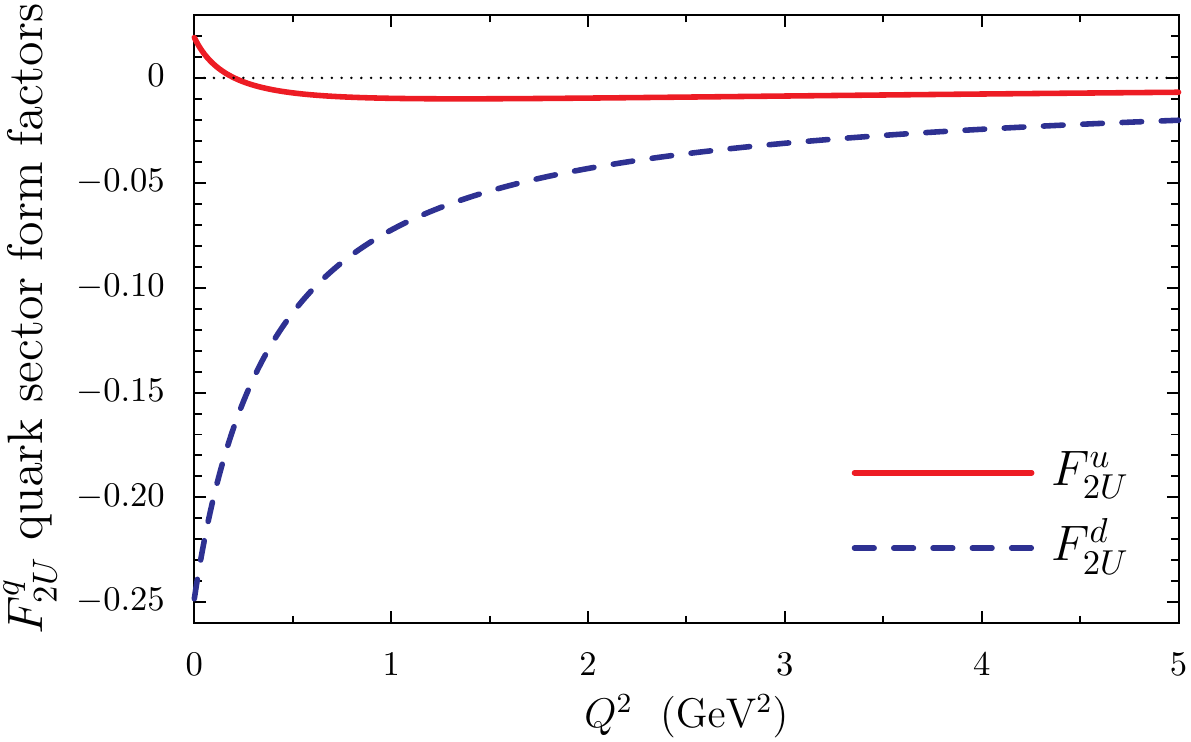}}
\caption{(Colour online) \textit{Upper panel:} Dressed up quark Dirac form factors, that also include pion cloud effects, separated into quark sectors. The solid line is the $u$-quark sector of the dressed up quark and the dashed line represents $d$-quark sector. \textit{Lower panel:} Dressed up quark Pauli form factors separated into quark sectors. The solid line is the $u$-quark sector and the dashed line the $d$-quark sector.}
\label{fig:constituentquarkflavoursectorformfactors}
\end{figure}

Results for the Dirac and Pauli dressed quark form factors, including pion loop effects, are given in Figs.~\ref{fig:constituentquarkformfactors}. The pion cloud softens the Dirac form factors, however its most important consequence is the non-zero Pauli form factor for the dressed quarks. At infinite $Q^2$ the dressed quark Dirac form factors now become
\begin{align}
F_{1U}(Q^2) &\stackrel{Q^2 \to \infty}{=} Z\,e_u, & F_{1D}(Q^2) &\stackrel{Q^2 \to \infty}{=} Z\,e_d,
\end{align}
whereas the Pauli form factors vanish for large $Q^2$. We find dressed quark anomalous magnetic moments of
\begin{align}
\kappa_U = 0.10 \qquad \text{and} \qquad \kappa_D = -0.17,
\label{eq:quarkanomalousresults}
\end{align}
defined as $\kappa_Q \equiv F_{2Q}(0)$. The quark charge and magnetic radii, defined with respect to the Sachs form factors and Eq.~\eqref{eq:radii}, take the values
\begin{align}
r^U_E &= 0.59\,\text{fm}, &  r^U_M &= 0.60\,\text{fm}, \\
r^D_E &= 0.73\,\text{fm}, &  r^D_M &= 0.67\,\text{fm}.
\end{align}
%

Decomposing the dressed quark form factors in quark/flavour sectors gives
\begin{align}
F_{1U}(Q^2) &= e_u\,F^u_{1U}(Q^2) + e_d\,F_{1U}^d(Q^2), \\
F_{1D}(Q^2) &= e_u\,F^u_{1D}(Q^2) + e_d\,F_{1D}^d(Q^2),
\end{align}
where the flavour sector dressed up quark form factors read
\begin{align}
\label{eq:F1uU}
&F^u_{1U} = Z\,\tfrac{1}{2}\lf[F_{1\o} + F_{1\rho}\rg]
+ \lf[3\,F_{1\o} - F_{1\rho}\rg]f_{1}^{(q)} + F_{1\rho}\, f_{1}^{(\pi)}, \\
&F^d_{1U} = Z\,\tfrac{1}{2}\lf[F_{1\o} - F_{1\rho}\rg]
+ \lf[3\,F_{1\o} + F_{1\rho}\rg]f_{1}^{(q)} - F_{1\rho}\, f_{1}^{(\pi)}, \\
&F^u_{2U} = \lf[3\,F_{1\o} - F_{1\rho}\rg]f_{2}^{(q)} + F_{1\rho}\, f_{2}^{(\pi)}, \\
\label{eq:F2dU}
&F^d_{2U} = \lf[3\,F_{1\o} + F_{1\rho}\rg]f_{2}^{(q)} - F_{1\rho}\, f_{2}^{(\pi)}.
\end{align}
The flavour sector dressed down quark form factors are given by
\begin{align}
F_{iD}^u &= F_{iU}^d \qquad \text{and} \qquad  F_{iD}^d = F_{iU}^u
\label{eq:chargesymmetry}
\end{align}
where $i=(1,\,2)$. Therefore, these results satisfy charge symmetry and are illustrated in Figs.~\ref{fig:constituentquarkflavoursectorformfactors}. For the quark sector anomalous magnetic moments we find
\begin{align}
\kappa^u_U = 0.02 \qquad \text{and} \qquad \kappa^d_U = -0.25,
\label{eq:quark_sector_amm_dressed_quarks}
\end{align}
and therefore the $d$ current quarks carry the bulk of the dressed up quark anomalous magnetic moment. This will have important implications for the nucleon form factors.

\section{Diquark and Meson Form Factors \label{sec:diquark_results}}
Critical to our picture of nucleon structure are diquark correlations inside the nucleon. An essential step therefore, in calculating the nucleon form factors, is to first determine the interaction of the virtual photon with the diquarks. A further reason to discuss the diquark form factors is that the scalar and axialvector diquarks are the $qq$ analogs of the $\pi$ and $\rho$ mesons.

\begin{figure}[tbp]
\centering\includegraphics[width=\columnwidth,clip=true,angle=0]{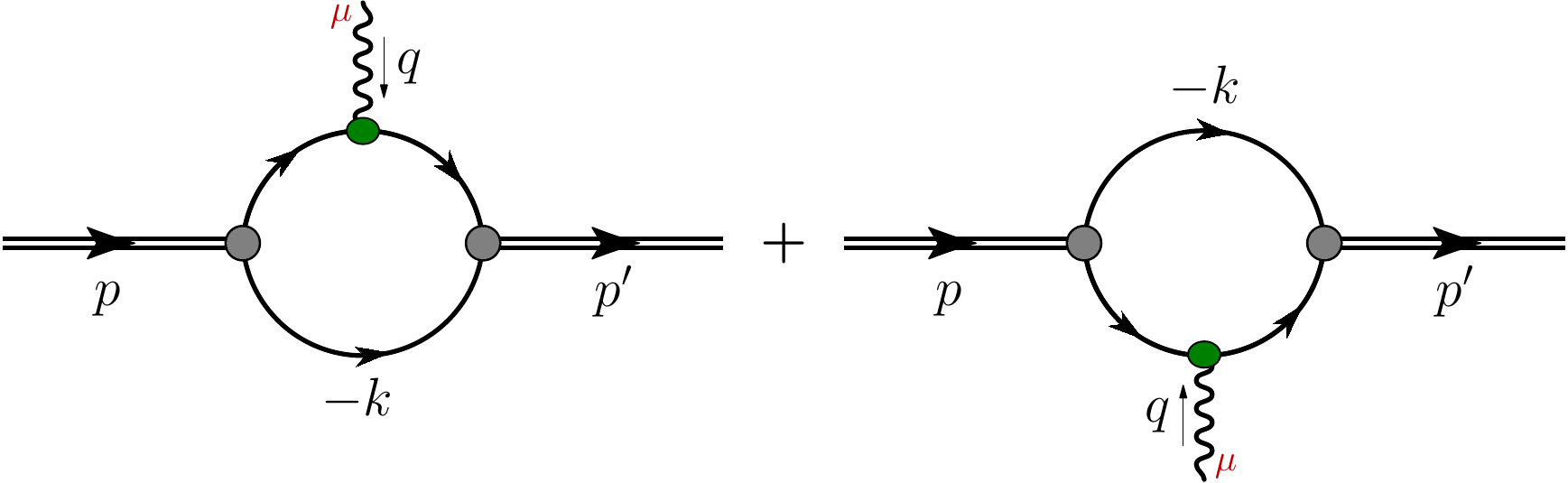}
\caption{(Colour online) Feynman diagrams that represent the diquark electromagnetic current. The shaded circles are the diquark Bethe-Salpeter vertices and the shaded oval is the quark-photon vertex. The Feynman diagrams for the meson form factors are analogous. However the flow of baryon number on one of the quark lines must be reversed.}
\label{fig:diquarkemcurrent}
\end{figure}

The electromagnetic current of a diquark is represented by the Feynman diagrams illustrated in Fig.~\ref{fig:diquarkemcurrent} and is expressed as
\begin{align}
&j^\mu(p',p) = i\int \frac{d^4k}{(2\pi)^4} \no \\
&\mathrm{Tr}\lf[\ol{\G}(p')\,S(p'+k)\,\L^\mu_{\g Q}(p',p)\,S(p+k)\,\G(p)\,S^T(-k)\rg],
\label{eq:diquarkemcurrent}
\end{align}
where the superscript $T$ indicates transpose. The Bethe-Salpeter vertices are represented by $\G(p)$ and are given in Eq.~\eqref{eq:homogeneousBSvertex}. The dressed quark-photon vertex $\L^\mu_{\g Q}(p',p)$ is given in Eqs.~\eqref{eq:quark_photon_vertex} and \eqref{eq:dressedquarkcurrent}. 

Hadron form factors will be determined using the three variants for the dressed quark form factors discussed in Sect.~\ref{sec:QPV} and illustrated, for the non-trivial variants, in Fig.~\ref{fig:constituentquarkformfactors}. Results obtained by treating the dressed quarks as pointlike will be labelled with a superscript (bare), while those obtained using the dressed quark form factors from the BSE, Eq.~\eqref{eq:bseconstituent}, will be labelled with a superscript (bse) and our full results, where the quark form factors also include pion loop corrections, Eqs.~\eqref{eq:f1U_quarkpion}--\eqref{eq:f2D_quarkpion}, will have no superscript label.

\begin{figure}[tbp]
\centering\includegraphics[width=\columnwidth,clip=true,angle=0]{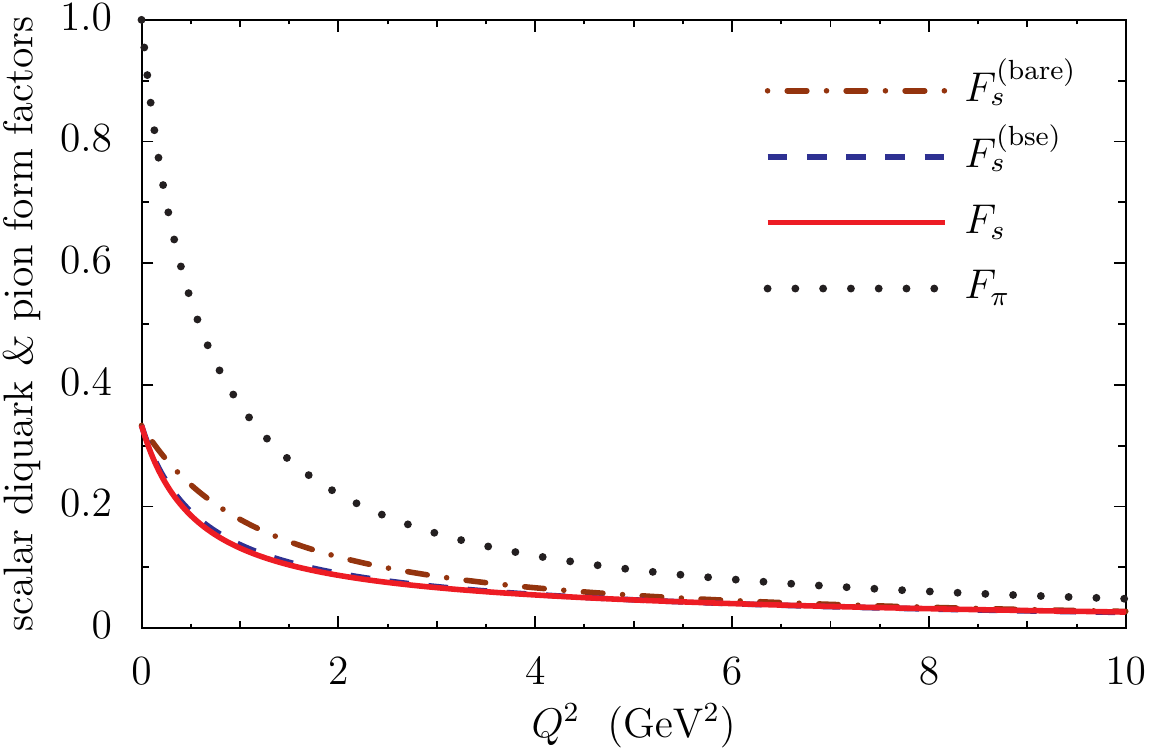}
\caption{(Colour online) Results for the scalar diquark and pion form factors. For the pion we just show the full results, however for the scalar diquark form factors we show the cases when the dressed quarks are pointlike, the quark-photon vertex is given by the BSE and finally when pion loop corrections are also included.}
\label{fig:scalardiquarkff}
\end{figure}

The electromagnetic current for a scalar diquark, or any on-shell spin-zero particle, has the general form
\begin{align}
j_s^\mu(p',p) = \lf(p'+p\rg)^\mu\,F_s(Q^2),
\end{align}
and is therefore parameterized by a single form factor. Evaluating Eq.~\eqref{eq:diquarkemcurrent} for the scalar diquark gives
\begin{multline}
F_s(Q^2) = \lf[F_{1U}(Q^2) + F_{1D}(Q^2)\rg]f_s^V(Q^2) \\ 
+ \lf[F_{2U}(Q^2) + F_{2D}(Q^2)\rg]f_s^T(Q^2),
\end{multline}
where $f_s^V$, $f_s^T$ are the scalar diquark body form factors associated with the vector and tensor photon couplings to the dressed quarks (see Eq.~\eqref{eq:dressedquarkcurrent}). Results for the scalar diquark form factor are given in Fig.~\ref{fig:scalardiquarkff} for the three variants of dressed quark form factors. Vertex corrections introduced by the BSE result in a softer form factor (dashed line) in comparison with results obtained using pointlike dressed quark form factors (dash-dotted line). Including pion loop corrections only slightly alters the scalar diquark form factor (solid line).

\begin{figure}[tbp]
\centering\includegraphics[width=\columnwidth,clip=true,angle=0]{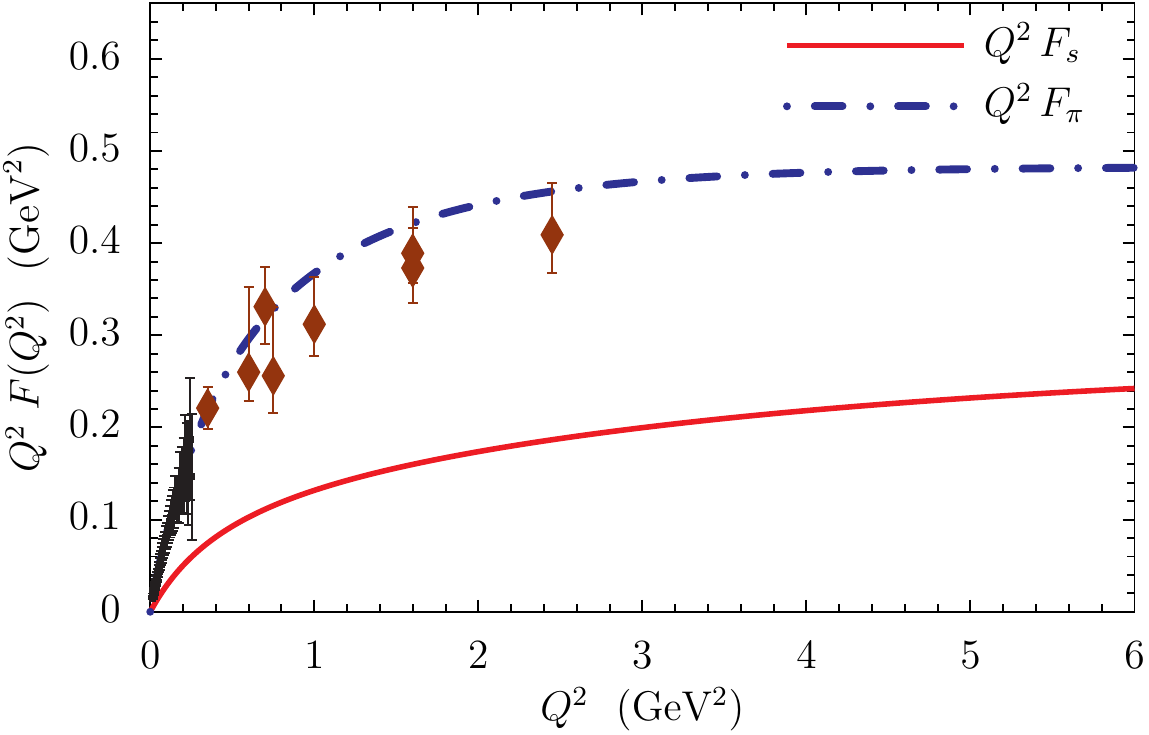}
\caption{(Colour online) Scalar diquark and pion form factors multiplied by $Q^2$. The pion form factor data is from Refs.~\cite{Amendolia:1986wj,Amendolia:1984nz,Horn:2006tm,Tadevosyan:2007yd,Blok:2008jy}.}
\label{fig:Q2scalardiquarkff}
\end{figure}

The $\bar{q}q$ analog of the scalar diquark is the pion, where for the $\pi^+$ the electromagnetic form factor is given by
\begin{multline}
F_\pi(Q^2) = \lf[F_{1U}(Q^2) - F_{1D}(Q^2)\rg]f_s^V(Q^2) \\ 
          + \lf[F_{2U}(Q^2) - F_{2D}(Q^2)\rg]f_s^T(Q^2).
\label{eq:pionformfactor}
\end{multline}
The body form factors in Eq.~\eqref{eq:pionformfactor} are the same as those for the scalar diquark, except they are now functions of the pion mass instead of the scalar diquark mass.\footnote{There is also a factor of two because of the different definition for the Bethe-Salpeter normalization given in Eq.~\eqref{eq:Zpi}, compared to that in Eq.~\eqref{eq:Zs}.} We do not include pion loop corrections on the dressed quarks in the case of the pion form factor, because at the hadronic level there is no three pion vertex. The full result for the pion form factor is given as the dotted curve in Fig.~\ref{fig:scalardiquarkff}. The scalar diquark and pion form factors multiplied by $Q^2$ are presented in Fig.~\ref{fig:Q2scalardiquarkff}, where good agreement with pion form factor data from  Refs.~\cite{Amendolia:1986wj,Amendolia:1984nz,Horn:2006tm,Tadevosyan:2007yd,Blok:2008jy} is seen. At large $Q^2$ both form factors plateau, where we find $Q^2\,F_\pi(Q^2) \to 0.48$ and  $Q^2\,F_s(Q^2) \to 0.30$. The pion form factor result is consistent with the perturbative QCD prediction~\cite{Lepage:1979zb,Farrar:1979aw}:
\begin{align}
Q^2\,F_\pi(Q^2) \stackrel{Q^2\to \infty}{\longrightarrow} 16\,\pi\,f_\pi^2\,\a_s(Q^2),
\label{eq:perturbativepionformfactor}
\end{align}
in the sense that the strong coupling constant, $\a_s(Q^2)$, corresponds to a constant in the NJL model and therefore $Q^2\,F_\pi(Q^2)$ should become constant as $Q^2 \to \infty$. Taking Eq.~\eqref{eq:perturbativepionformfactor} literally, our pion form factor result implies that $\a_s(Q^2) = 1.12$, which using a NNLO result for the running coupling~\cite{Gluck:1998xa} would correspond to an NJL model scale of $Q_0^2 \sim 0.18\,$GeV$^2$, which is consistent with previous estimates~\cite{Cloet:2005rt,Cloet:2005pp,Cloet:2006bq}. Our calculated pion form factor reaches its plateau by $Q^2 \simeq 6\,$GeV$^2$, which corresponds to the same scale at which the Dyson-Schwinger equation results
of Ref.~\cite{Chang:2013nia} reach a maximum, after which the result of Ref.~\cite{Chang:2013nia} decreases because of the logarithmic running of $\a_s(Q^2)$ in QCD.

\begin{table}[tbp]
\addtolength{\tabcolsep}{5pt}
\addtolength{\extrarowheight}{2.2pt}
\begin{tabular}{l|ccccccc}
\hline\hline
                      & $r^{\text{(bare)}}_E$ & $r^{\text{(bse)}}_E$ & $r_E$ & $r^{\text{exp}}_E$   \\
\hline
scalar diquark        & 0.46              &  0.62            & 0.63  &                   \\
pion                  & 0.46              &  0.62            & 0.62  & 0.663 $\pm$ 0.006 \\
\hline\hline
\end{tabular}
\caption{Charge radii for the scalar diquark and pion, 
each shown for the three variants for the dressed quark form factors.
The experimental value for the pion is from 
Ref.~\cite{Dally:1982zk,Amendolia:1986wj}. All radii are in units of fm.}
\label{tab:pion_scalar_diquark_radii}
\end{table}

Results for the scalar diquark and pion charge radii are given in Tab.~\ref{tab:pion_scalar_diquark_radii}, for the three variants of the dressed quark form factors. The charge radius of the pion and scalar diquark are found to be very similar, where the pion radius is approximately 5\% smaller than the experimental value from Refs.~\cite{Dally:1982zk,Amendolia:1986wj}. 
 
The electromagnetic current for an axialvector diquark, or any on-shell spin-one particle, has the general form~\cite{Frankfurt:1977vc}
\begin{align}
j_a^{\mu,\a\b}(p',p) &= \lf[g^{\a\b}F_{1a}(Q^2) - \frac{q^\a q^\b}{2\,M_a^2}\,F_{2a}(Q^2)\rg]\lf(p'+p\rg)^\mu \no \\
&\hs{14mm}
- \lf(q^\a g^{\mu\b} - q^\b g^{\mu\a}\rg)F_{3a}(Q^2),
\label{eq:axialcurrent}
\end{align}
where the Lorentz indices $\mu$, $\a$, $\b$ represent the polarizations of the photon, initial axialvector diquark and final axialvector diquark, respectively. The Lorentz covariant form factors of Eq.~\eqref{eq:axialcurrent} are often re-expressed as the Sachs-like charge, magnetic and quadruple form factors for a spin-one particle, given by
\begin{align}
\label{eq:gc}
G_C(Q^2) &= F_1(Q^2) + \frac{2}{3}\,\eta\,G_Q(Q^2), \\
G_M(Q^2) &= F_3(Q^2), \\
\label{eq:gq}
G_Q(Q^2) &= F_1(Q^2) + \lf(1 + \eta\rg)F_2(Q^2) - F_3(Q^2),
\end{align}
where $\eta = \frac{Q^2}{4\,m_H^2}$ and $m_H$ is the relevant hadron mass. At $Q^2 = 0$ these form factors give, respectively, 
the charge, magnetic moment and quadruple moment of a spin-one particle, in units of $e$, $e/(2\,m_H)$ and $e/m_H^2$. The charge, magnetic and quadrupole radii -- $\lf<r_C^2\rg>$, $\lf<r_M^2\rg>$, $\lf<r_Q^2\rg>$ -- are defined with respect to these Sachs-like form factors.

\begin{figure}[t]
\centering\includegraphics[width=\columnwidth,clip=true,angle=0]{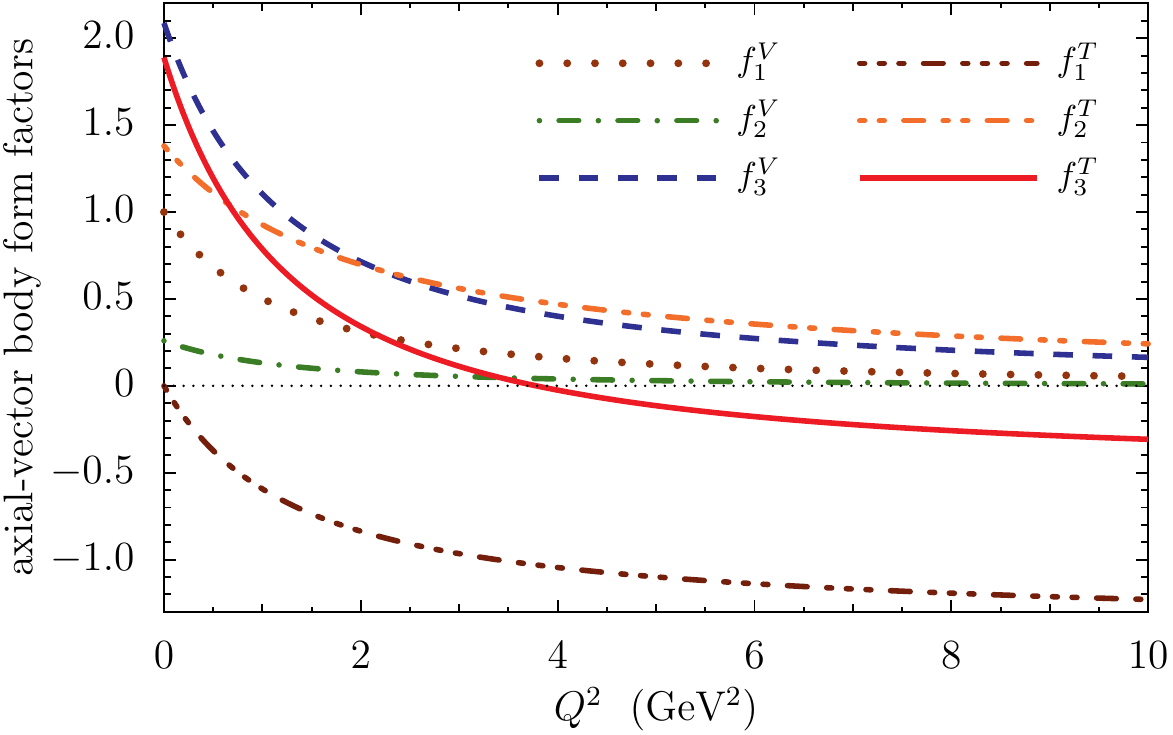}
\caption{(Colour online) Axial--vector diquark body form factors. These body form factors must still be multiplied by the appropriate dressed quark form factors to obtain the axialvector diquark form factors.}
\label{fig:axialdiquarkbodyformfactors}
\end{figure}

Evaluating the Feynman diagrams of Fig.~\ref{fig:diquarkemcurrent}, using the axialvector diquark Bethe-Salpeter vertex given in Eq.~\eqref{eq:homogeneousBSvertex} and the quark-photon vertex of Eq.~\eqref{eq:quark_photon_vertex} gives, for an axialvector diquark with quark content $\{ud\}$, the form factor result
\begin{align}
F^{\{ud\}}_{ia}(Q^2) &= \lf[F_{1U}(Q^2) + F_{1D}(Q^2)\rg]f_{i}^V(Q^2) \no \\ 
&\,+ \lf[F_{2U}(Q^2) + F_{2D}(Q^2)\rg]f_{i}^T(Q^2),
\label{eq:axialvectorffs}
\end{align}
where $i \in 1,\,2,\,3$ correspond to the form factors in Eq.~\eqref{eq:axialcurrent}. Expressions for axialvector diquarks of the $\{uu\}$ and $\{dd\}$ type are simply given by Eq.~\eqref{eq:axialvectorffs} with the appropriate substitution of the dressed quark form factors. The vector and tensor body form factors, $f_{i}^V$ and $f_{i}^T$ , are illustrated in Fig.~\ref{fig:axialdiquarkbodyformfactors}. A notable feature of these form factors is that charge conservation implies $f_1^V(0) = 1$ and $f_1^T(0) = 0$. The magnetic moment equals $f_3^V(0) = 2.09$ which, because of relativistic effects, is slightly larger than the canonical value of $\mu_1 = 2$ for a spin-one particle. For the quadrupole moment  the body form factors imply $\mathcal{Q} = -0.83$ which, because of relativistic effects, is about 17\% smaller than the canonical value of $\mathcal{Q} = -1$.

\begin{figure}[tbp]
\subfloat{\centering\includegraphics[width=\columnwidth,clip=true,angle=0]{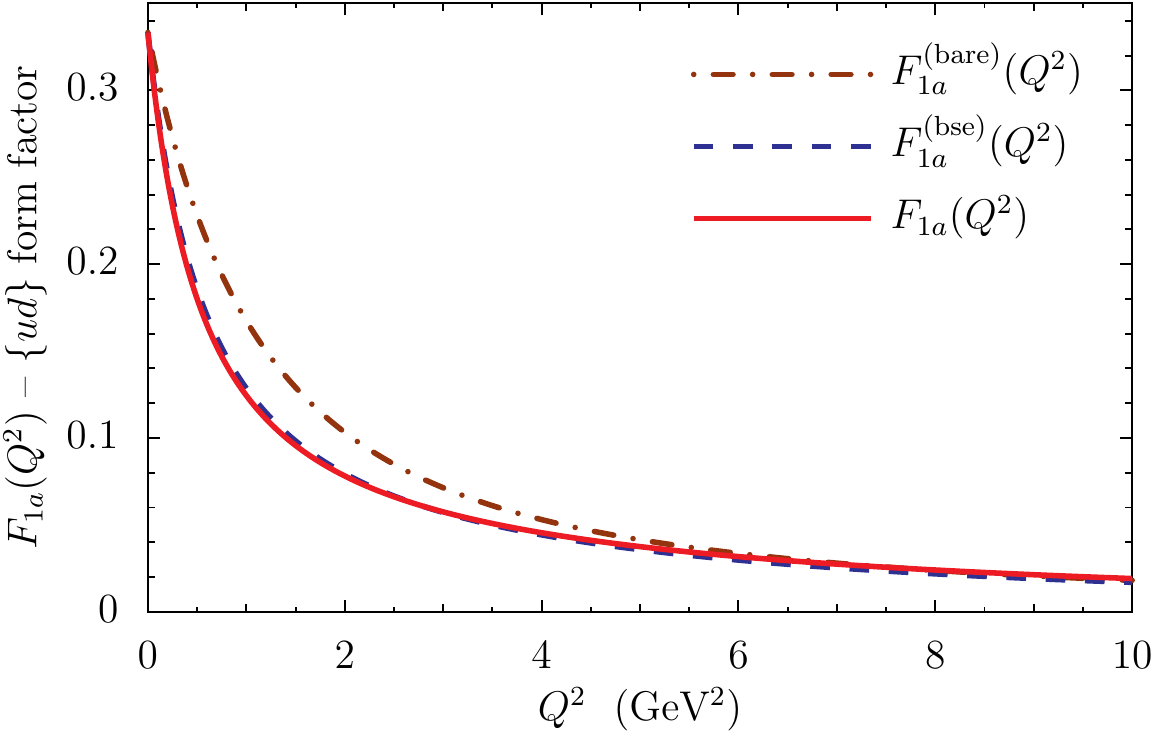}}\\
\subfloat{\centering\includegraphics[width=\columnwidth,clip=true,angle=0]{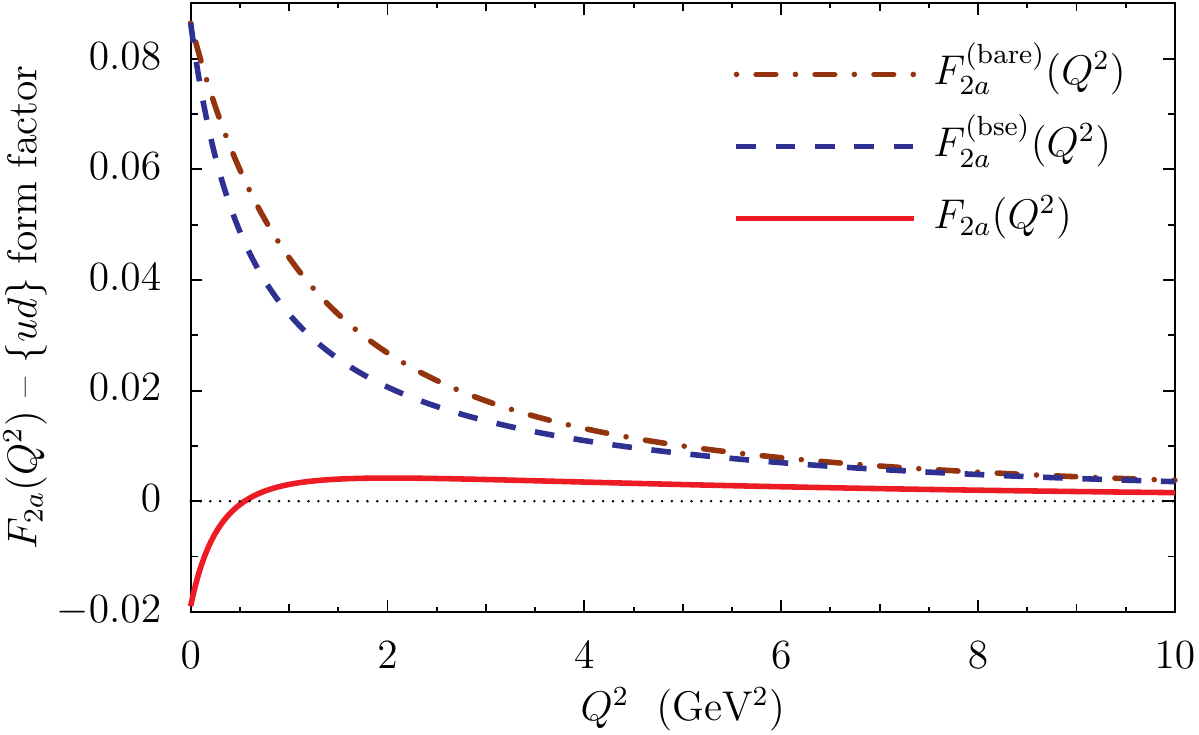}}\\
\subfloat{\centering\includegraphics[width=\columnwidth,clip=true,angle=0]{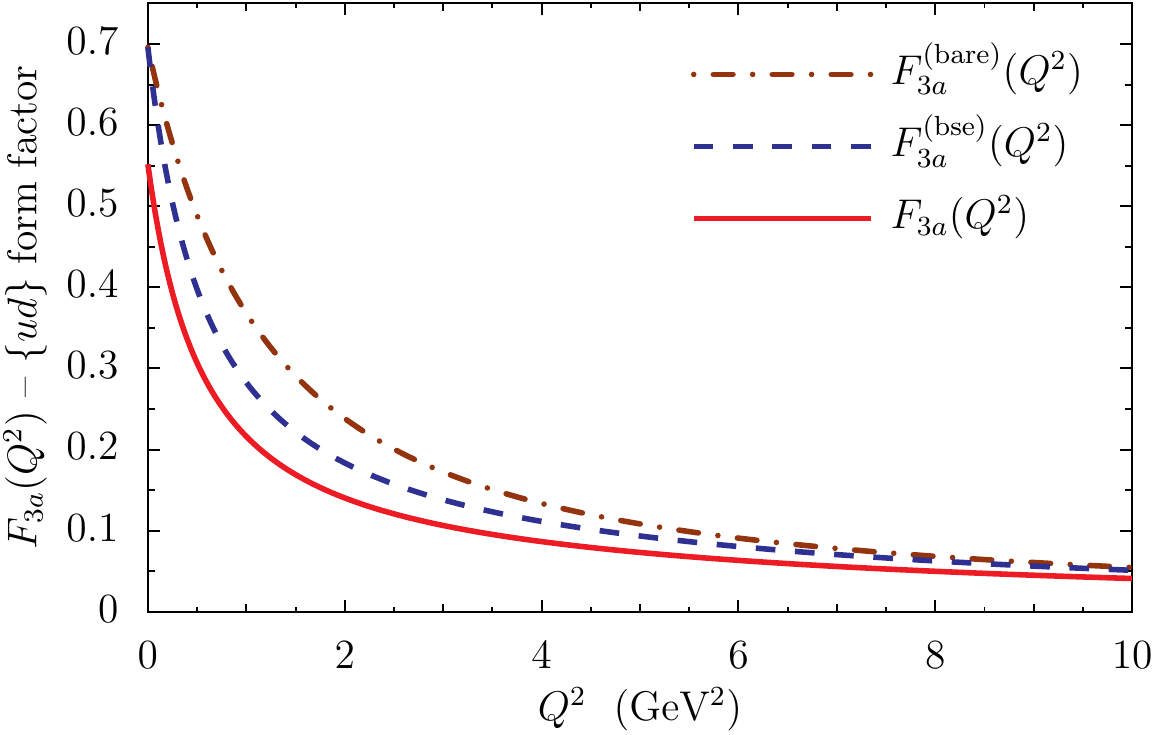}}
\caption{(Colour online) Form factors for an axialvector diquark with quark content $\{ud\}$.}
\label{fig:axialdiquarkffs}
\end{figure}

Results for the form factors of Eq.~\eqref{eq:axialcurrent} for an axial--vector diquark with quark content $\{ud\}$ are presented in Figs.~\ref{fig:axialdiquarkffs}. In each case the vertex dressing from the quark-photon inhomogeneous BSE results in a softening of form factors, compared to the case of pointlike dressed quarks. Although the pion loop effects leave $F_{1a}$ almost unchanged, both $F_{2a}$ and $F_{3a}$ receive sizeable negative corrections. The origin of these corrections can be traced back to Eq.~\eqref{eq:axialcurrent} and the results in Fig.~\ref{fig:axialdiquarkbodyformfactors}. The tensor body from factors $f_2^T$ and $f_3^T$ are large and positive for small $Q^2$. 
This, together with the large negative anomalous magnetic moment of the dressed down quark (see Eq.~\eqref{eq:quarkanomalousresults}), 
results in sizeable corrections to $F_{2a}$ and $F_{3a}$ from pion loop effects.

\begin{table*}[t]
\addtolength{\tabcolsep}{4.9pt}
\addtolength{\extrarowheight}{2.2pt}
\begin{tabular}{l|ccccccccccccccccccc}
\hline\hline
 & $\mu^{\text{(bse)}}$ & $\mu$  & & $\mathcal{Q}^{\text{(bse)}}$ & $\mathcal{Q}$  & & $r^{\text{(bse)}}_C$ & $r_C$ & & $r^{\text{(bse)}}_M$ & $r_M$ & & $r^{\text{(bse)}}_Q$ & $r_Q$  \\[0.2ex]
\hline
$\{uu\}$ axialvector diquark  & \ph{-}2.78  & \ph{-}3.14  &&      -1.10 &      -1.20  && 0.65 & 0.76  & & 0.61 & 0.74 & & 0.61 & 0.74   \\
$\{ud\}$ axialvector diquark  & \ph{-}0.70  & \ph{-}0.55  &&      -0.28 &      -0.24  && 0.37 & 0.38  & & 0.60 & 0.62 & & 0.61 & 0.64   \\
$\{dd\}$ axialvector diquark  &      -1.39  &      -2.04  && \ph{-}0.55 & \ph{-}0.73  && 0.65 & 0.84  & & 0.61 & 0.80 & & 0.62 & 0.79   \\
rho plus                       & \ph{-}2.08  & \ph{-}2.57  &&      -0.87 &      -1.06  && 0.67 & 0.82  & & 0.62 & 0.77 & & 0.62 & 0.77   \\
\hline\hline

\end{tabular}
\caption{Results for the magnetic moment, quadruple moment 
and the charge, magnetic and quadruple radius of
the axialvector diquarks and $\rho^+$ meson.
In each case we present results for various levels of sophistication for
the constituent quark form factors. 
All radii are in units of fm, the magnetic moment has units $e/(2\,m_H)$ and
the quadruple moment $e/m_H^2$, where $m_H$ is the 
mass of the relevant diquark or meson.}
\label{tab:rho_axial_diquark_moments}
\end{table*}

\begin{figure}[t]
\subfloat{\centering\includegraphics[width=\columnwidth,clip=true,angle=0]{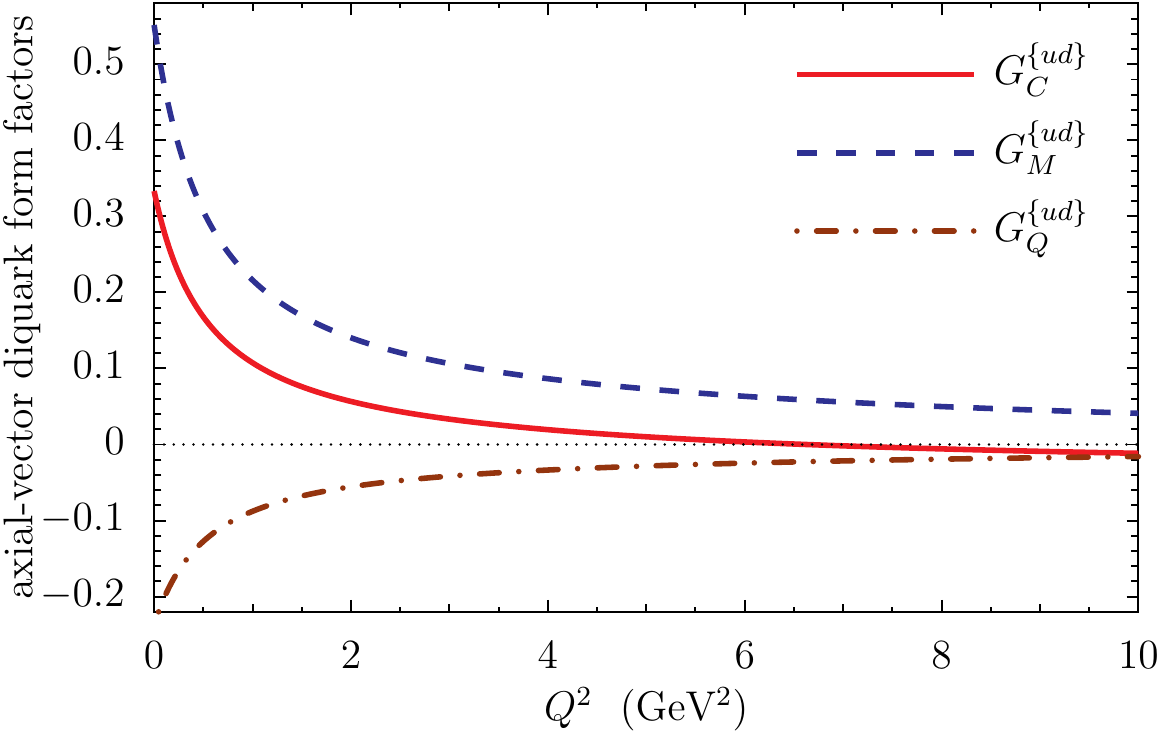}} \\
\subfloat{\centering\includegraphics[width=\columnwidth,clip=true,angle=0]{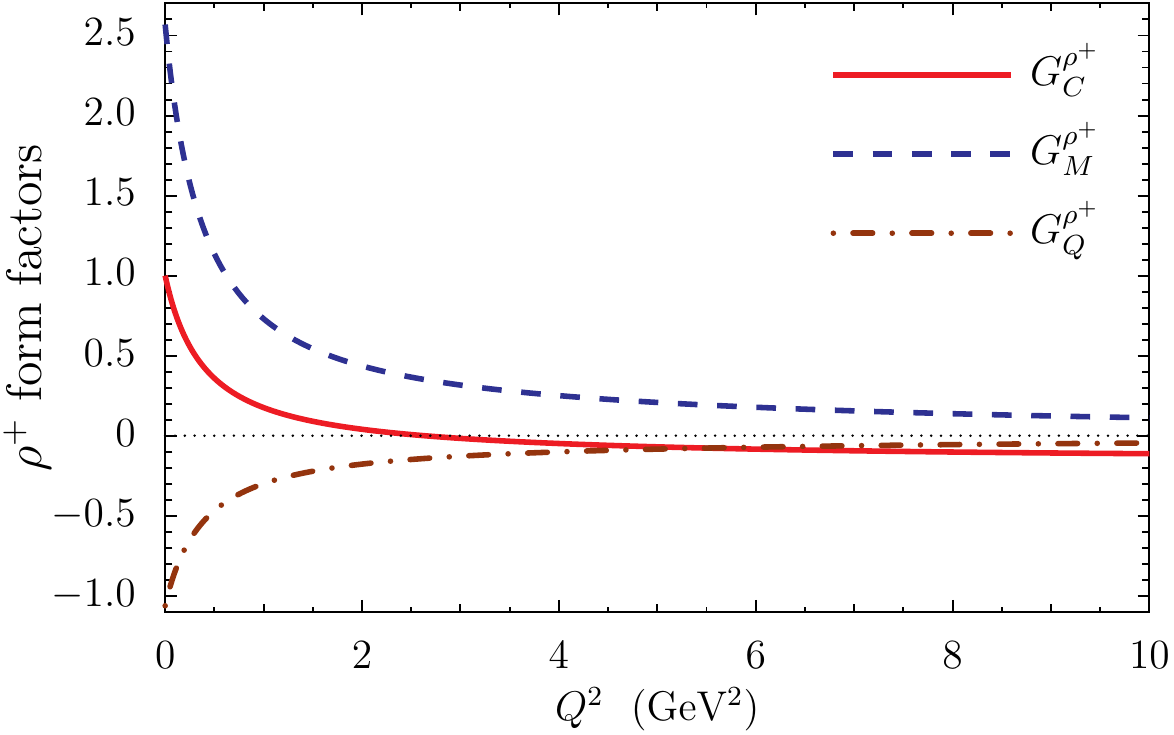}}
\caption{(Colour online) \textit{Upper panel:} results for the charge, magnetic and quadruple form factors of a $ud$ axialvector diquark.
\textit{Lower panel}: results for the charge, magnetic and 
quadruple form factors of a $\rho^+$ meson.}
\label{fig:axialvectorsachsformfactor}
\end{figure}

The Lorentz covariant form factors for the $\rho^+$ meson, associated with the current of Eq.~\eqref{eq:axialcurrent}, are given by
\begin{align}
F_{i\rho}(Q^2) &= \lf[F_{1U}(Q^2) - F_{1D}(Q^2)\rg]f_{i}^V(Q^2) \\ 
&\,+ \lf[F_{2U}(Q^2) - F_{2D}(Q^2)\rg]f_{i}^T(Q^2),
\label{eq:rhoformfactors}
\end{align}
where $i \in 1,\,2,\,3$ and the body form factors are now functions of the rho mass instead of the axialvector diquark mass.
Results for the Sachs-like spin-one form factors defined in Eqs.~\eqref{eq:gc}--\eqref{eq:gq} are illustrated in Fig.~\ref{fig:axialvectorsachsformfactor} for a $\{ud\}$--type axialvector diquark and the $\rho^+$ meson. In these figures we only show the full results which include pion cloud effects. The zero in the charge form factors occurs at $Q^2 \simeq 6.6\,$GeV$^2$ for $G^{\{ud\}}_{C}$ and for $G^{\rho^+}_{C}$ at $Q^2 \simeq 2.6\,$GeV$^2$.

Static properties of the $\rho^+$ and the axialvector diquarks are given Tab.~\ref{tab:rho_axial_diquark_moments} for variants of the dressed quark form factors. We find that pion loop effects have a substantial impact on the static properties of the axialvector diquarks and rho mesons. For example, the pion cloud increases the magnitude of the $\rho^+$ magnetic moment by 24\% and the quadrupole moment by 22\%, 
while for the $\{ud\}$--type axialvector diquarks we find a reduction of the magnetic moment by 21\% and the magnitude of  the quadrupole moment by 14\%. The sign difference between these corrections for the $\rho^+$ and $\{ud\}$--type axialvector diquark arises because the dressed down quark form factors enter the respective currents with the opposite sign -- see Eqs.~\eqref{eq:axialvectorffs} and \eqref{eq:rhoformfactors} -- and the dressed down quark has a large anomalous magnetic moment. For the $\rho^+$ meson the pion cloud uniformly increases the charge, magnetic and quadrupole radii by approximately 16\%, whereas for the $\{ud\}$--type axialvector diquarks the pion cloud has little effect on the charge and quadrupole radii but increases the magnetic radius by 38\%.

As an interesting check on the large $Q^2$ behaviour of our rho or axialvector diquark form factor results, 
we make a comparison with the relations derived in Ref.~\cite{Brodsky:1992px}. That is, at large timelike or spacelike momenta, the ratio of the form factors for a spin-one particle should behave as
\begin{align}
G_C(Q^2):G_M(Q^2):G_Q(Q^2) = \lf(1 - \tfrac{2}{3}\eta\rg):\,2\,:\,-1,
\label{eq:hillerbrodsky}
\end{align}
where corrections are of the order $\L_{\text{QCD}}/Q$ and $\L_{\text{QCD}}/M_\rho$. For our spin-one results we find that the $G_C/G_Q$ constraint is satisfied to better than $15\%$ for $Q^2 = 10\,$GeV$^2$, to better than $3\%$ for $Q^2=100\,$GeV$^2$ and for $Q^2 > 1000\,$GeV$^2$ our result takes the value given in Eq.~\eqref{eq:hillerbrodsky}. The calculated ratios $G_C/G_M$ and $G_M/G_Q$ 
saturate within $15\%$ of the values in Eqs.~\eqref{eq:hillerbrodsky}. However this deviation is well within the leading correction of $\L_{\text{QCD}}/M_\rho \sim 0.3$.

\begin{figure}[t]
\centering\includegraphics[width=\columnwidth,clip=true,angle=0]{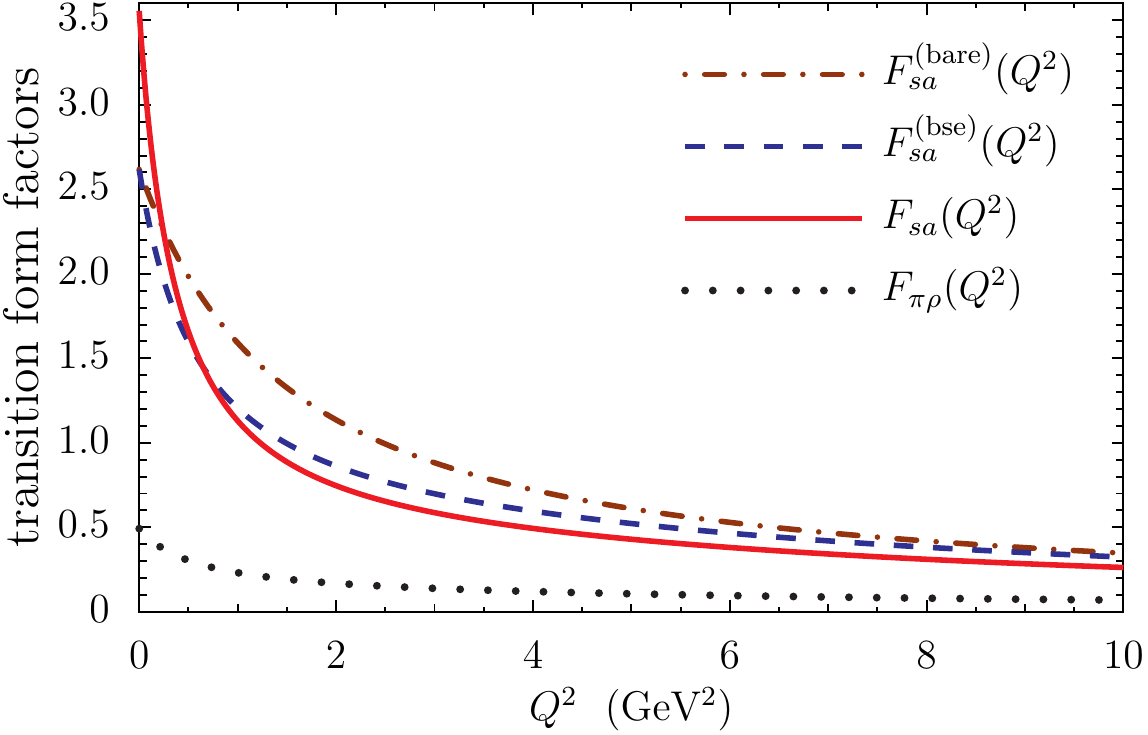}
\caption{(Colour online) Results for the 
scalar $\leftrightarrow$ axialvector diquark and $\pi \leftrightarrow \rho$
electromagnetic transition form factors.}
\label{fig:mixingdiquarkformfactor}
\end{figure}

The remaining diquark electromagnetic current that contributes to the nucleon form factors is the transition current between scalar and axialvector diquarks. This current has the form
\begin{align}
j^{\mu,\a}_{sa}(p',p) &= \pm\ \frac{1}{M_s+M_a}\ i\ve^{\a\mu\s\l}p_\s'p_\l\,F_{sa}(Q^2),
\label{eq:satransition}
\end{align}
where the plus sign indicates a scalar $\to$ axialvector transition and the reverse process has the minus sign. The Lorentz indices $\mu$ and $\a$ represent the polarizations of the photon and the axialvector diquark. Evaluating the Feynman diagram of Fig.~\ref{fig:diquarkemcurrent} for this transition process gives 
\begin{align}
F_{sa}(Q^2) &= \lf[F_{1U}(Q^2) - F_{1D}(Q^2)\rg]f_{sa}^V(Q^2) \no \\
&\,+ \lf[F_{2U}(Q^2) - F_{2D}(Q^2)\rg]f_{sa}^T(Q^2),
\end{align}
where $f_{sa}^V(Q^2)$ and $f_{sa}^T(Q^2)$ are the vector and tensor body form factors. The electromagnetic transition form factor describing the $\g^*\pi^+ \to \rho^+$ process is given by
\begin{align}
F_{\pi\rho}(Q^2) &= \lf[F_{1U}(Q^2) + F_{1D}(Q^2)\rg]f_{sa}^V(Q^2) \no \\
&\,+ \lf[F_{2U}(Q^2) + F_{2D}(Q^2)\rg]f_{sa}^T(Q^2),
\end{align}
where body form factors are now functions of the $\pi$ and $\rho$ masses. Results for $F_{sa}$ and $F_{\pi\rho}$ are presented 
in Fig.~\ref{fig:mixingdiquarkformfactor}. The vertex dressing from the BSE produces a softer form factor, and for the 
diquark transition the large isovector combination of the constituent quark Pauli form factors, arising from the pion cloud, gives a sizeable correction for $Q^2 \lesssim 1\,$GeV$^2$. Results for the transition moment and transition radius are given in Tab.~\ref{tab:transition_form_factors}.

\begin{table}[t]
\addtolength{\tabcolsep}{9.5pt}
\addtolength{\extrarowheight}{2.2pt}
\begin{tabular}{l|ccccccc}
\hline\hline
                            & $\kappa_T^{\text{(bse)}}$ & $\kappa_T$ &  & $r_T^{\text{(bse)}}$ & $r_T$   \\
\hline
$s \leftrightarrow a$       & 2.66                  & 3.61       &  &  0.75            & 0.99  \\
$\pi \leftrightarrow \rho$  & 0.62                  & 0.49       &  &  0.54            & 0.54  \\
\hline\hline
\end{tabular}
\caption{Results for the transition moment, defined as $\kappa_T \equiv F(0)$, the transition radius (which is normalized by $\kappa_T$), for scalar $\leftrightarrow$ axialvector diquark and pion $\leftrightarrow$ rho transitions. Radii are in units of fm.}
\label{tab:transition_form_factors}
\end{table}

\section{Nucleon Form Factor Results \label{sec:nucleon_results}}
The Feynman diagrams that contribute to the nucleon's electromagnetic current 
are illustrated in Fig.~\ref{fig:nucleon_em_current_feynman_diagrams}, 
where the coupling of the photon to the dressed quarks and diquarks has been discussed 
in Sects.~\ref{sec:QPV} and \ref{sec:diquark_results}, respectively. Using a 
quark-photon vertex of the form given in Eq.~\eqref{eq:quark_photon_vertex} demarcates the nucleon form factors into flavour sectors defined by the dressed quarks, such that
\begin{align}
\label{eq:protondirac}
F_{ip}(Q^2) &= F_{ip}^U(Q^2) + F_{ip}^D(Q^2), \\
F_{in}(Q^2) &= F_{in}^U(Q^2) + F_{in}^D(Q^2),
\end{align}
where $i=(1,\,2)$. The dressed quark flavour sector nucleon form factors 
are given by the product of dressed quark form factors 
(e.g. Eqs.~\eqref{eq:f1U_quarkpion}--\eqref{eq:f2D_quarkpion}) with the nucleon 
body form factors, such that
\begin{align}
\label{eq:FpQ}
F_{ip}^Q &= F_{1Q}\,f_{ip}^{Q,V} + F_{2Q}\,f_{ip}^{Q,T}, \\
\label{eq:FnQ}
F_{in}^Q &= F_{1Q}\,f_{in}^{Q,V} + F_{2Q}\,f_{in}^{Q,T},
\end{align}
where $Q=(U,\,D)$ and the $Q^2$ dependence of each form factor has been omitted. The superscript $V$ indicates a vector body form factor and the superscript $T$ a tensor body form factor, which arise from the quark current of Eq.~\eqref{eq:dressedquarkcurrent}.

\begin{figure}[t]
\subfloat{\centering\includegraphics[width=\columnwidth,clip=true,angle=0]{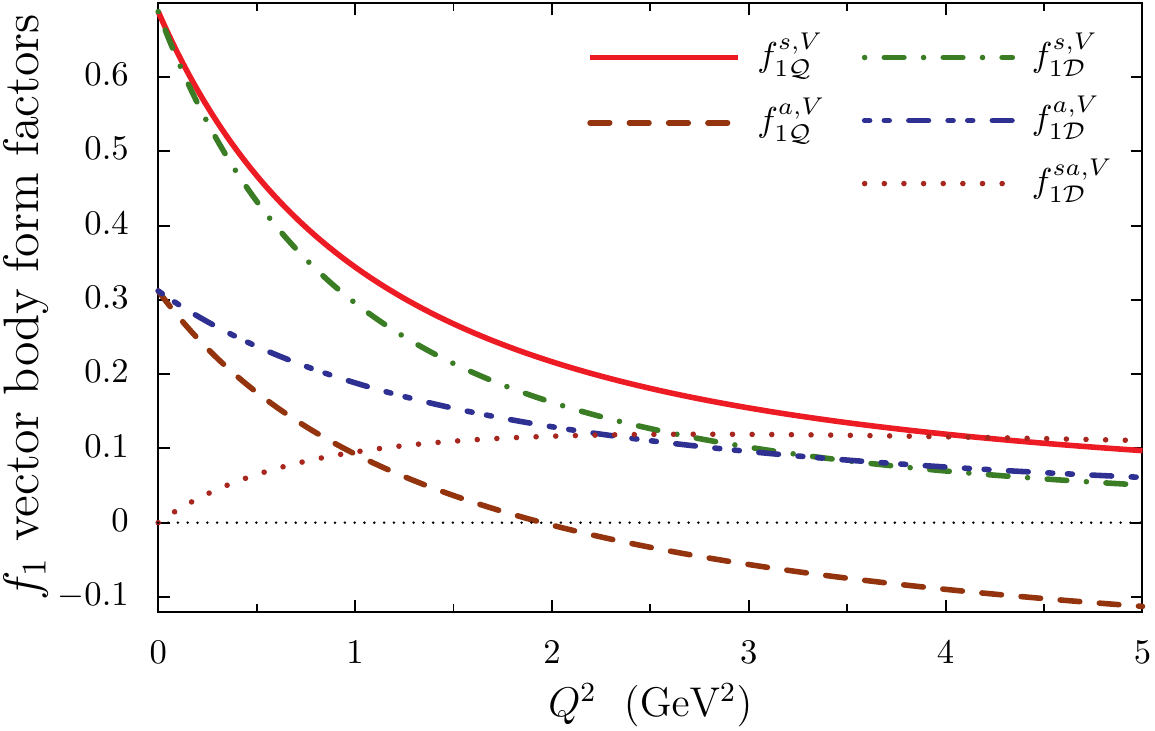}}\\
\subfloat{\centering\includegraphics[width=\columnwidth,clip=true,angle=0]{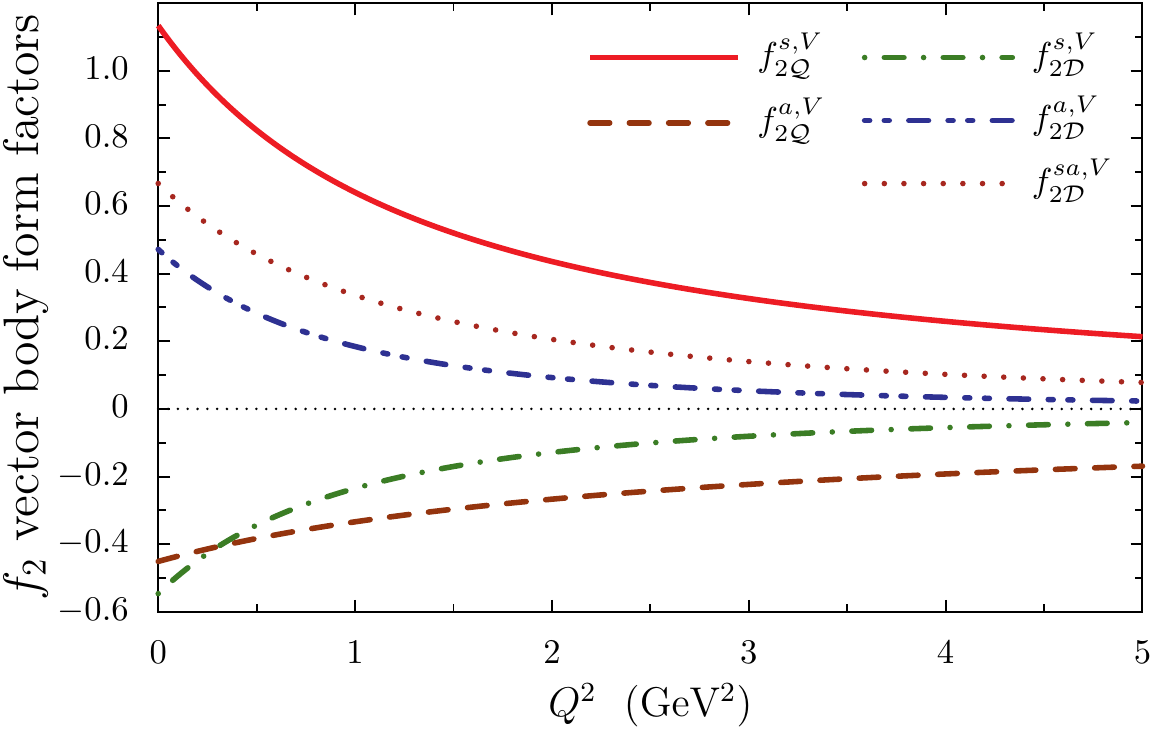}}
\caption{(Colour online) Nucleon Dirac (\textit{upper panel}) and Pauli (\textit{lower panel}) body form factors which result from a vector coupling to the quarks in the Feynman diagrams of Fig.~\ref{fig:nucleon_em_current_feynman_diagrams}. To obtain their contribution to the nucleon form factors these results must be multiplied by the appropriate isospin factors, as in Eqs.~\eqref{eq:f1U_V} and \eqref{eq:f1D_V}, and the dressed quark Dirac form factors.}
\label{fig:vector_body_form_factors}
\end{figure}

\begin{figure}[t]
\subfloat{\centering\includegraphics[width=\columnwidth,clip=true,angle=0]{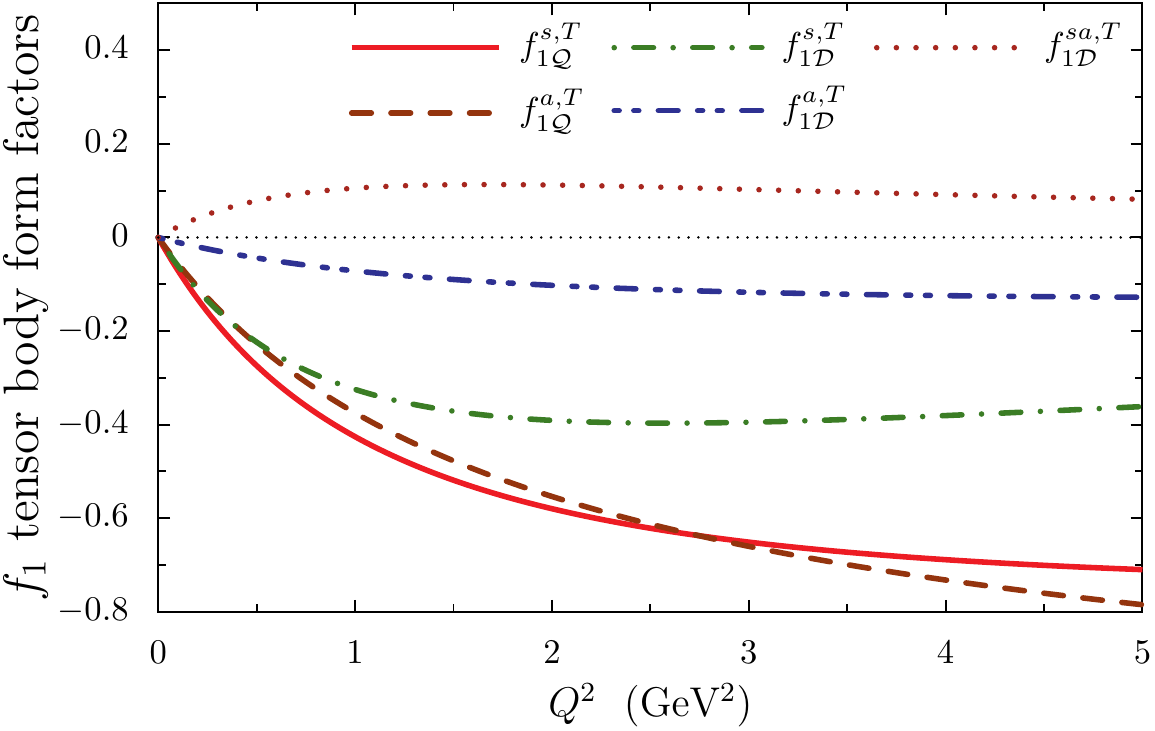}}\\
\subfloat{\centering\includegraphics[width=\columnwidth,clip=true,angle=0]{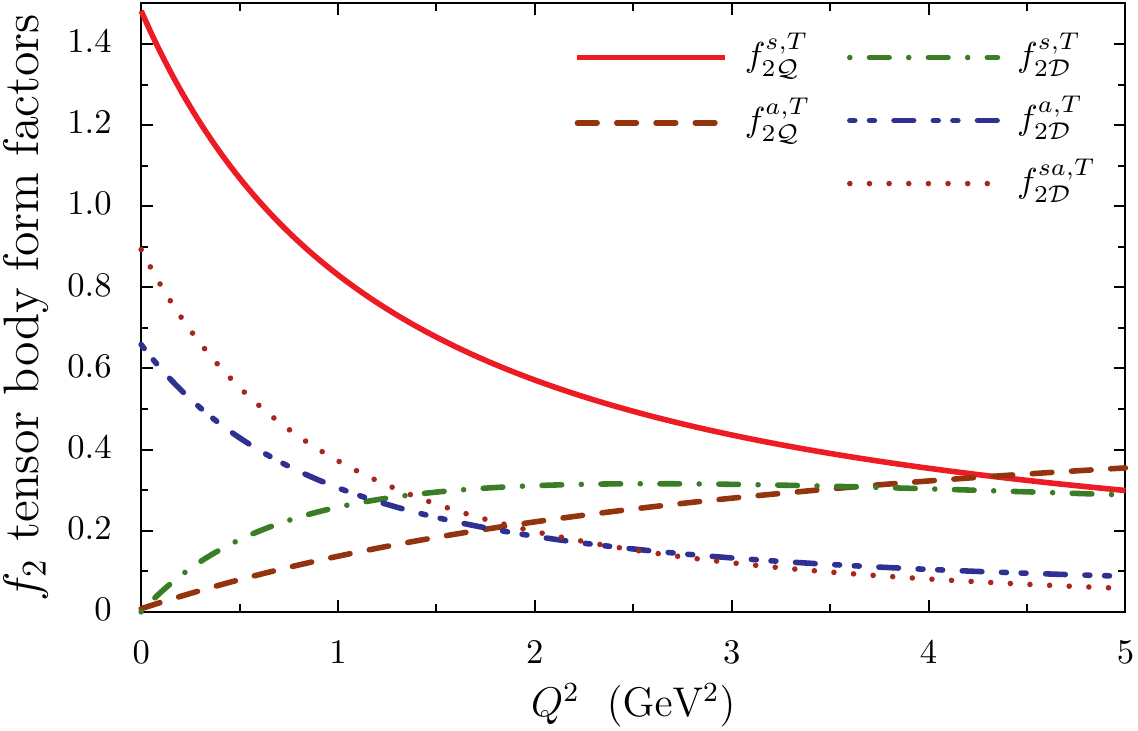}}
\caption{(Colour online) Nucleon Dirac (\textit{upper panel}) and Pauli (\textit{lower panel}) body form factors which result from a tensor coupling to the quarks in the Feynman diagrams of Fig.~\ref{fig:nucleon_em_current_feynman_diagrams}. To obtain their contribution to the  nucleon form factors these results must be multiplied by the appropriate isospin factors, as in Eqs.~\eqref{eq:f1U_V} and \eqref{eq:f1D_V} and the dressed quark Pauli form factors.}
\label{fig:tensor_body_form_factors}
\end{figure}

\begin{table*}[tbp]
\addtolength{\tabcolsep}{4.5pt}
\addtolength{\extrarowheight}{2.0pt}
\begin{tabular}{c|cccccccccc|cccc}
\hline\hline
                  & $f^{s,V}_{i\calQ}$ & $f^{a,V}_{i\calQ}$ & $f^{s,V}_{i\calD}$ & $f^{a,V}_{i\calD}$ & $f^{sa,V}_{i\calD}$ &
                    $f^{s,T}_{i\calQ}$ & $f^{a,T}_{i\calQ}$ & $f^{s,T}_{i\calD}$ & $f^{a,T}_{i\calD}$ & $f^{sa,T}_{i\calD}$ & 
                    $f^{U,V}_{ip}$ & $f^{D,V}_{ip}$ & $f^{U,T}_{ip}$ & $f^{D,T}_{ip}$ \\[0.6ex]
\hline
Dirac   & 0.688 & \ph{-}0.312 & \ph{-}0.688 & 0.312 &  0    & 0     & 0     & 0   & 0     &  0     & 2\ph{.00} & \ph{-}1\ph{.00} & 0\ph{.00} & \ph{-}0\ph{.00}\\
Pauli   & 1.134 &      -0.451 &      -0.546 & 0.472 & 0.666 & 1.482 & 0.008 & 0.0 & 0.659 &  0.893 & 1.61      & -1.07           & 3.10      & -0.29\\
\hline\hline
\end{tabular}
\caption{Nucleon Dirac and Pauli body form factors evaluated at $Q^2 = 0$. The subscript $i=1,2$ corresponds to either the first or second row of the Table. An entry with only one significant figure takes that exact value because of charge conservation. The last four columns give results for the vector and tensor versions of Eqs.~\eqref{eq:f1U_V}--\eqref{eq:f1D_V} at $Q^2 = 0$. To obtain nucleon form factor results at $Q^2 = 0$ these results must be multiplied the appropriate quark charge for the vector coupling diagrams and by the appropriate dressed quark anomalous magnetic moment for the tensor coupling diagrams.}
\label{tab:nucleon_body_form_factors}
\end{table*}

The proton body form factors in Eq.~\eqref{eq:FpQ}, which represent the sum of the six Feynman diagrams of Fig.~\ref{fig:nucleon_em_current_feynman_diagrams}, have the structure
\begin{align}
\label{eq:f1U_V}
\hs*{-1.5mm}f_{ip}^{U,V} &= f^{s,V}_{i\calQ} + \tfrac{1}{3}f^{a,V}_{i\calQ} + f^{s,V}_{i\calD} + \tfrac{5}{3}f^{a,V}_{i\calD} + \tfrac{1}{\sqrt{3}}f^{sa,V}_{i\calD}, \\
\label{eq:f1D_V}
\hs*{-1.5mm}f_{ip}^{D,V} &= \hs{11mm}    \tfrac{2}{3}f^{a,V}_{i\calQ} + f^{s,V}_{i\calD} + \tfrac{1}{3}f^{a,V}_{i\calD} 
- \tfrac{1}{\sqrt{3}}f^{sa,V}_{i\calD}.
\end{align}
For equal current quark masses the neutron body form factors in Eq.~\eqref{eq:FnQ} are given by
\begin{align}
f_{in}^{D,V} = f_{ip}^{U,V} \quad \text{and} \quad f_{in}^{U,V} = f_{ip}^{D,V},
\label{eq:neutronbody}
\end{align}
and therefore the nucleon body form factors satisfy the constraints imposed by charge symmetry. Expressions for the nucleon tensor body form factors are obtained from Eqs.~\eqref{eq:f1U_V}--\eqref{eq:neutronbody} with $V \to T$. The nomenclature for these nucleon body form factors is: a subscript $\calQ$ implies that the photon couples directly to a quark (\textit{quark diagram}) and a subscript $\calD$ implies that the photon couples to (a quark inside) a diquark (\textit{diquark diagram}); a superscript $s$ indicates
that the diagram contains only a scalar diquark, while the superscript $a$ only an axialvector diquark and the superscript $sa$ implies the sum of the two diagrams where a photon induces a transition between scalar and axialvector diquarks. The numerical coefficients in Eqs.~\eqref{eq:f1U_V} and \eqref{eq:f1D_V} arise from the isospin structure of the proton Faddeev and the quark-photon vertices, given in Eqs.~\eqref{eq:faddeevvertex} and \eqref{eq:quark_photon_vertex}, respectively.

Nucleon body form factor results for each diagram 
in Fig.~\ref{fig:nucleon_em_current_feynman_diagrams}, 
as expressed by Eqs.~\eqref{eq:f1U_V}--\eqref{eq:f1D_V}, are presented 
in Fig.~\ref{fig:vector_body_form_factors} for the vector coupling to the dressed 
quarks and in Fig.~\ref{fig:tensor_body_form_factors} for the tensor coupling. 
Table~\ref{tab:nucleon_body_form_factors} gives 
the $Q^2=0$ values of the nucleon body form factors. Charge conservation for the 
vector coupling implies that in this case diagrams with the same quark--diquark 
content must be equal at $Q^2 = 0$. Furthermore, with the normalization used here, 
the sum of quark diagrams and of diquark diagrams must each equal one in 
the vector case. For the vector coupling, charge conservation also forbids the 
scalar--axialvector diquark diagram ($sa$) from contributing to the charge. 
However, this diagram does give an important contribution to the nucleon anomalous magnetic moment. For 
the vector coupling diagrams the only object with a magnetic moment is the 
axialvector diquark. Thus the non-zero values for the other $f_2$ body form factor 
diagrams, in the lower panel of Fig.~\ref{fig:vector_body_form_factors}, 
indicate that the associated pieces of the nucleon wave function have sizable 
$p$ and $d$ wave components. Therefore the nucleon wave function contains 
a significant amount of quark orbital angular momentum.

Figure~\ref{fig:tensor_body_form_factors} and Table~\ref{tab:nucleon_body_form_factors} 
demonstrate that the tensor coupling diagrams do not contribute to the nucleon charge, 
which is consistent with constraints imposed by the Ward--Takahashi identity for 
the nucleon electromagnetic current. However, these diagrams do have an important 
impact on the anomalous magnetic moment. The $Q^2$ behavior of the form factors 
is also influenced by the tensor coupling diagrams. However, once multiplied by 
the dressed quark Pauli form factors, their contribution diminishes rapidly 
with $Q^2$, being of little importance for $Q^2 \gtrsim 1\,$GeV$^2$.

In Sect.~\ref{sec:NJL} the nucleon Faddeev equation was solved by first making a pole approximation for the diquark $t$-matrices -- see for example Eqs.~\eqref{eq:scalarpropagatorpoleform} and \eqref{eq:axialpropagatorpoleform}. For a consistent nucleon form factor calculation we must therefore approximate all two-body $t$-matrices by their pole form, which also includes the quark-photon vertex obtained from the inhomogeneous BSE illustrated in Fig.~\ref{fig:quarkphotonvertex}. Expressing $\Pi_{VV}(q^2) = q^2\,\hat{\Pi}_{VV}(q^2)$ in Eq.~\eqref{eq:bseformfactors} and expanding $\hat{\Pi}_{VV}(q^2)$ about either the rho or omega pole mass, we obtain the following  results for the pole forms of the BSE form factors:
\begin{align}
F_{1i}(Q^2)  &= \frac{1}{1 + Q^2/ m_i^2}, \qquad i \in \o,\,\rho,
\label{eq:vmdformfactors}
\end{align}
which is the familiar vector meson dominance result. The dressed quark form factors therefore maintain the vector meson pole structure 
in the time-like region obtained in the original BSE results of Eq.~\eqref{eq:bseformfactors}. For the nucleon form factor calculations the result in Eq.~\eqref{eq:vmdformfactors} will replace the full BSE result of Eq.~\eqref{eq:bseformfactors} used in the dressed quark form factors, for example, in Eq.~\eqref{eq:bseconstituent} and Eqs.~\eqref{eq:f1U_quarkpion}--\eqref{eq:f2D_quarkpion}.

\begin{figure}[t]
\subfloat{\centering\includegraphics[width=\columnwidth,clip=true,angle=0]{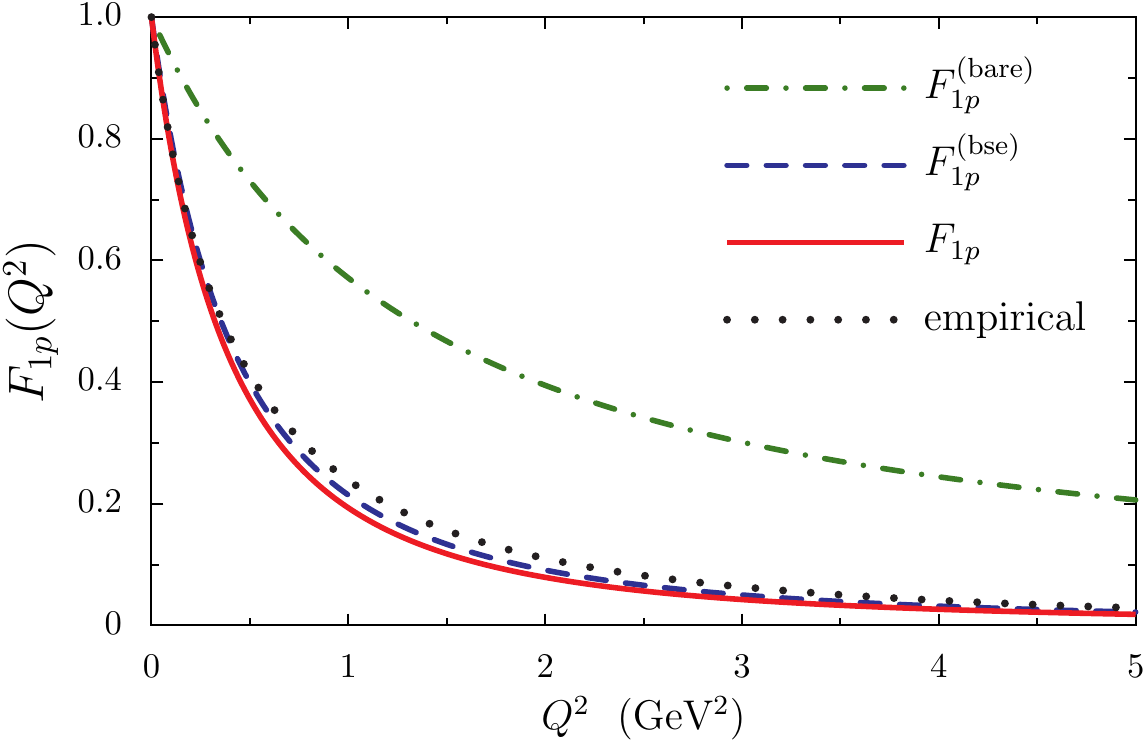}} \\
\subfloat{\centering\includegraphics[width=\columnwidth,clip=true,angle=0]{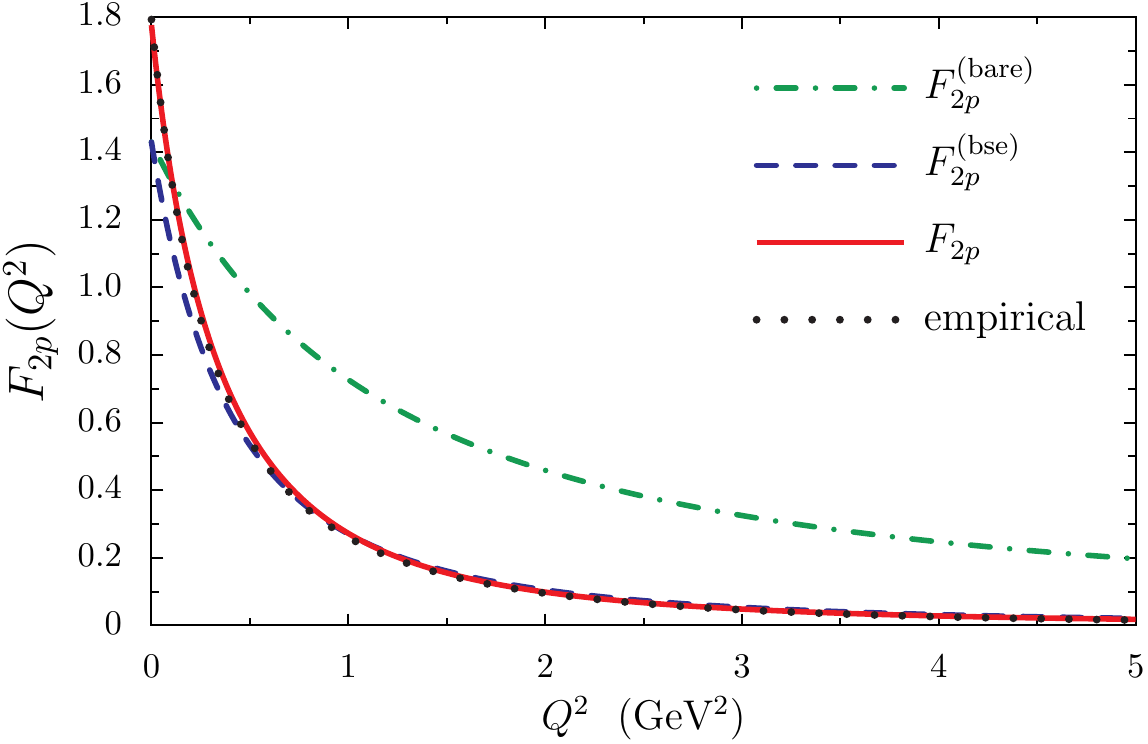}}
\caption{(Colour online) Results for the proton Dirac (\textit{upper panel}) and Pauli (\textit{lower panel}) form factors. In each case the \textit{dot-dashed} curve (superscript (bare)) gives the result when the constituent quark form factors are those of an elementary Dirac particle, the \textit{dashed} curve (superscript (bse)) includes the quark-photon vertex dressing effects from the BSE and the \textit{solid} curve is the full result which also includes pion loop effects. The dotted curve is the empirical result from Ref.~\cite{Kelly:2004hm}.}
\label{fig:proton_form_factors}
\end{figure}

\begin{figure}[t]
\subfloat{\centering\includegraphics[width=\columnwidth,clip=true,angle=0]{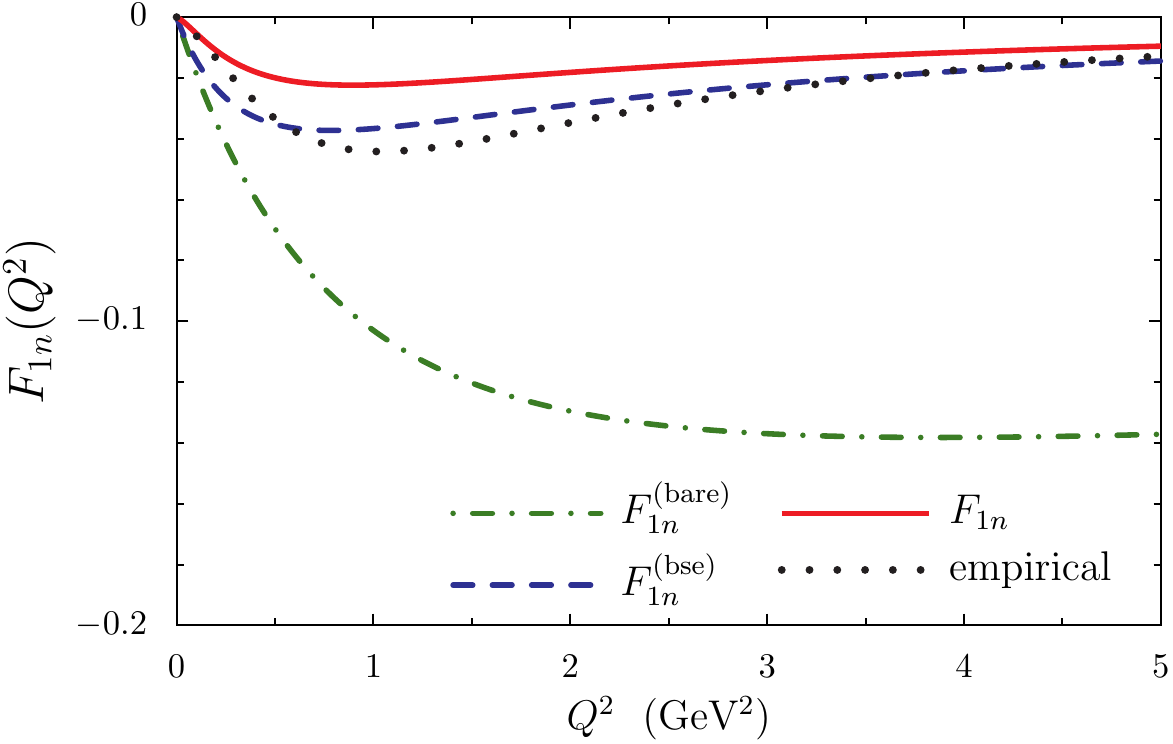}}\\
\subfloat{\centering\includegraphics[width=\columnwidth,clip=true,angle=0]{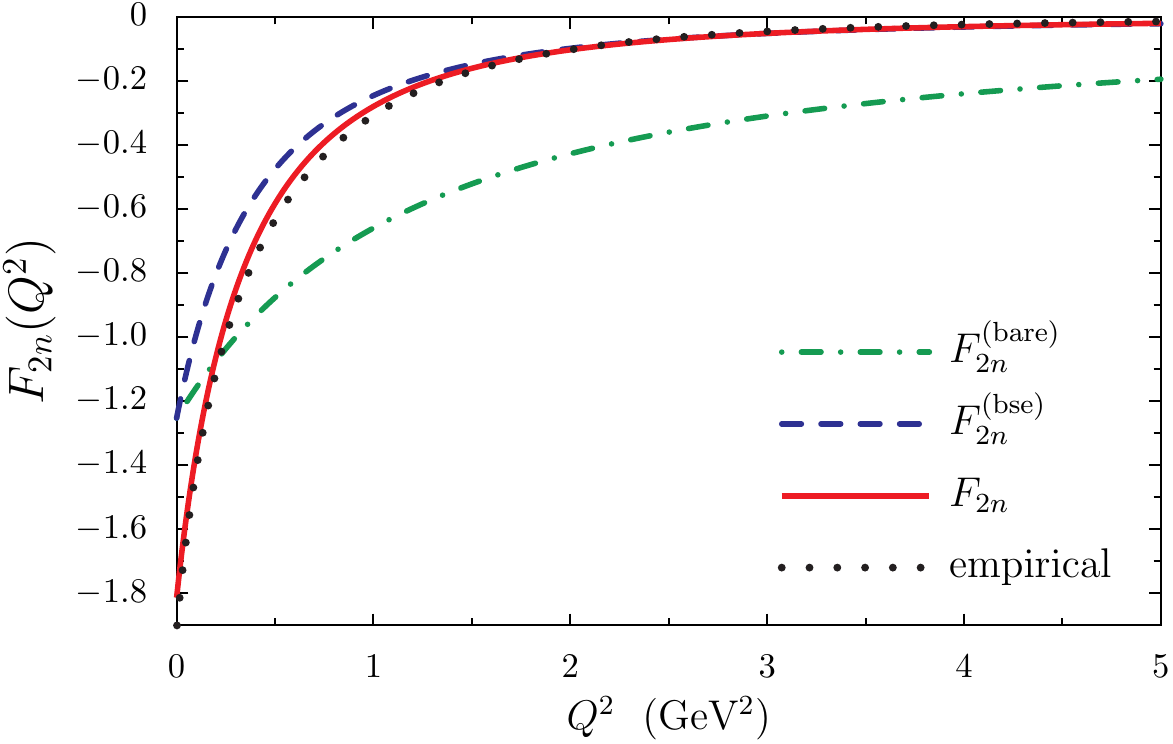}}
\caption{(Colour online) Results for the neutron Dirac (\textit{upper panel}) and Pauli (\textit{lower panel}) form factors. In each case the \textit{dot-dashed} curve (superscript (bare)) gives the results when the constituent quark form factors are those of an elementary Dirac particle, the \textit{dashed} curve (superscript (bse)) includes the quark-photon vertex dressing effects from the BSE and the \textit{solid} curve is the full result which also includes pion loop effects. The dotted curve is the empirical result from Ref.~\cite{Kelly:2004hm}.}
\label{fig:neutron_form_factors}
\end{figure}

\begin{figure}[t]
\subfloat{\centering\includegraphics[width=\columnwidth,clip=true,angle=0]{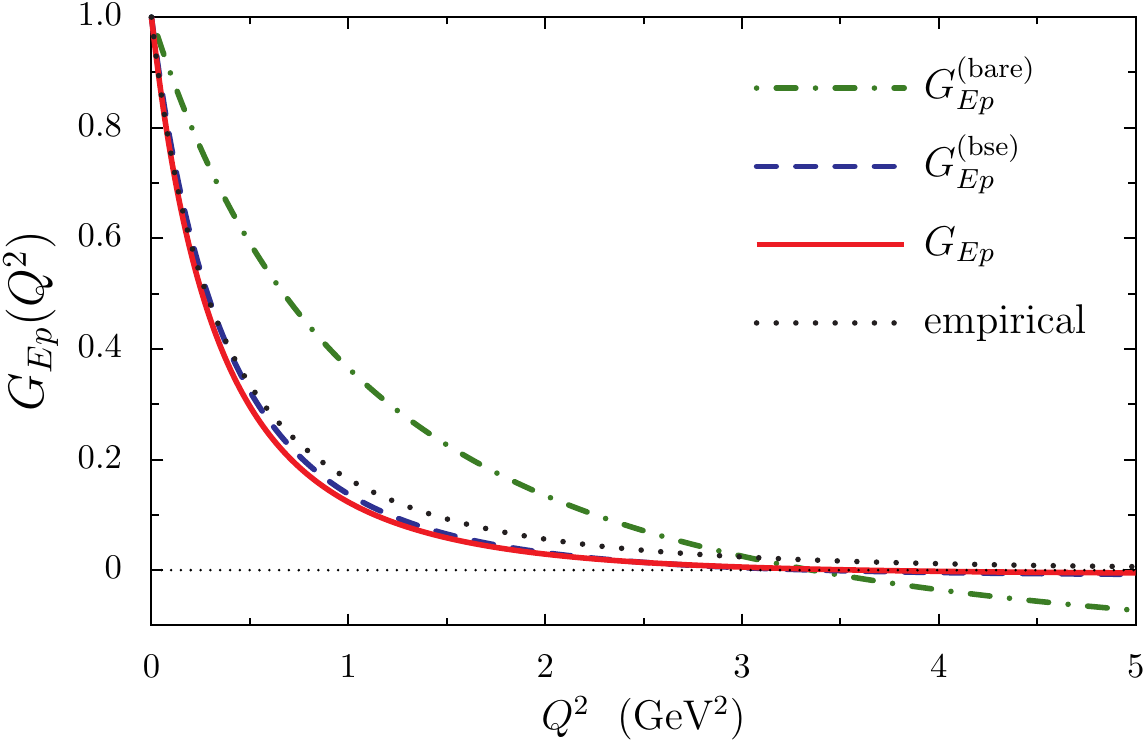}}\\
\subfloat{\centering\includegraphics[width=\columnwidth,clip=true,angle=0]{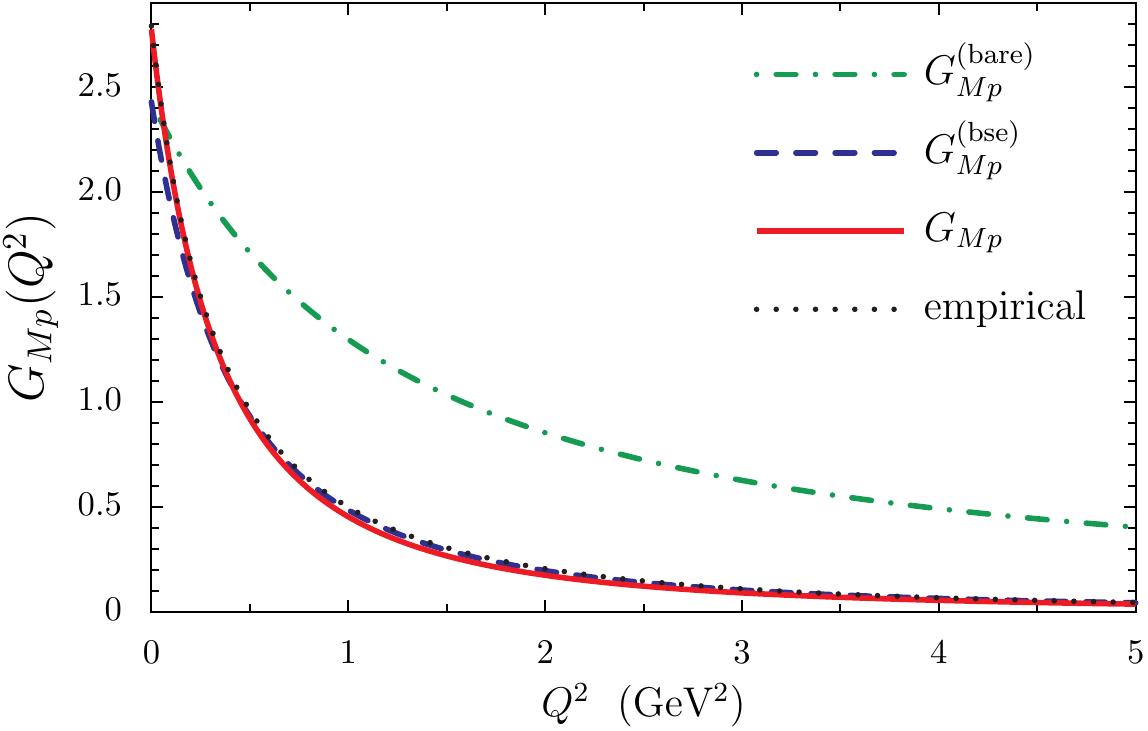}}
\caption{(Colour online) Results for the proton Sachs electric (\textit{upper panel}) and magnetic (\textit{lower panel}) form factors. In each case the dot-dashed curve (superscript (bare)) is the result when the constituent quark form factors are those of an elementary Dirac particle, the dashed curve (superscript (bse)) includes the quark-photon vertex dressing effects from the BSE and the solid curve is the full result which also includes pion loop effects. The dotted curve is the empirical result from Ref.~\cite{Kelly:2004hm}.}
\label{fig:proton_sachs}
\end{figure}

\begin{figure}[t]
\subfloat{\centering\includegraphics[width=\columnwidth,clip=true,angle=0]{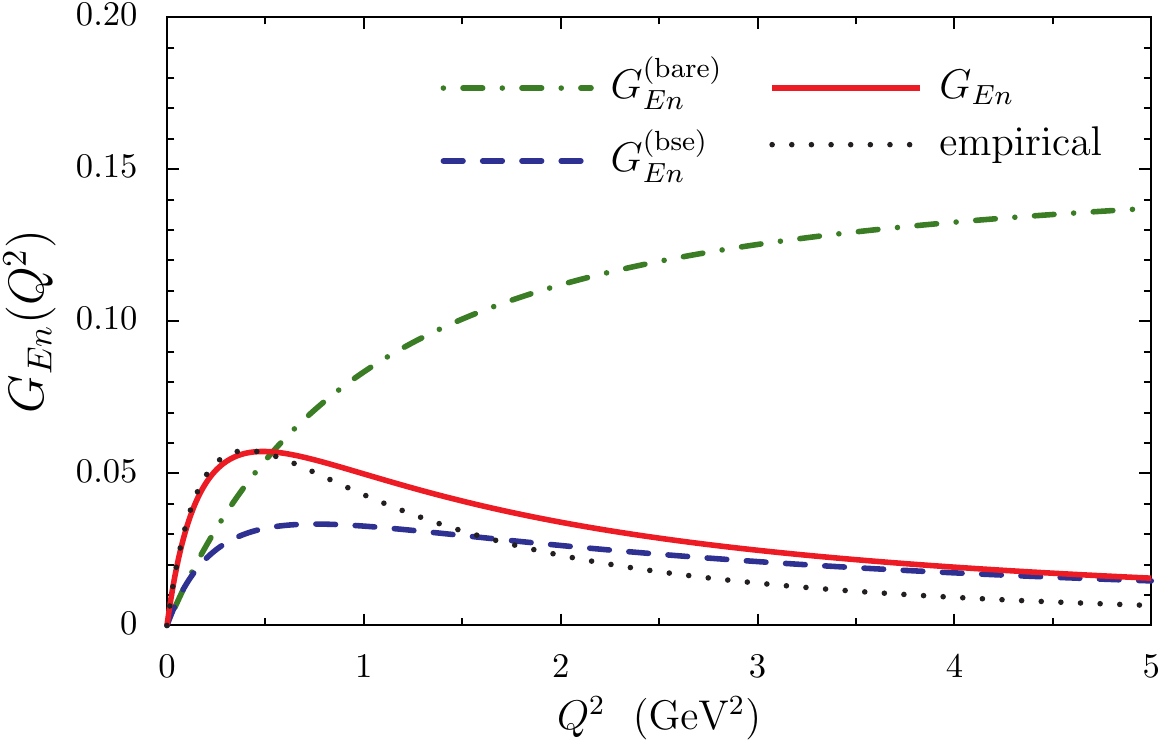}}\\
\subfloat{\centering\includegraphics[width=\columnwidth,clip=true,angle=0]{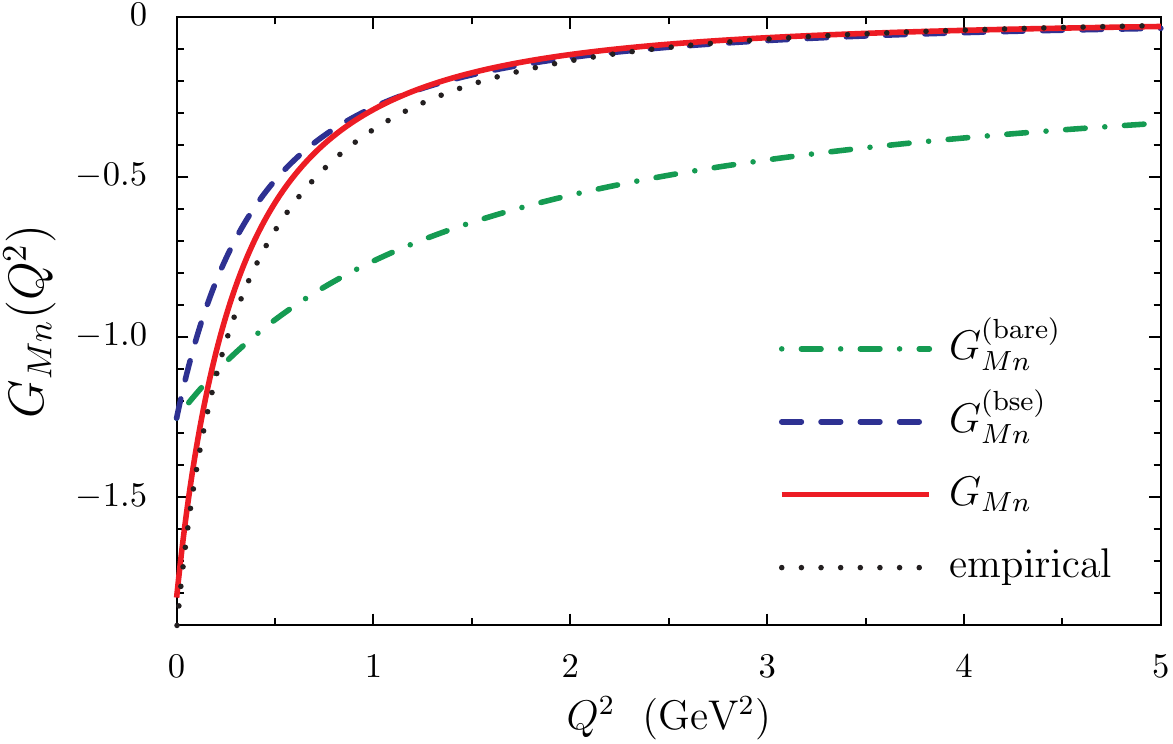}}
\caption{Colour online) Results for the neutron Sachs electric (\textit{upper panel}) and magnetic (\textit{lower panel}) form factors. In each case the dot-dashed curve (superscript (bare)) is the results when the constituent quark form factors are those of an elementary Dirac particle, the dashed curve 
(superscript (bse)) includes the quark-photon vertex dressing effects from the BSE and the solid curve is the full result which also includes pion loop effects. The dotted curve is the empirical result from Ref.~\cite{Kelly:2004hm}.}
\label{fig:neutron_sachs}
\end{figure}

Dirac and Pauli form factor results for the proton and neutron are presented 
in Fig.~\ref{fig:proton_form_factors} and \ref{fig:neutron_form_factors}, 
respectively, while results for the Sachs form factors are given 
in Fig.~\ref{fig:proton_sachs} and \ref{fig:neutron_sachs}. The three curves 
in each figure represent results for the three variants of the dressed 
quark form factors used in Eqs.~\eqref{eq:protondirac}--\eqref{eq:FnQ}. 
The dot-dashed curve is the result where the dressed quarks  are treated as pointlike and therefore their Dirac form factors are constants equal to the quark charges and the Pauli form factors are zero. These results are labelled with the superscript (bare). Results for the nucleon form factors that include the dressing of the quark-photon vertex by vector mesons, generated by Eq.~\eqref{eq:vmdformfactors}, are illustrated by the dashed lines, with the superscript (bse). Finally, we use dressed quark form factors that also incorporate effects from pion loops, which generate a non-zero Pauli form factor for the dressed quarks. These results are illustrated as the solid lines (without a superscript label).

The full results for the nucleon form factors, including pion loop effects, 
display good agreement with the empirical parametrizations from 
Ref.~\cite{Kelly:2004hm}, which are illustrated as the dotted curves 
in Figs.~\ref{fig:proton_form_factors} through \ref{fig:neutron_sachs}. 
Both the proton and neutron Dirac form factor are slightly softer than the empirical 
parametrizations, whereas the Pauli form factors are in almost perfect agreement. 
The dressing of the quark-photon vertex by the pole form of the 
BSE (Eq.~\eqref{eq:vmdformfactors}) results in a significant softening of all 
nucleon form factors, proving critical for a realistic $Q^2$ dependence of 
the form factors. Pion loop corrections result in a further 50\% reduction of 
the neutron Dirac form factor for low to moderate $Q^2$ and significantly enhance 
the nucleon Pauli form factors for $Q^2 \lesssim 1\,$GeV$^2$. 
These enhancements correspond to increases in the magnitude of the proton and 
neutron anomalous magnetic moments by $25\%$ and $45\%$, as indicated in Table~\ref{tab:nucleon_kappa_radii}. For the proton and neutron magnetic moments we find $\mu_p = 2.78\,\mu_N$ and $\mu_n = -1.81\,\mu_N$, which agree well with the experimental values of $\mu_p = 2.793\,\mu_N$ and $\mu_n =-1.913\,\mu_N$~\cite{Beringer:1900zz}. To obtain the physical result $\lf|\kappa_n\rg| > \kappa_p$ for the nucleon anomalous magnetic 
moments, we find that the dressed quark anomalous magnetic moments of 
Eq.~\eqref{eq:quarkanomalousresults} are critical. In particular, $\kappa_U$ must 
be positive and $\kappa_D$ negative, with $\lf|\kappa_D\rg| > \kappa_U$. 
We obtain $\lf|\kappa_D\rg| > \kappa_U$ because the second diagram 
in Fig.~\ref{fig:quarkpionphotonvertex} only contributes to the dressed down quark 
anomalous magnetic moment (c.f. Eqs.~\eqref{eq:f2U_quarkpion} and \eqref{eq:f2D_quarkpion}), 
giving an additional negative contribution.

%

Results for the charge and magnetic radii, defined by Eqs.~\eqref{eq:charge_radius} and \eqref{eq:magnetic_radius}, are given in Table~\ref{tab:nucleon_kappa_radii} for the two cases where the dressed quark form factors are given by Eq.~\eqref{eq:vmdformfactors}  and where pion loop effects are also included. The pion loop effects result in a $65\%$ increase in magnitude of the neutron charge radius, a $19\%$ increase in its magnetic radius, while the proton charge radius increases by 6\% and the magnetic radius by 12\%. All nucleon radii agree well with the empirical values taken from Ref.~\cite{Kelly:2004hm}. A recent global fit to data~\cite{Zhan:2011ji} found the proton charge and magnetic radius to be
\begin{align}
r_{Ep} &= 0.875 \pm 0.008(\text{exp}) \pm 0.006(\text{fit})~\text{fm}, \\
r_{Mp} &= 0.867 \pm 0.009(\text{exp}) \pm 0.018(\text{fit})~\text{fm},
\end{align}
and a recent Mainz experiment found~\cite{Bernauer:2010wm}
\begin{align}
r_{Ep} &= 0.879\ph{1}(5)_{\text{stat}}(4)_{\text{syst}}(2)_{\text{model}}(4)_{\text{group}}~\text{fm}, \\
r_{Mp} &= 0.777(13)_{\text{stat}}(9)_{\text{syst}}(5)_{\text{model}}(2)_{\text{group}}~\text{fm}.
\end{align}
Our proton results agree well with those of Ref.~\cite{Zhan:2011ji}. The origin of 
the sizeable discrepancy between the two experimental results for the proton magnetic 
radius is discussed, for example, in Ref.~\cite{Sick:2012zz}. In addition, in 
view of the muonic hydrogen controversy~\cite{Pohl:2010zza}, the experimental 
errors quoted in both places appear to be rather low.

\begin{table*}[tbp]
\addtolength{\tabcolsep}{7.6pt}
\addtolength{\extrarowheight}{2.2pt}
\begin{tabular}{l|cccccccccccccc}
\hline\hline
& $\mu^{\text{(bse)}}$ & $\mu$ & $\mu^{\text{exp}}$ && $r_E^{\text{(bse)}}$ & $r_E$ & $r_E^{\text{exp}}$ && $r_M^{\text{(bse)}}$ & $r_M$ & $r_M^{\text{exp}}$ \\
\hline
proton  & \ph{-}2.43 & \ph{-}2.78 & \ph{-}2.793 && \ph{-}0.81 & \ph{-}0.86 & \ph{-}0.863$\pm$0.004 && 0.76 & 0.84 & 0.848$\pm$0.003\\
neutron &      -1.25 &      -1.81 &      -1.913 &&      -0.20 &      -0.34 &      -0.335$\pm$0.055 && 0.74 & 0.88 & 0.907$\pm$0.016\\
\hline\hline
\end{tabular}
\caption{Results for the nucleon magnetic moments and radii, with dressed quark form factors given by Eqs.~\eqref{eq:bseconstituent} and \eqref{eq:vmdformfactors}, labelled with a superscript (bse), and results that also include pion cloud effects at the dressed quark level (these results do not carry a superscript label). Experimental results, labelled with a superscript exp, are taken from Ref.~\cite{Kelly:2004hm}.}
\label{tab:nucleon_kappa_radii}
\end{table*}

\begin{figure}[tbp]
\subfloat{\centering\includegraphics[width=\columnwidth,clip=true,angle=0]{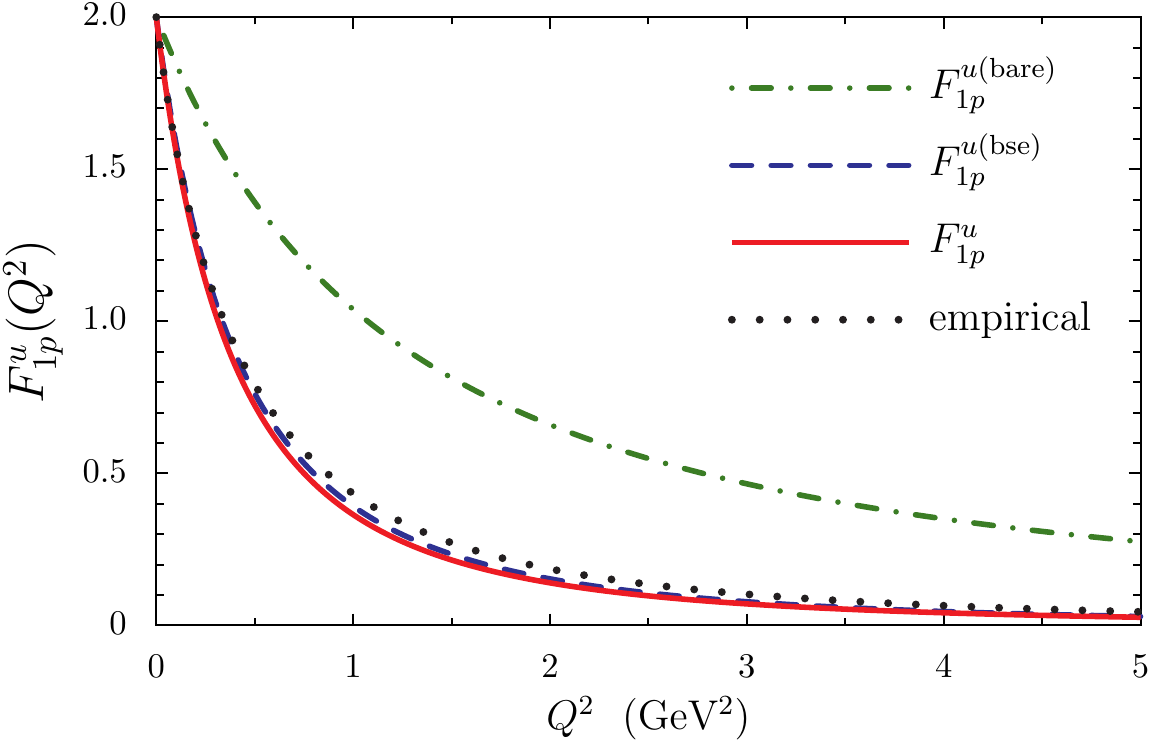}} \\
\subfloat{\centering\includegraphics[width=\columnwidth,clip=true,angle=0]{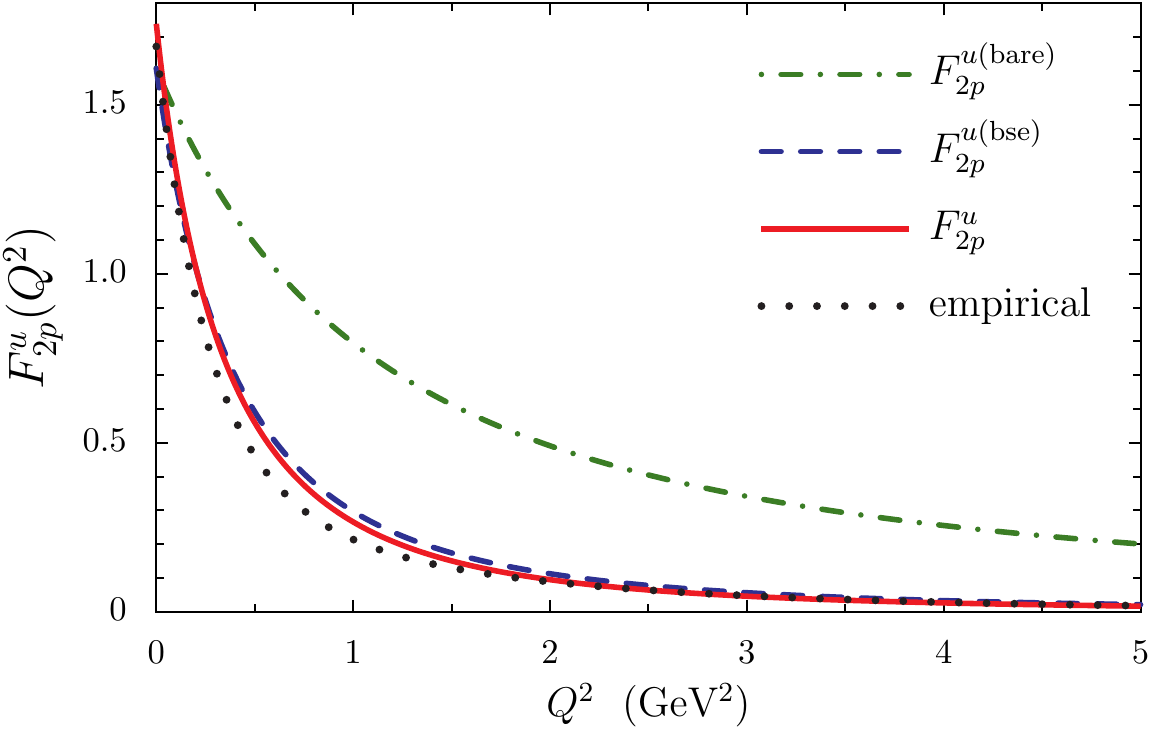}}
\caption{(Colour online) Proton up quark sector Dirac and Pauli form factors. The empirical results are obtained using Ref.~\cite{Kelly:2004hm} and Eq.~\eqref{eq:flavourseparation}.}
\label{fig:proton_up_form_factors}
\end{figure}

\begin{figure}[tbp]
\subfloat{\centering\includegraphics[width=\columnwidth,clip=true,angle=0]{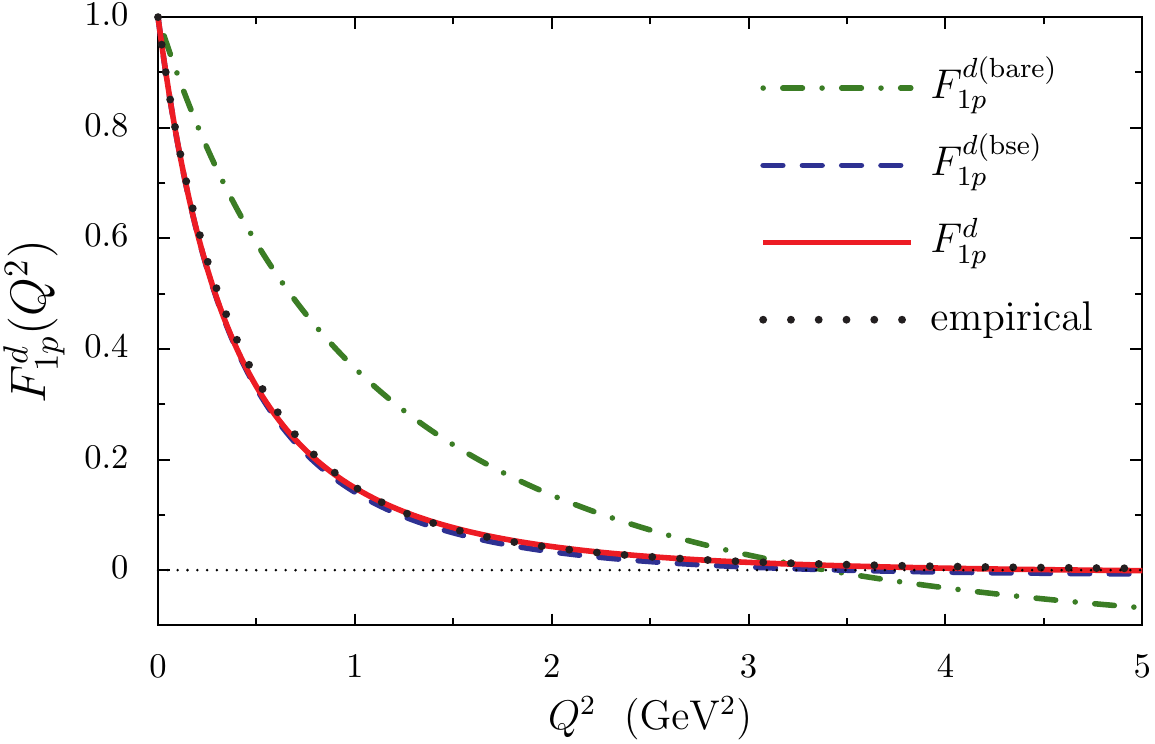}}\\
\subfloat{\centering\includegraphics[width=\columnwidth,clip=true,angle=0]{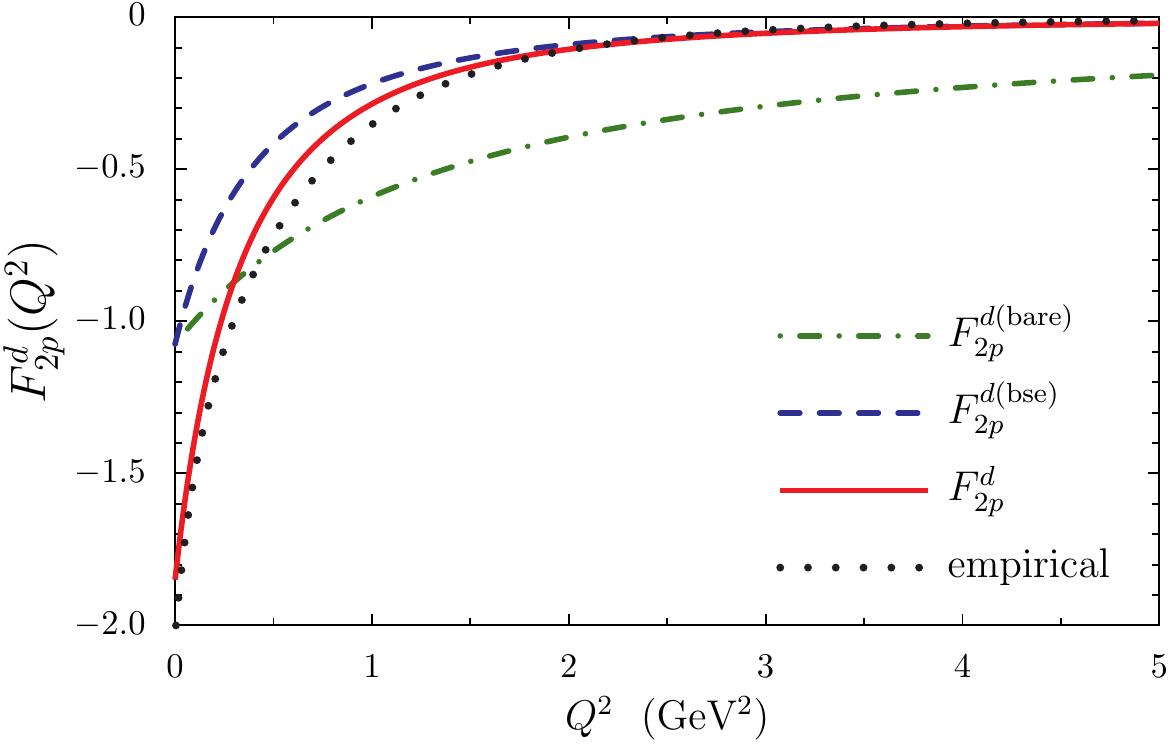}}
\caption{(Colour online) Proton down quark sector Dirac and Pauli form factors. The empirical results are obtained using Ref.~\cite{Kelly:2004hm} and Eq.~\eqref{eq:flavourseparation}.}
\label{fig:proton_down_form_factors}
\end{figure}

The flavour sector nucleon form factors defined by the dressed quarks, 
as given in
Eqs.~\eqref{eq:FpQ}--\eqref{eq:FnQ}, 
do not satisfy the standard charge symmetry relations, that is
\begin{align}
\frac{F^U_{ip}}{e_u} &\neq \frac{F^D_{in}}{e_d} &&\text{and}&  \frac{F^D_{ip}}{e_d} &\neq \frac{F^U_{in}}{e_u}.
\end{align}
where $i=(1,\,2)$.\footnote{Here we must divide out the quark charges because they are included in the definition of the dressed quark form factor, see Eqs.~\eqref{eq:FpQ}--\eqref{eq:FnQ}.}
The reason for this lies not with the nucleon 
body form factors, c.f. Eq.~\eqref{eq:neutronbody}, but 
with the form factors of the dressed quarks. 
Dressed quarks are quasi-particles that contain an infinite number of
$u$ and $d$ current quarks. 
Hence a dressed up quark form factor, for example, contains
contributions from both $u$ and $d$ current quarks.
To obtain the nucleon quark sector form factors, 
defined in general in Eq.~\eqref{eq:quark_sectorform_factors}, the 
dressed quark form factors must be expressed in 
their quark sector form as given in Eqs.~\eqref{eq:F1uU}--\eqref{eq:F2dU}.
The nucleon quark sector form factors are therefore given by
\begin{align}
F^q_{ip} &= F_{1Q}^q\,f_{ip}^{Q,V} + F_{2Q}^q\,f_{ip}^{Q,T}, \\
F^q_{in} &= F_{1Q}^q\,f_{in}^{Q,V} + F_{2Q}^q\,f_{in}^{Q,T},
\end{align}
where $i=(1,\,2)$, $q=(u,\,d)$ and there is an implied sum 
over $Q=(U,\,D)$. These results satisfy 
the charge symmetry constraints
\begin{align}
F^u_{in} &= F^d_{ip} &&\text{and} & F^d_{in} &= F^u_{ip}.
\end{align}

\begin{figure}[tbp]
\subfloat{\centering\includegraphics[width=\columnwidth,clip=true,angle=0]{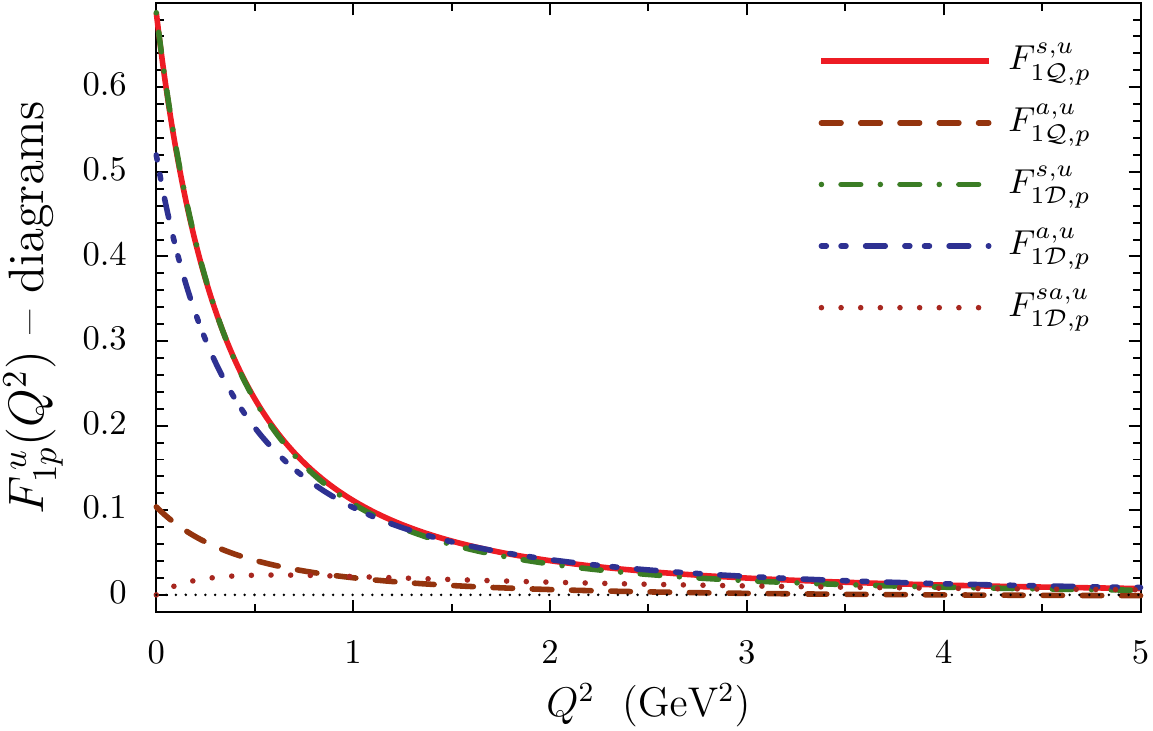}} \\
\subfloat{\centering\includegraphics[width=\columnwidth,clip=true,angle=0]{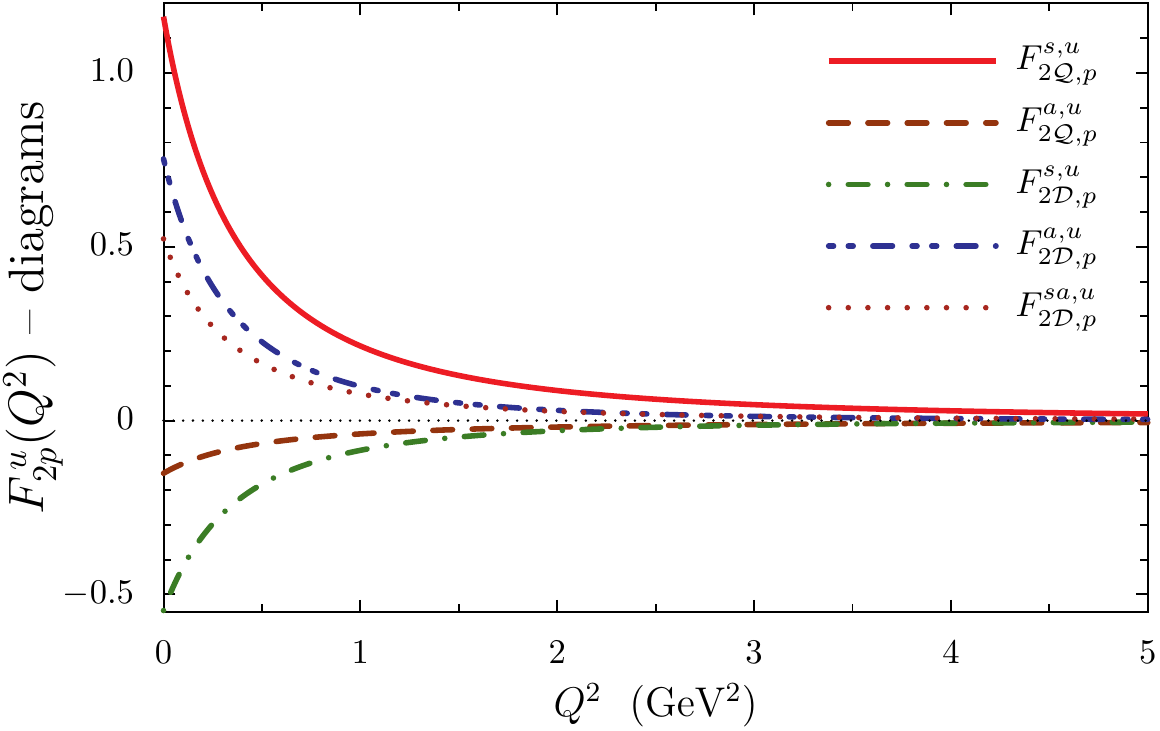}}
\caption{(Colour online) Total contributions to the proton $u$-sector form factors from each Feynman diagram in Fig.~\ref{fig:nucleon_em_current_feynman_diagrams}. These results include both the vector and tensor coupling contributions and the sum gives the total $u$-sector Dirac and Pauli proton form factors (solid lines in Fig.~\ref{fig:proton_up_form_factors}).}
\label{fig:flavour_sector_up_diagrams}
\end{figure}

\begin{figure}[tbp]
\subfloat{\centering\includegraphics[width=\columnwidth,clip=true,angle=0]{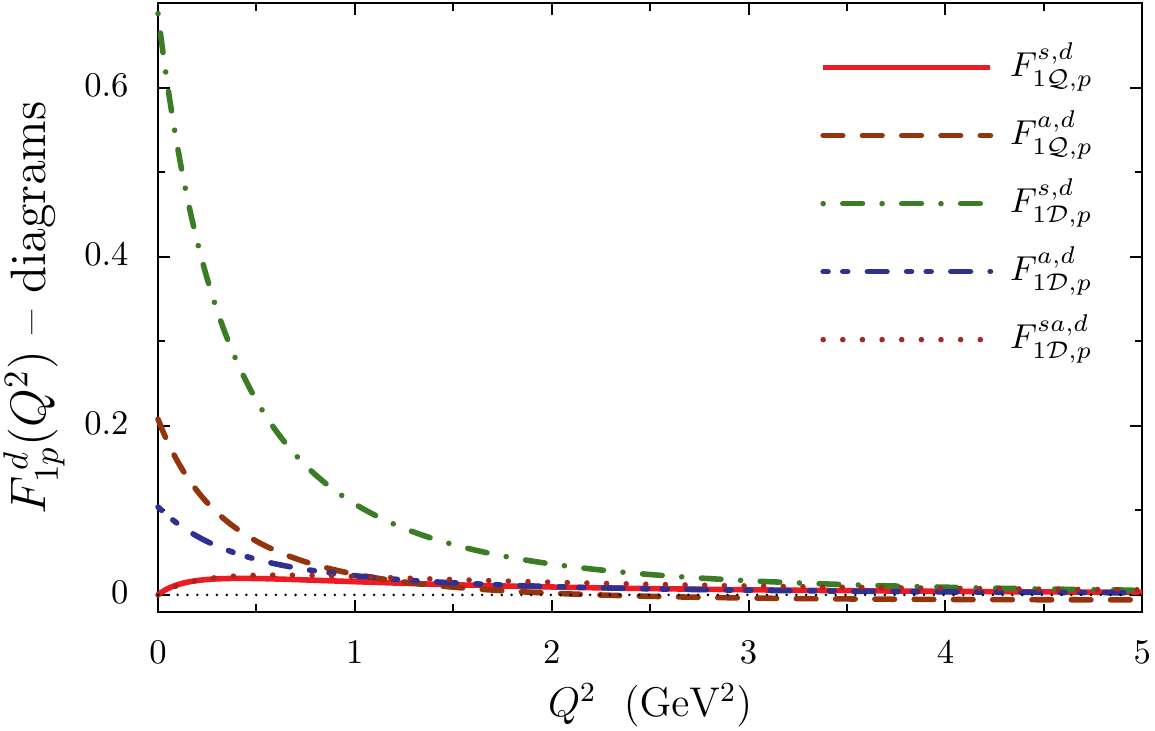}}\\
\subfloat{\centering\includegraphics[width=\columnwidth,clip=true,angle=0]{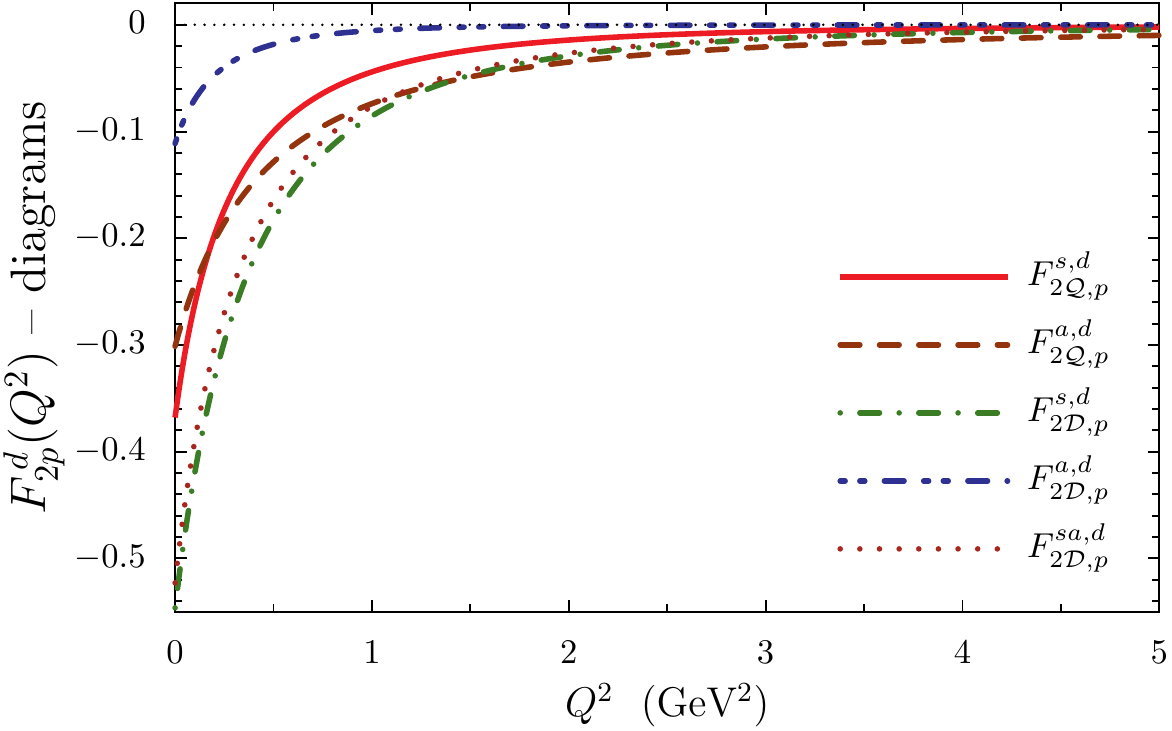}}
\caption{(Colour online) Total contributions to the proton $d$-sector form factors from each Feynman diagram in Fig.~\ref{fig:nucleon_em_current_feynman_diagrams}. These results include both the vector and tensor coupling contributions and the sum gives the total $d$-sector Dirac and Pauli proton form factors (solid lines in Fig.~\ref{fig:proton_down_form_factors}).}
\label{fig:flavour_sector_down_diagrams}
\end{figure}

\begin{table*}[t]
\addtolength{\tabcolsep}{7.0pt}
\addtolength{\extrarowheight}{2.2pt}
\begin{tabular}{c|cccccccccccccc}
\hline\hline
$q$ & $\kappa^{q,\text{(bse)}}$ & $\kappa^q$   & $\kappa^{q,\text{exp}}$  & & $r_E^{q,\text{(bse)}}$ & $r_E^q$ & $r_E^{q,\text{exp}}$ & & $r_M^{q,\text{(bse)}}$ & $r_M^q$  & $r_E^{q,\text{exp}}$ \\
\hline
$u$ sector & \ph{-}1.61 & \ph{-}1.74 & \ph{-}1.673 && 0.79 & 0.82 & 0.829$\pm$0.097 && \ph{-}0.77 & 0.83 & 0.816$\pm$0.087\\
$d$ sector &      -1.07 &      -1.85 &      -2.033 && 0.75 & 0.71 & 0.720$\pm$0.118 &&      -0.53 & 0.98 & 1.048$\pm$0.319\\
\hline\hline
\end{tabular}
\caption{Results for the quark sector contribution to 
the proton anomalous magnetic moments and radii, 
with constituent quark form factors given by
Eqs.~\eqref{eq:bseconstituent} and \eqref{eq:vmdformfactors} (labelled with (bse)) 
and results that also include the pion cloud. 
The experimental values for the quark sector anomalous 
magnetic moments and radii are obtained from from 
Ref.~\cite{Kelly:2004hm} using Eq.~\eqref{eq:flavourseparation}.}
\label{tab:nucleon_kappa_radii_quark_sector}
\end{table*}

Quark sector proton form factor results are presented in Figs.~\ref{fig:proton_up_form_factors} and \ref{fig:proton_down_form_factors}, for the three stages of sophistication in the description of the dressed quark form factors. Empirical results, shown by the dotted line, were  obtained from Ref.~\cite{Kelly:2004hm} using 
Eq.~\eqref{eq:flavourseparation}. While the agreement between our full results, 
which include pion loop effects, and the empirical parametrization is very good, 
{}for the $u$ quark sector we find that our Dirac form factor is slightly too soft 
and the Pauli form factor a little too hard. For the $d$ quark sector  
the Dirac form factor is in excellent agreement with the empirical parametrization, 
whereas the Pauli form factor is slightly too soft. As we shall see, such small 
differences can produce apparently large effects in the combination required 
to compute $G_E$.

\begin{table*}[t]
\addtolength{\tabcolsep}{4.9pt}
\addtolength{\extrarowheight}{2.2pt}
\begin{tabular}{c|cccccc|c|ccccccc}
\hline\hline
    & $r_1^{\text{(bse)}}$ & $r_1$ & $r_1^{\text{exp}}$ & $r_2^{\text{(bse)}}$ & $r_2$ & $r_2^{\text{exp}}$ & & $r_1^{q,\text{(bse)}}$ & $r_1^q$ & $r_1^{q,\text{exp}}$ & $r_2^{q,\text{(bse)}}$ & $r_2^q$ & $r_2^{q,\text{exp}}$ \\
\hline
proton  & 0.75 & 0.79 & 0.791 & 0.77 & 0.85 & 0.879 & $u$-sector & 0.76 & 0.79 & 0.795 & 0.77 & 0.88 & 0.841 \\
neutron & 0.20 & 0.09 & 0.119 & 0.76 & 0.88 & 0.911 & $d$-sector & 0.80 & 0.80 & 0.809 & 0.76 & 0.88 & 0.938 \\
\hline\hline
\end{tabular}
\caption{Results for radii defined by Eq.~\eqref{eq:radii}, for the proton and neutron Dirac and Pauli form factors, and for the quark sector proton Dirac and Pauli form factors. In each case we show results where the dressed quark form factors are given by Eqs.~\eqref{eq:bseconstituent} and \eqref{eq:vmdformfactors} (labelled with (bse)) and results that also include the pion cloud. The empirical values are obtained from Ref.~\cite{Kelly:2004hm} and for the quark sector results using Eq.~\eqref{eq:flavourseparation}.}
\label{tab:nucleon_radii_quark_sector}
\end{table*}

An interesting feature of these results is the role of the pionic corrections to the quark sector Pauli form factors. In contrast to the usual proton and neutron Pauli form factors, which each receive significant corrections from the pion cloud, for the quark sector form factors only $F^d_{2p}$ receives sizeable pionic corrections. For example, pion loop effects increase the magnitude of the $d$ sector anomalous magnetic moment by 73\%, whereas the $u$ quark sector only receives an 8\% correction. This result is a consequence of the Pauli quark sector form factors for the dressed quarks, where from Eq.~\eqref{eq:quark_sector_amm_dressed_quarks} we see that the $d$ quark sector contribution to the dressed up quark anomalous magnetic moment has a magnitude twelve times larger than the $u$ sector contribution, and the proton consists of two dressed up quarks and one dressed down quark. When compared with experiment the $d$-sector anomalous magnetic moment is 10\% too small and the $u$-sector 4\% too large.

Table~\ref{tab:nucleon_kappa_radii_quark_sector} presents results for the quark 
sector contribution to the proton anomalous magnetic moments and radii. 
We find that the pion cloud has only a minor impact on the $d$-sector 
charge radius and the $u$-sector radii, 
whereas the $d$-sector magnetic radius actually changes sign once pion loop 
effects are included. Again the origin of this lies with the large value of 
$\kappa_U^d$ in Eq.~\eqref{eq:quark_sector_amm_dressed_quarks}. With pion 
cloud corrections included, all our results for the charge and magnetic 
quark sector radii agree well with experiment.

Table~\ref{tab:nucleon_radii_quark_sector} gives results for the Dirac and Pauli radii, defined by Eq.~\eqref{eq:radii}, for the proton and neutron and, for the proton, the corresponding quark sector radii. The agreement with the empirical results of Ref.~\cite{Kelly:2004hm} is very good for the proton and neutron radii. For the proton quark sector radii, the Dirac radii results are in good agreement; however, the $u$ quark sector Pauli radius is slightly larger than experiment and the $d$ quark sector is 7\% smaller.

\begin{table*}[tp]
\addtolength{\tabcolsep}{6.0pt}
\addtolength{\extrarowheight}{2.2pt}
\begin{tabular}{c|cccccc|cccccccc}
\hline\hline
$q$       & $F^{s,q}_{1\calQ,p}$ & $F^{a,q}_{1\calQ,p}$ & $F^{s,q}_{1\calD,p}$ & $F^{a,q}_{1\calD,p}$ & $F^{sa,q}_{1\calD,p}$ & $F^{q}_{1p}$
          & $F^{s,q}_{2\calQ,p}$ & $F^{a,q}_{2\calQ,p}$ & $F^{s,q}_{2\calD,p}$ & $F^{a,q}_{2\calD,p}$ & $F^{sa,q}_{2\calD,p}$ & $F^{q}_{2p}$ \\[0.6ex]
\hline
$u$-sector & 0.69 & 0.10 & 0.69 & 0.52 & 0 & 2 & \ph{-}1.16 & -0.15 & -0.55 & \ph{-}0.75 & \ph{-}0.52 &  \ph{-}1.73 \\
$d$-sector & 0    & 0.21 & 0.69 & 0.10 & 0 & 1 &      -0.37 & -0.30 & -0.55 &      -0.11 &      -0.52 &       -1.85 \\
\hline\hline
\end{tabular}
\caption{Contributions to the nucleon quark-sector form factors from the various diagrams at $Q^2 = 0$. The vector contributions are obtained from the appropriate body form factors at $Q^2 = 0$ multiplied by isospin factors and quark charges. Therefore these results do not change with the various approximations for the dressed quark form factors. The tensor contributions are only non-zero if the dressed quarks have an anomalous magnetic moment, and in this framework this occurs solely from pion loop effects. Rows with an entry of ``0'' are identically zero because of charge conservation.}
\label{tab:nucleon_quark_sector_diagrams_Q20}
\end{table*}

\begin{figure}[tbp]
\centering\includegraphics[width=\columnwidth,clip=true,angle=0]{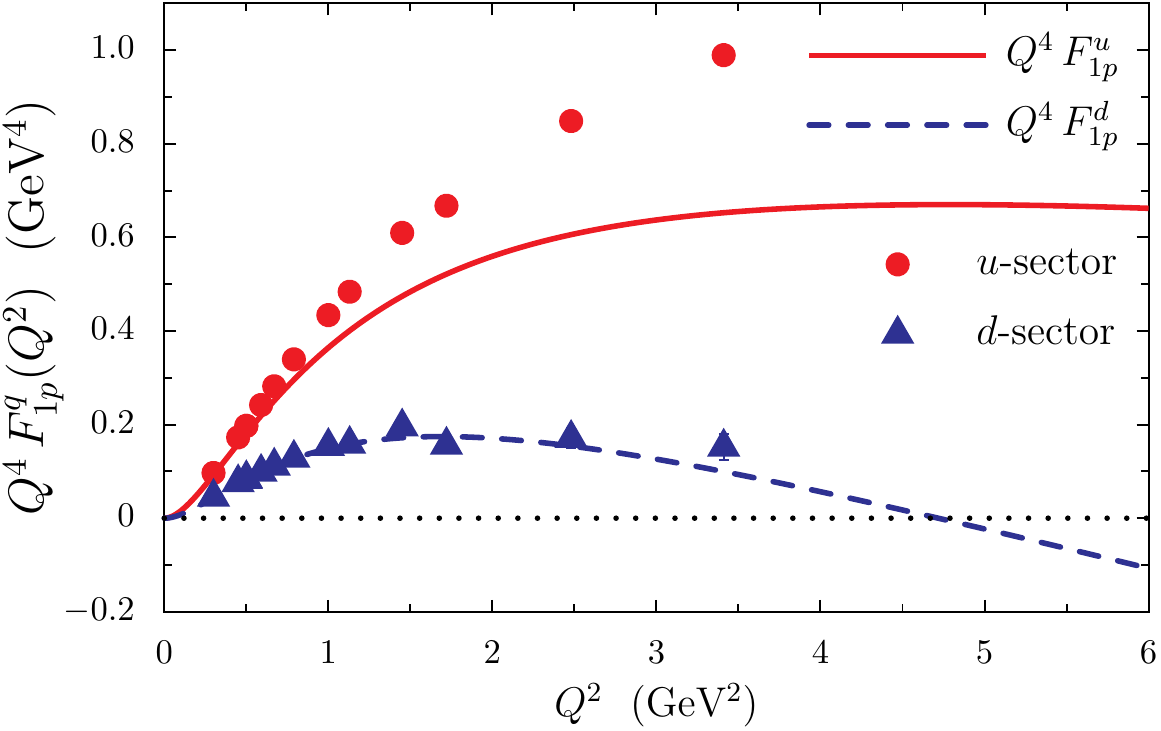}
\caption{(Colour online) Quark sector contributions to the proton Dirac form factor multiplied by $Q^4$. Experimental data are taken from Ref.~\cite{Cates:2011pz}.}
\label{fig:Q4F1_form_factors}
\end{figure}

Figures~\ref{fig:flavour_sector_up_diagrams} and \ref{fig:flavour_sector_down_diagrams} present results for the total contribution of each diagram in Fig.~\ref{fig:nucleon_em_current_feynman_diagrams} to the proton quark sector form factors. That is, the proton quark sector form factors are decomposed into
\begin{align}
F^q_{ip} &= F^{s,q}_{i\calQ,p} + F^{a,q}_{i\calQ,p} + F^{s,q}_{i\calD,p} + F^{a,q}_{i\calD,p} + F^{sa,q}_{i\calD,p},
\end{align}
where $i=(1,\,2)$, $q=(u,\,d)$ and each function represents the total contribution to each quark sector for the Feynman diagrams in Fig.~\ref{fig:nucleon_em_current_feynman_diagrams}.  Table~\ref{tab:nucleon_quark_sector_diagrams_Q20} gives results for the quark sector diagrams of Fig.~\ref{fig:nucleon_em_current_feynman_diagrams} evaluated at $Q^2 = 0$. For the Dirac form factors we see the dominance of the scalar diquark in the proton wave function, where these diagrams carry 69\% of both the $u$ and $d$ quark sector charges. Axialvector diquarks also play an important role for the $u$ quark sector form factors, carrying 26\% of the charge and 35\% of the anomalous magnetic moment. In the $d$ quark sector, $F^{s,d}_{2\calQ,p}$ would be zero without the effect of the pion cloud. The latter produces a contribution that constitutes 20\% of the $d$ sector anomalous magnetic moment.

\begin{figure}[tbp]
\centering\includegraphics[width=\columnwidth,clip=true,angle=0]{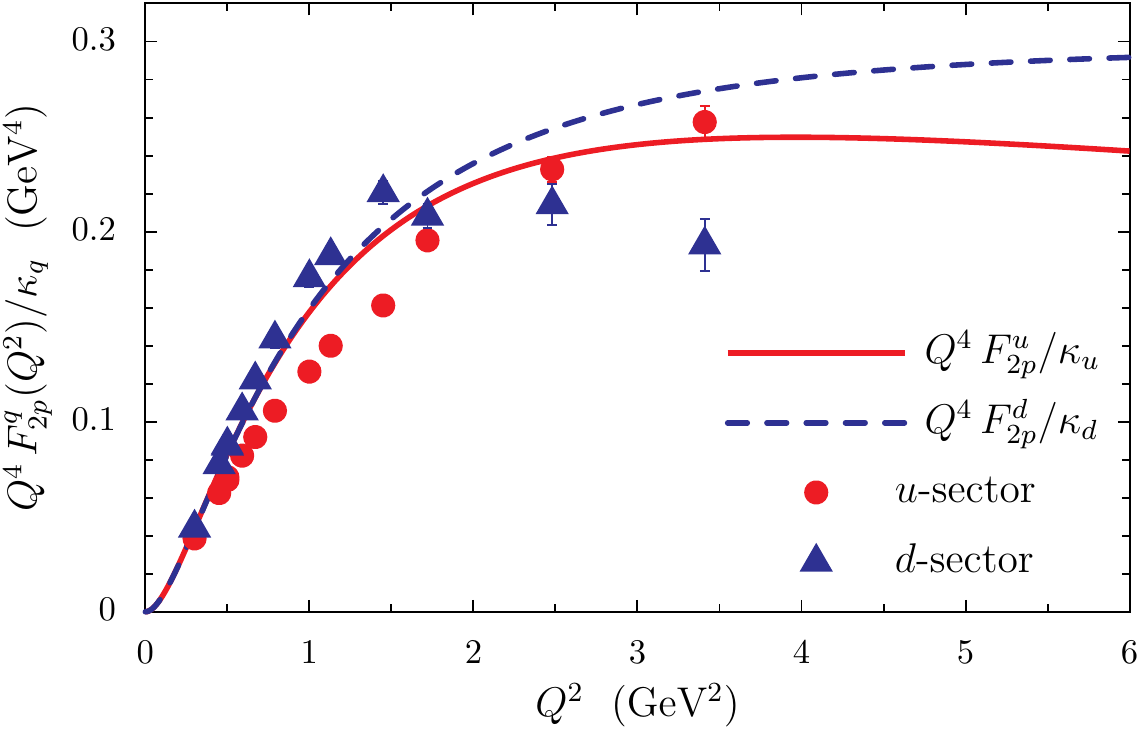}
\caption{(Colour online) Quark sector contributions to the proton 
Pauli form factor multiplied by $Q^4$. Experimental data are taken from Ref.~\cite{Cates:2011pz}.}
\label{fig:Q4F2_form_factors}
\end{figure}

Recent accurate neutron form factor data has enabled a precise experimental determination of the quark sector proton form factors, using Eq.~\eqref{eq:protonflavoursector}. The experimental quark sector results from Ref.~\cite{Cates:2011pz}, along with our results, are presented in Fig.~\ref{fig:Q4F1_form_factors} for the Dirac form factors and in Fig.~\ref{fig:Q4F2_form_factors} for the Pauli form factors. Prima facie, these experimental results 
are remarkable. For $Q^2$ beyond 1-2 GeV$^2$ the $d$ quark sector of the proton 
Dirac form factor is much softer than the $u$ quark sector. On the other hand, for 
the Pauli quark sector form factors, it is the $u$ quark sector that is softer for 
low $Q^2$. However, at around $Q^2 \sim 1.5\,$GeV$^2$ there is a cross-over and 
the $d$ quark sector form factor starts approaching zero more rapidly.

The empirical results illustrated in Fig.~\ref{fig:Q4F1_form_factors} are straightforward to understand within our framework. The dominant contributions to the quark sector Dirac form factors come from the two Feynman diagrams which involve only a quark and a scalar diquark. This is clear from the upper panels of Figs.~\ref{fig:flavour_sector_up_diagrams} and \ref{fig:flavour_sector_down_diagrams}. The upper panel in Figs.~\ref{fig:constituentquarkflavoursectorformfactors} demonstrates that the current $d$ quarks that contribute to $F_{1p}^d$ must primarily come from the dressed down quark, and these contributions are suppressed by order $1/Q^2$ relative to the current $u$ quarks from the quark diagram that contributes to $F_{1p}^u$. Thus the dominance of scalar diquark correlations in the nucleon clearly provides a very natural explanation of the data in Fig.~\ref{fig:Q4F1_form_factors}.

The zero-crossing in our result for $F_{1p}^d$ at $Q^2 \simeq 4.7\,$GeV$^2$ is also straightforward to understand. We first note that the large $Q^2$ behaviour of the form factors is governed by the quark diagrams in Fig.~\ref{fig:nucleon_em_current_feynman_diagrams},  because when the photon couples to a quark inside a diquark, the diquark form factors 
provide at least an additional factor of $1/Q^2$ relative to the quark diagrams. Considering only pointlike quarks, which is sufficient to study the large $Q^2$ behaviour, we have for the proton quark sector form factors
\begin{align}
\label{eq:FpUlargeQ2}
F_{ip}^u &~\stackrel{Q^2\to \infty}{\lra}~ f^{s,V}_{i\calQ} + \tfrac{1}{3}f^{a,V}_{i\calQ},\\[0.2ex]
F_{ip}^d &~\stackrel{Q^2\to \infty}{\lra}~ \hs*{11mm} \tfrac{2}{3}f^{a,V}_{i\calQ},
\end{align}
where $i=(1,\,2)$; c.f. Eqs.~\eqref{eq:f1U_V} and \eqref{eq:f1D_V}. Therefore the large $Q^2$ behaviour of $F_{1p}^d$ is governed by the nucleon body form factor $f^{a,V}_{1\calQ}$ (see Fig.~\ref{fig:vector_body_form_factors}), which becomes negative at large $Q^2$ and therefore $F_{1p}^d$ has a zero-crossing. Note that the empirical parameterizations of Ref.~\cite{Kelly:2004hm} also have a zero in $F_{1p}^d$ at $Q^2 \simeq 7.9\,$GeV$^2$.

\begin{figure}[tbp]
\subfloat{\centering\includegraphics[width=0.7\columnwidth,clip=true,angle=0]{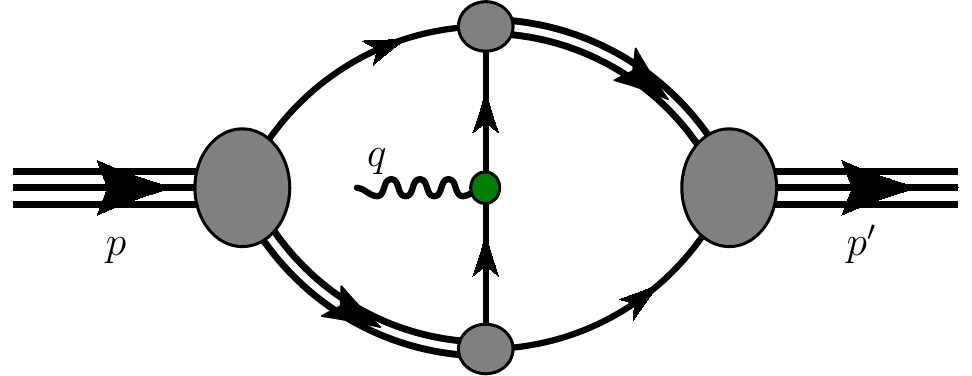}}
\caption{Exchange type diagrams that do not appear in our present form factor calculation because the static approximation is used for the quark exchange kernel.} 
\label{fig:exchange_diagrams}
\end{figure}

Understanding the $Q^2$ dependence of the proton Pauli quark sector form factors 
is more subtle within our model. Analogous to the Dirac form factor example, $F_{2p}^u$ receives a large contribution from the scalar quark diagram $f^{s,V}_{2\calQ}$, however, many other contributions are negative. In contrast all diagrams add constructively to the $F_{2p}^d$ form factor, which also receives a significant contribution from the pion cloud. 
Therefore at low to moderate $Q^2$ we find $F_{2p}^u/\kappa_u \sim F_{2p}^d/\kappa_d$, with reasonable agreement with the data. However, at larger $Q^2$ the two quark diagrams in Eq.~(\ref{eq:FpUlargeQ2}) partially cancel, giving $F_{2p}^u/\kappa_u < F_{2p}^d/\kappa_d$, which is opposite to the behaviour observed in the data. The suppression of $F_{2p}^d$ with respect to $F_{2p}^u$  at large $Q^2$ was found in Ref.~\cite{Cloet:2008re}, where a major difference from the framework used here is that we make the static approximation to the quark exchange kernel and therefore exchange type diagrams, as illustrated in Fig.~\ref{fig:exchange_diagrams}, are absent from our form  factor calculation. This is the likely reason for the discrepancy with experiment at large $Q^2$ observed in Fig.~\ref{fig:Q4F2_form_factors}.

\begin{figure}[tbp]
\centering\includegraphics[width=\columnwidth,clip=true,angle=0]{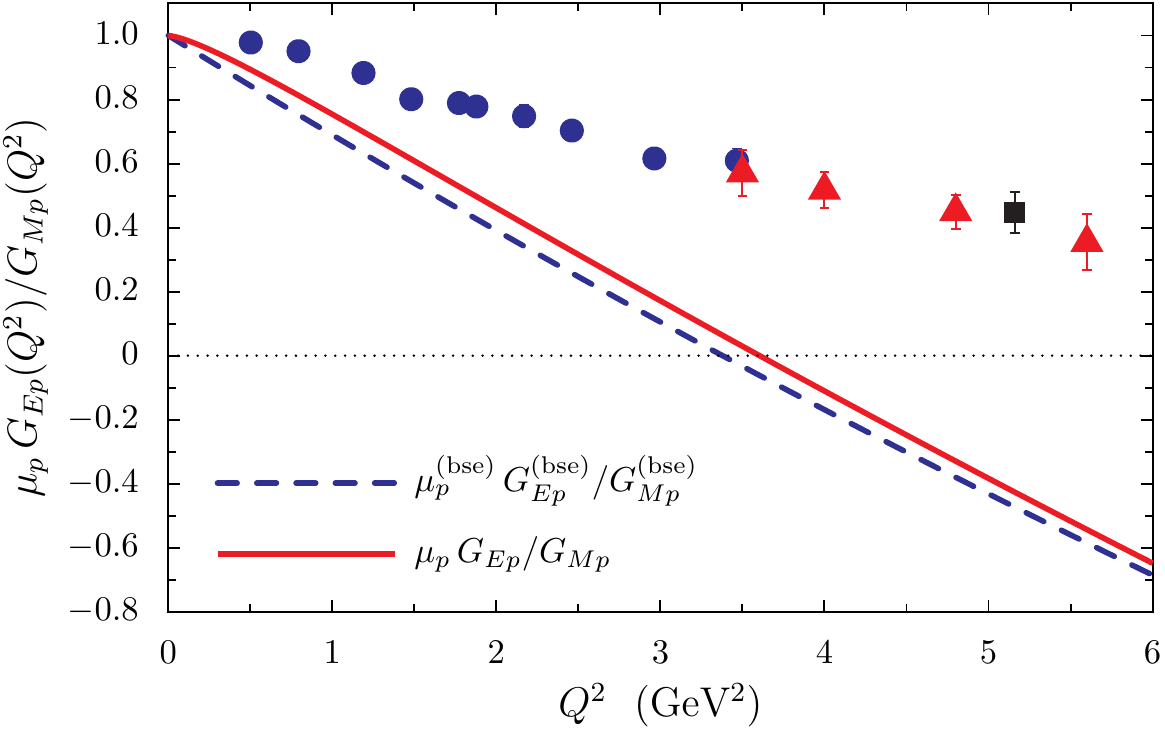}
\caption{(Colour online) Proton Sachs form factor ratio. Data are from Refs.~\cite{Jones:1999rz,Gayou:2001qd,Gayou:2001qt,Puckett:2010ac,Puckett:2011xg}.}
\label{fig:proton_sachs_ratio}
\end{figure}

Detailed results for the proton and neutron Sachs form factors are given in 
Appendix~\ref{sec:sachs}. Of contemporary
interest is the proton Sachs form factor ratio, $G_{Ep}/G_{Mp}$, for which our result 
is  presented in Fig.~\ref{fig:proton_sachs_ratio}. We find that this ratio 
decreases almost linearly with $Q^2$ but the slope we obtain is significantly 
larger than the experimental results obtained via the polarization transfer experiments,  
leading to a zero-crossing at $Q^2 \approx 3.7$ GeV$^2$.
So far no such zero-crossing has been seen in the data but if it were to occur 
it would have to be in the domain $Q^2 \gtrsim 8\,$GeV$^2$. 
The zero in the $G_{Ep}/G_{Mp}$ ratio found here results from a zero in $G_{Ep}$
and, as we have already noted,
the cancellation between $F_1$ and $F_2$ in the linear combination needed for 
$G_E$ means that even relatively small differences between the experimental and 
theoretical values of the individual from factors can be magnified there.
We find that this zero actually arises from the $u$ quark sector, 
as illustrated in the upper panel of Fig.~\ref{fig:sachs_u_sector}. 
This zero has its origin in the quark diagram with the scalar diquark spectator, 
which becomes negative at around $Q^2 \simeq 1.8\,$GeV$^2$ and dominates at large $Q^2$. 
This can be seen in the  upper panel of Fig.~\ref{fig:sachs_u_sector_diagrams}. 
A possible reason for the discrepancy with data for the $G_{Ep}/G_{Mp}$ ratio is 
the omission of exchange diagram contributions (illustrated in Fig.~\ref{fig:exchange_diagrams}), 
which do not appear in the model described herein. The running of the quark mass function
in QCD may also play an important role~\cite{Cloet:2013gva}.

\begin{figure}[tbp]
\centering\includegraphics[width=\columnwidth,clip=true,angle=0]{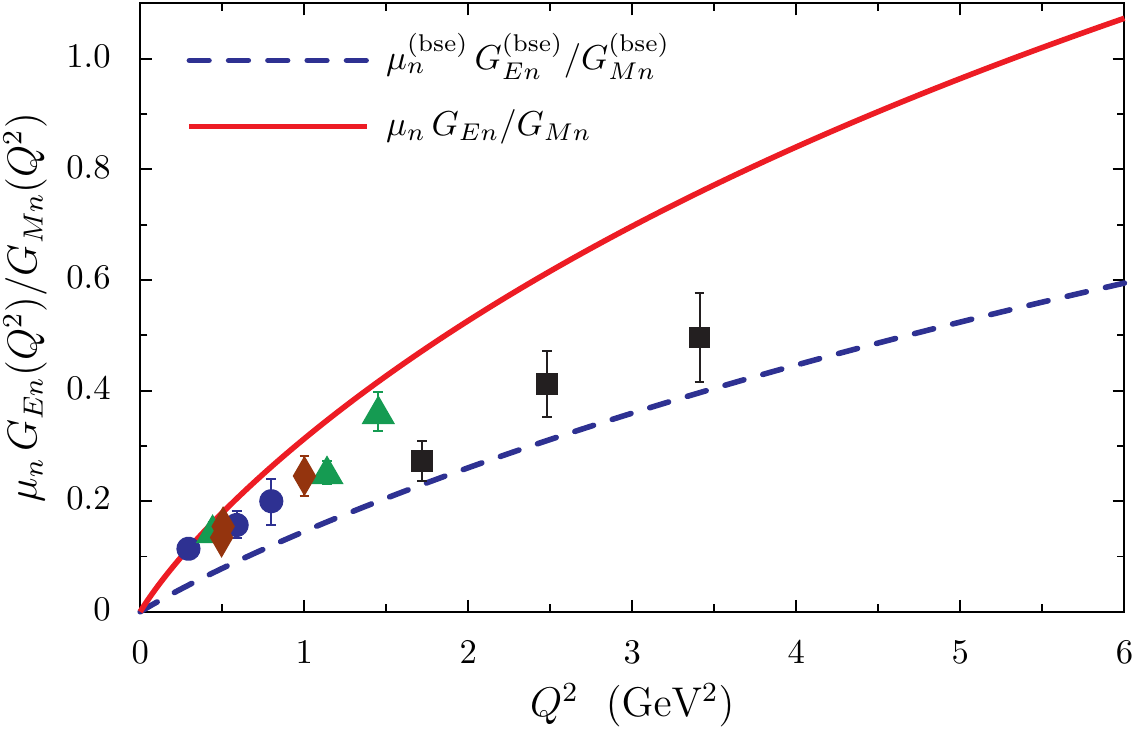}
\caption{(Colour online) Neutron Sachs form factor ratio.
 Data are from Refs.~\cite{Riordan:2010id}.}
\label{fig:neutron_sachs_ratio}
\end{figure}

Results for the neutron Sachs form factor ratio, $G_{En}/G_{Mn}$,  are presented in Fig.~\ref{fig:neutron_sachs_ratio}. For $Q^2 \lesssim 1.5\,$GeV$^2$ our results that include pion loop corrections agree well with data. However, at larger $Q^2$ our ratio continues to grow too rapidly to be consistent with data. Our result for $G_{En}/G_{Mn}$ does not possess a zero-crossing for any $Q^2$ value. This is in contrast to the results of Ref.~\cite{Cloet:2008re} which find a zero-crossing at $Q^2 \simeq 11\,$GeV$^2$.

\section{Conclusion \label{sec:conclusion}}
We have presented calculations of the nucleon form factors using a covariant and 
confining NJL model, which is a Poincar\'e covariant quantum field theory with many 
of the properties of QCD at low to moderate energies. The model satisfies current 
conservation exactly and because the framework is covariant the form factors 
are determined without the need to specify a reference frame. Poincar\'e covariance
also demands non-zero quark orbital angular momentum in the proton wave function, and this
is reflected in our results by large contributions to the nucleon Pauli form factors
from quark-diquark components of the nucleon wave function that only carry charge 
(see Fig.~\ref{fig:vector_body_form_factors} and related discussion).

A unique feature of these results is the parameter-free self-consistent inclusion of pion loop 
effects, as a perturbation to the ``quark core'' results obtained from 
the solution of a relativistic Faddeev equation. These pion cloud effects play 
a vital role for $Q^2 \lesssim 1\,$GeV$^2$. For example, the pion cloud increases 
the magnitude of  the proton and neutron anomalous magnetic moments by 25\% and 45\%, 
respectively, giving final results of $\k_p = 1.78$ and $\k_n = -1.81$, which 
are in rather good agreement with the empirical values. 

In the limit of equal current quark masses our model satisfies charge symmetry and 
therefore the proton quark sector form factors can be unambiguously determined. 
For the quark sector radii we find that $r^u_E$ is 16\% larger than $r^d_E$, 
whereas for the magnetic radii $r^d_M$ is 18\% larger than $r^u_M$. 
The quark sector magnetic radius result can be understood because pion loop effects 
induce a $d$ quark sector anomalous magnetic moment for the dressed up quark 
twelve times larger than the $u$ quark sector contribution. 
For the quark sector form factors, pion cloud effects are largely concentrated in 
the $d$ quark sector. For example, $r^d_M$ actually changes sign when pionic 
effects are included and the value of $G_{Mp}^d(0)$ increases by a factor of ten  
because of pion loop effects.

An area of particular interest which has been identified in our study is 
the interplay between the respective roles of diquark 
correlations and pion effects. This is most dramatically illustrated by the 
comparison of Figs.~\ref{fig:Q4F1_form_factors} and \ref{fig:Q4F2_form_factors}. 
In the first we see the crucial importance on the 
behavior of the Dirac form factor of the 
dominance of scalar diquarks, when they can contribute. The smaller role 
of these scalar diquarks in the $d$-quark case naturally explains the 
suppression of the $d$-quark sector at larger values of the momentum transfer.
On the other hand, in the case of the Pauli form factor the axialvector diquarks and pion make
significant contributions to the $d$-quark sector and this effectively counteracts 
the effect of the scalar diquark correlations. These are subtle but crucial 
aspects of the observed form factors.

Finally, looking to the future, an important near term goal must be to apply 
the framework developed here to the study of nucleon 
transition form factors, for example, nucleon to $\Delta$ and nucleon to 
Roper transitions. This will elucidate the role of pion loop effects in 
these transitions and help to expose the nature of diquark correlations in the 
structure of baryons. The results presented herein and earlier work on nucleon 
PDFs~\cite{Cloet:2005pp,Cloet:2007em} will also serve an a critical starting point 
for forthcoming studies of generalized parton distribution functions~\cite{Diehl:2003ny,Belitsky:2005qn}.

\appendix
\section{Conventions \label{sec:notations}}
We use the conventions of Ref.~\cite{Itzykson:1980rh}. For example the metric tensor has the form:
\begin{align}
g^{\mu\nu} = \text{diag}\lf[1,\,-1,\,-1,\,-1\rg]
\end{align}
and the totally anti-symmetric Levi-Civita tensor is normalized such that $\ve^{0123}=1$.
Some important Dirac matrices are defined as
\begin{align}
\g_5 &= i\g^0\g^1\g^2\g^3 = -\frac{i}{4}\ve_{\mu\nu\rho\sigma}\g^\mu\g^\nu\g^\rho\g^\s,  \\
\s^{\mu\nu} &= \frac{i}{2}\lf[\g^\mu,\,\g^\nu\rg],
\end{align}
and therefore
\begin{align}
\mathrm{Tr}\lf[\g_5\g^\mu\g^\nu\g^\rho\g^\s\rg] = -4i\,\ve^{\mu\nu\rho\sigma} = 4i\,\ve_{\mu\nu\rho\sigma}.
\end{align}

\section{Proof of Electromagnetic Gauge Invariance \label{sec:gaugeinvariance}}
Electromagnetic gauge invariance manifests as electromagnetic current conservation, which for the nucleon is embodied in the statement
\begin{align}
q_\mu\,j^\mu_N(p',p) = 0,
\end{align}
where $q = p'-p$ and $j^\mu_N(p',p)$ is the nucleon electromagnetic current. The nucleon electromagnetic current is given by the six Feynman diagrams represented by Fig.~\ref{fig:nucleon_em_current_feynman_diagrams} and therefore has the form
\begin{align}
&j^\mu_N(p',p) = j^{s,\mu}_{\mathcal{Q}}(p',p) + j^{a,\mu}_{\mathcal{Q}}(p',p) + j^{s,\mu}_{\mathcal{D}}(p',p)  \no \\
&\hs{8mm}
+ j^{a,\mu}_{\mathcal{D}}(p',p) + j^{s\to a,\mu}_{\mathcal{D}}(p',p) + j^{a\to s,\mu}_{\mathcal{D}}(p',p),
%
\end{align}
where the nomenclature is explained in Sect.~\ref{sec:nucleon_results}. The individual Feynman diagram contributions to the nucleon electromagnetic current are
\begin{align}
j^{s,\mu}_{\mathcal{Q}}(p',p) &= -Z_N\,\ol{\G}_s \int\frac{d^4k}{\lf(2\pi\rg)^4} \no \\
&\hs{9mm} 
\times S(\ell')\,\L^\mu_{\g Q}(\ell',\ell)\,S(\ell)\,\tau_s(k)\,\G_s, \\
j^{a,\mu}_{\mathcal{Q}}(p',p) &= -Z_N\,\ol{\G}^\a_a \int\frac{d^4k}{\lf(2\pi\rg)^4} \no \\
&\hs{3mm} 
\times S(\ell')\ \L^\mu_{\g Q}(\ell',\ell)\,S(\ell)\,\tau_{a,\a\b}(k)\,\G_a^\b, \\
j^{s,\mu}_{\mathcal{D}}(p',p) & = -Z_N\,i\ol{\G}_s \int\frac{d^4k}{\lf(2\pi\rg)^4}  \no \\
&\hs{10mm} 
\times S(k)\,\tau_s(\ell')\, \L^{\mu}_s(\ell',\ell)\,\tau_s(\ell)\,\G_s, \\
j^{a,\mu}_{\mathcal{D}}(p',p) & = -Z_N\,i\ol{\G}_a^\l \int\frac{d^4k}{\lf(2\pi\rg)^4} \no \\
&\hs{-4mm} 
\times 
S(k)\,\tau_{a,\l\b}(\ell')\,
\L^{\mu,\a\b}_{a}(\ell',\ell)\,\tau_{a,\a\s}(\ell)\,\G_a^\s,  \\
j^{s\to a,\mu}_{\mathcal{D}}(p',p) & = -Z_N\,i\ol{\G}_a^\l \int\frac{d^4k}{\lf(2\pi\rg)^4} \no \\
&\hs{2mm} 
\times 
S(k)\,\tau_{a,\l\a}(\ell')\,
\L^{\mu,\a}_{s\to a}(\ell',\ell)\,\tau_s(\ell)\,\G_s,  \\
j^{a\to s,\mu}_{\mathcal{D}}(p',p)  & = -Z_N\,i\ol{\G}_s \int\frac{d^4k}{\lf(2\pi\rg)^4} \no \\
&\hs{1mm} 
\times 
S(k)\,\tau_s(\ell')\,
\L^{\mu,\a}_{a\to s}(\ell',\ell)\,\tau_{a,\a\l}(\ell)\,\G_a^\l, 
\end{align}
where $\ell' = p'- k$, $\ell = p - k$ and we have also dropped the momentum dependence of the Faddeev vertices. The quark-photon vertex is labelled by $\L^\mu_{\g Q}$ and the various diquark-photon vertices are represented by $\L^{\mu}_s$, $\L^{\mu,\a\b}_{a}$, $\L^{\mu,\a}_{s\to a}$ and $\L^{\mu,\a}_{a\to s}$. These vertices satisfy the following Ward--Takahashi identities:

\begin{align}
%
%
q_\mu\,\L^\mu_{\g Q}(\ell',\ell) &= \hat{Q}_q\lf[S^{-1}(\ell') - S^{-1}(\ell)\rg], \\
q_\mu\,\L^{\mu}_s(\ell',\ell) &= -i\hat{Q}_s\lf[\tau_s^{-1}(\ell')-\tau_s^{-1}(\ell)\rg], \\
q_\mu\,\L^{\mu,\a\b}_{a,ij}(\ell',\ell)
&= -i\hat{Q}^{ij}_a\lf[\tau_a^{-1,\a\b}(\ell')-\tau_a^{-1,\a\b}(\ell)\rg], \\
%
%
q_\mu\,\L^{\mu,\a}_{s \to a}(\ell',\ell) &= 0, \\
q_\mu\,\L^{\mu,\a}_{a \to s}(\ell',\ell) &= 0,
\end{align}
where the charge operators have the form
\begin{align}
\hat{Q}_q &= \frac{1}{6} + \frac{\tau_3}{2}, & \hat{Q}_s &= \frac{1}{3}, &
\hat{Q}^{ij}_a &= \frac{1}{3}\,\d_{ij} + i\ve_{ij3}.
\end{align}
The indices on $\hat{Q}^{ij}_a$ represent isospin, where $i$ is the initial diquark
and $j$ the final diquark. The contraction of the current with $q_\mu$ therefore gives
\begin{align}
q_\mu\,j^{s,\mu}_{\mathcal{Q}} &= -\hat{Q}_q\,\ol{\G}_s \int_k
\lf[S(\ell') - S(\ell)\rg]\tau_s(k)\,\G_s, \\
q_\mu\,j^{a,\mu}_{\mathcal{Q}} &= -\hat{Q}_q\,\ol{\G}^\a_a \int_k
\lf[S(\ell')-S(\ell)\rg]\tau_{a,\a\b}(k)\,\G_a^\b, \\
q_\mu\,j^{s,\mu}_{\mathcal{D}} &= -\hat{Q}_s\ol{\G}_s \int_k
S(k)\lf[\tau_s(\ell')-\tau_s(\ell)\rg]\,\G_s, \\
q_\mu\,j^{a,\mu}_{\mathcal{D}} &= -\hat{Q}^{ij}_a\,\ol{\G}_a^{\a,j} \int_k
S(k)\lf[\tau_{a,\a\b}(\ell')-\tau_{a,\a\b}(\ell)\rg]\,\G_a^{\b,i}.
\end{align}
where we have used $\tau_{a,\a\b}\,\tau^{-1,\l\s}_a = \d_{\a\l}\d_{\b\s}$.
Therefore
\begin{align}
&q_\mu\,j^{s,\mu}_{\mathcal{Q}} + q_\mu\,j^{s,\mu}_{\mathcal{D}}
= -\lf(\hat{Q}_q + \hat{Q}_s\rg) \no \\
&\hs{19mm}
\times \ol{\G}_s(p')\lf[\Pi_{Ns}(p') - \Pi_{Ns}(p)\rg]\G_s(p), \\[0.2ex]
&q_\mu\,j^{a,\mu}_{\mathcal{Q}} + q_\mu\,j^{a,\mu}_{\mathcal{D}} =
-\lf(\hat{Q}_q\,\d^{ij} + \hat{Q}^{ij}_a\rg) \no \\
&\hs{10mm}
\times \ol{\G}^{\a,j}_a(p')\lf[\Pi_{Na,\a\b}(p')  - \Pi_{Na,\a\b}(p)\rg]\G_a^{\b,i}(p), \\[0.4ex]
&q_\mu\,j^{s\to a,\mu}_{\mathcal{D}} + q_\mu\,j^{a \to s,\mu}_{\mathcal{D}} = 0.
\end{align}
In matrix notation we therefore have
\begin{align}
q_\mu \, j^\mu_N = 
\ol{\G}_{N}(p')\,\hat{Q}_N\lf[\Pi_N(p') - \Pi_N(p)\rg]\G_N(p), 
\end{align}
where
\begin{align}
\hat{Q}_N &= \begin{bmatrix}\hat{Q}_q + \hat{Q}_s & 0 \\ 0 &  \hat{Q}_q\,\d^{ij} + \hat{Q}^{ij}_a \end{bmatrix},
\end{align}
and $\Pi_N(p)$ is defined in Eq.~\eqref{eq:nucleon_bubble_matrix}.
It is straightforward to show
\begin{align}
\hat{Q}_N\,\G_N(p) &= Q_N\,\G_N(p),
\end{align}
where
\begin{align}
Q_N =\begin{cases}
      1 & \text{proton~~}\\
      0 & \text{neutron}
      \end{cases}.
\end{align}

Therefore
\begin{align}
q_\mu \, j^\mu_{N} &= Q_N\,\ol{\G}_{N}(p')\lf[\Pi_N(p')  - \Pi_N(p)\rg]\G_N(p).
\end{align}
The Faddeev equations for $\G_N(p)$ and $\ol{\G}_{N}(p)$ are
\begin{align}
\G_N(p) &= Z\,\Pi_N(p)\,\G_N(p), \\ 
\ol{\G}_{N}(p) &= \ol{\G}_{N}(p)\,\Pi_N(p)\,Z,
\end{align}
where $Z$ is the quark exchange kernel. Therefore
\begin{align}
q_\mu \, j^\mu_{N} = 
Q_N\,\ol{\G}_{N}(p')\lf[Z^{-1}  - Z^{-1}\rg]\G_N(p) = 0,
\end{align}
as required by current conservation.

\section{Sachs Form Factors \label{sec:sachs}}
In the non-relativistic limit the electric and magnetic Sachs form factors are rigorously related 
to the charge and magnetization densities via a 3-dimensional Fourier transform. In a Poincar\'e covariance 
quantum field theory this relation breaks down, but such a correspondence may still be a useful
tool, at least for large distances. The Sachs form factors appear in the Rosenbluth 
parameterization of the elastic scattering differential cross-section, given by
\begin{align}
\frac{d \s}{d\O} = \frac{\s_{\text{Mott}}}{1+\tau}\lf[G_E^2(Q^2) + \frac{\tau}{\epsilon}\,G_M^2(Q^2)\rg],
\end{align}
where $\tau = Q^2/(4\,M_N^2)$, $\s_{\text{Mott}}$ represents that cross-section for the scattering of the electron from a pointlike scalar particle and $\epsilon$ is the longitudinal polarization of the virtual photon that mediates the interaction in Born approximation. For the Rosenbluth separation technique one considers the reduced cross-section, namely
\begin{align}
\s_R \equiv  \epsilon\, \frac{d \s}{d\O}\ \frac{1+\tau}{\s_{\text{Mott}}} = \epsilon\,G_E^2(Q^2) + \tau\,G_M^2(Q^2).
\end{align}
Therefore $\s_R$ is linearly dependent on $\epsilon$, so a linear fit to the reduced cross-section at fixed $Q^2$ but a range of $\epsilon$ values give $G_E^2(Q^2)$ as the slope and $\tau\,G_M^2(Q^2)$ as the $y$-axis intercept. At large $Q^2$ the reduced cross-section is dominated by $\tau\,G_M^2(Q^2)$ making an accurate extraction of $G_E^2(Q^2)$ increasingly more difficult.

\begin{figure}[tbp]
\subfloat{\centering\includegraphics[width=\columnwidth,clip=true,angle=0]{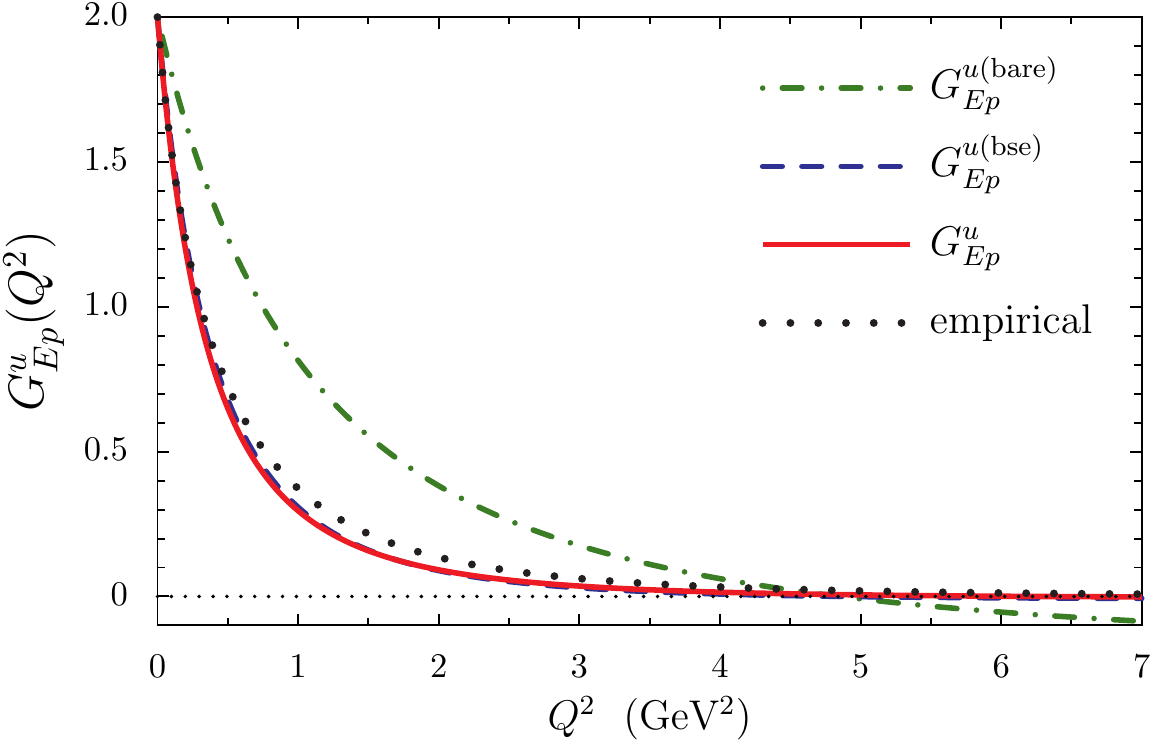}}\\
\subfloat{\centering\includegraphics[width=\columnwidth,clip=true,angle=0]{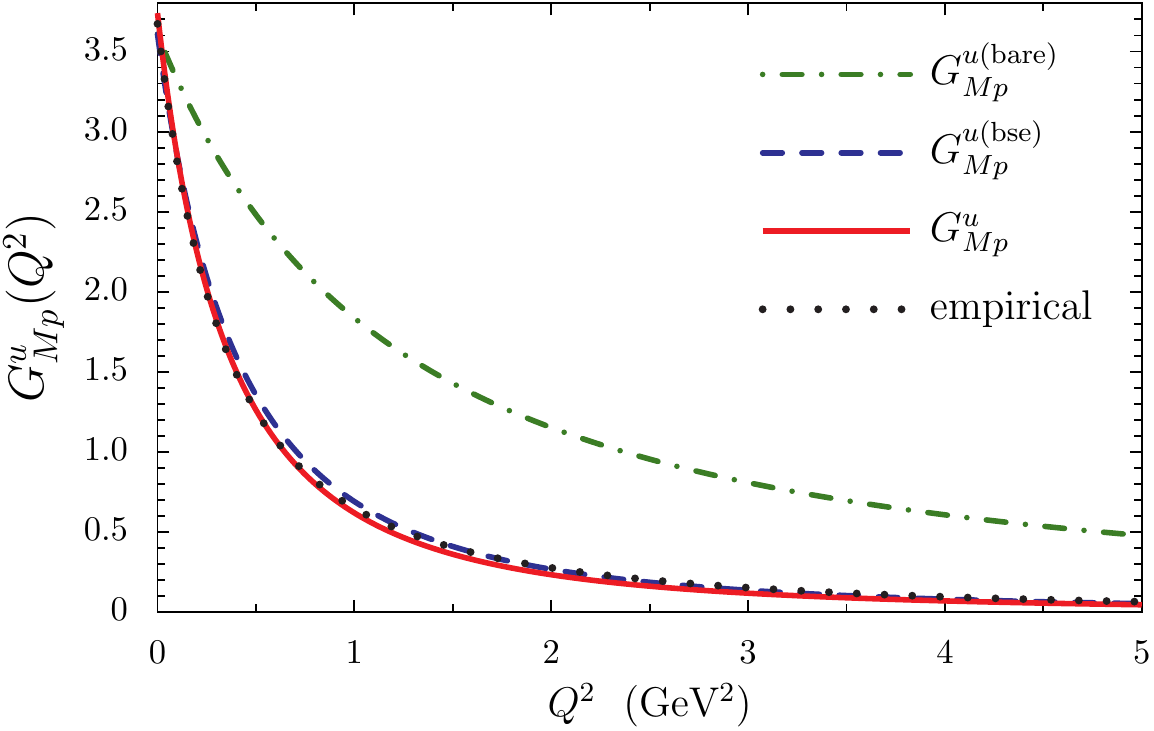}}
\caption{(Colour online) Proton up quark sector Sachs form factors. The empirical results are obtained 
using Ref.~\cite{Kelly:2004hm} and Eq.~\eqref{eq:flavourseparation}.}
\label{fig:sachs_u_sector}
\end{figure}

\begin{figure}[tbp]
\subfloat{\centering\includegraphics[width=\columnwidth,clip=true,angle=0]{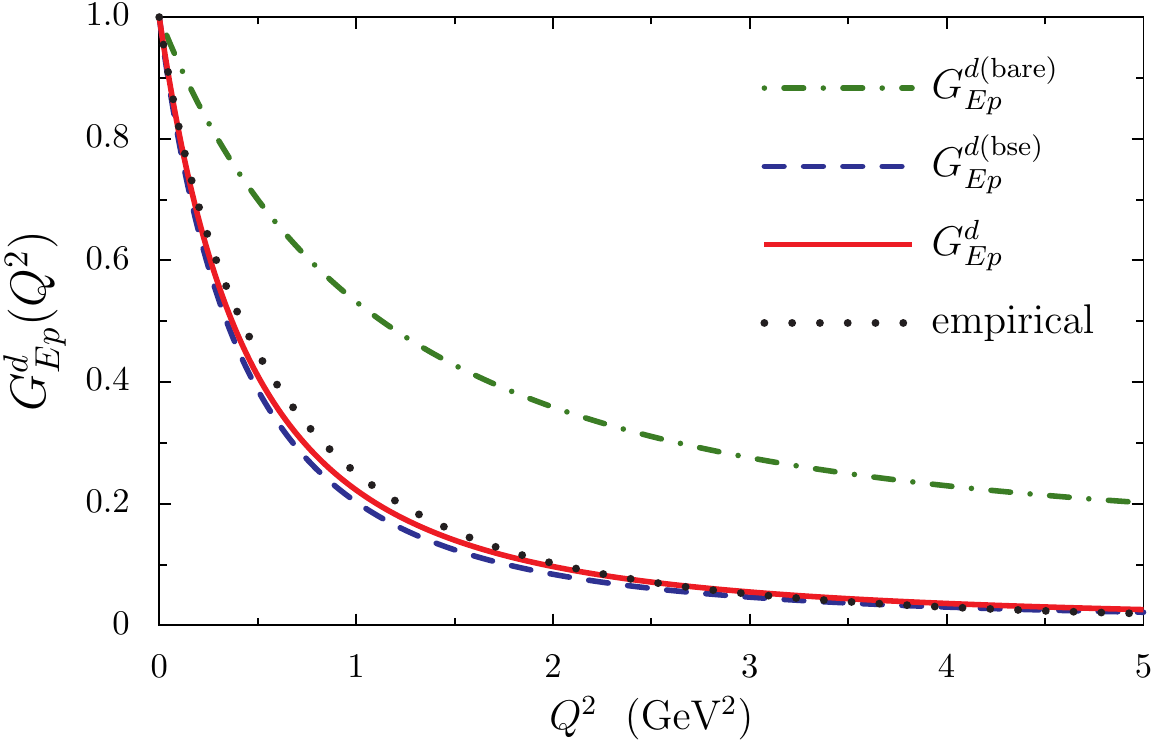}}\\
\subfloat{\centering\includegraphics[width=\columnwidth,clip=true,angle=0]{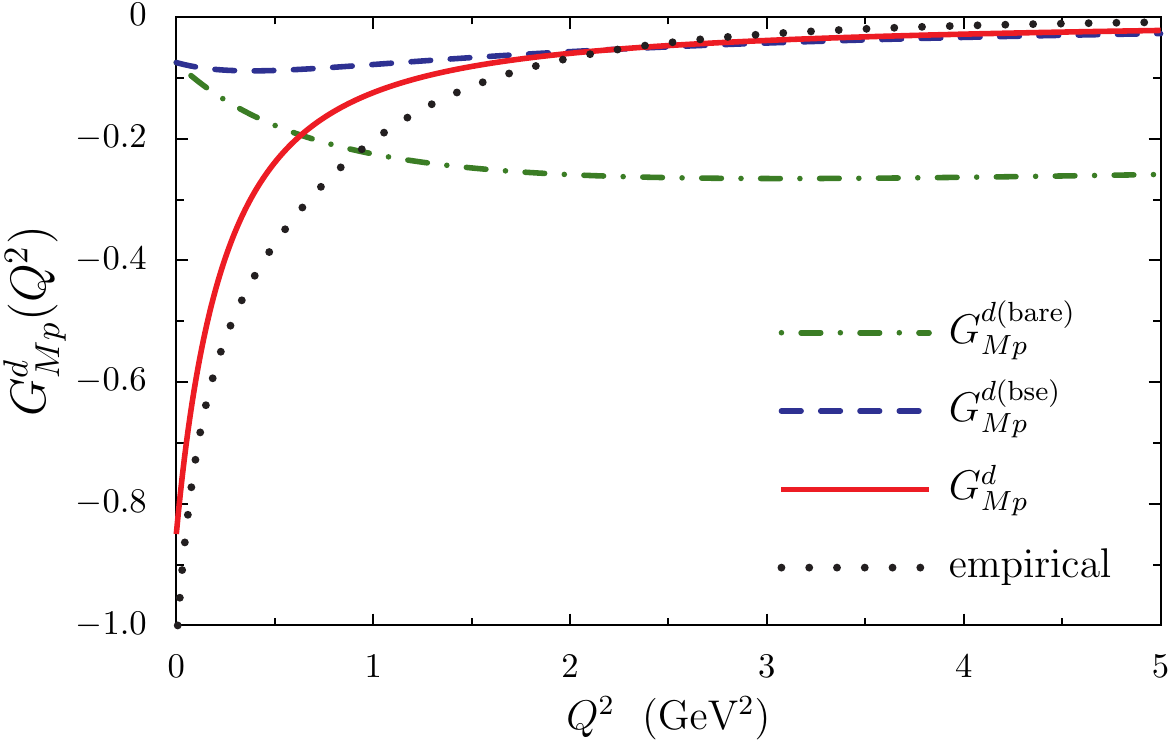}}
\caption{(Colour online) Neutron up quark sector Sachs form factors. The empirical results are obtained 
using Ref.~\cite{Kelly:2004hm} and Eq.~\eqref{eq:flavourseparation}.}
\label{fig:sachs_d_sector}
\end{figure}

\begin{figure}[tbp]
\subfloat{\centering\includegraphics[width=\columnwidth,clip=true,angle=0]{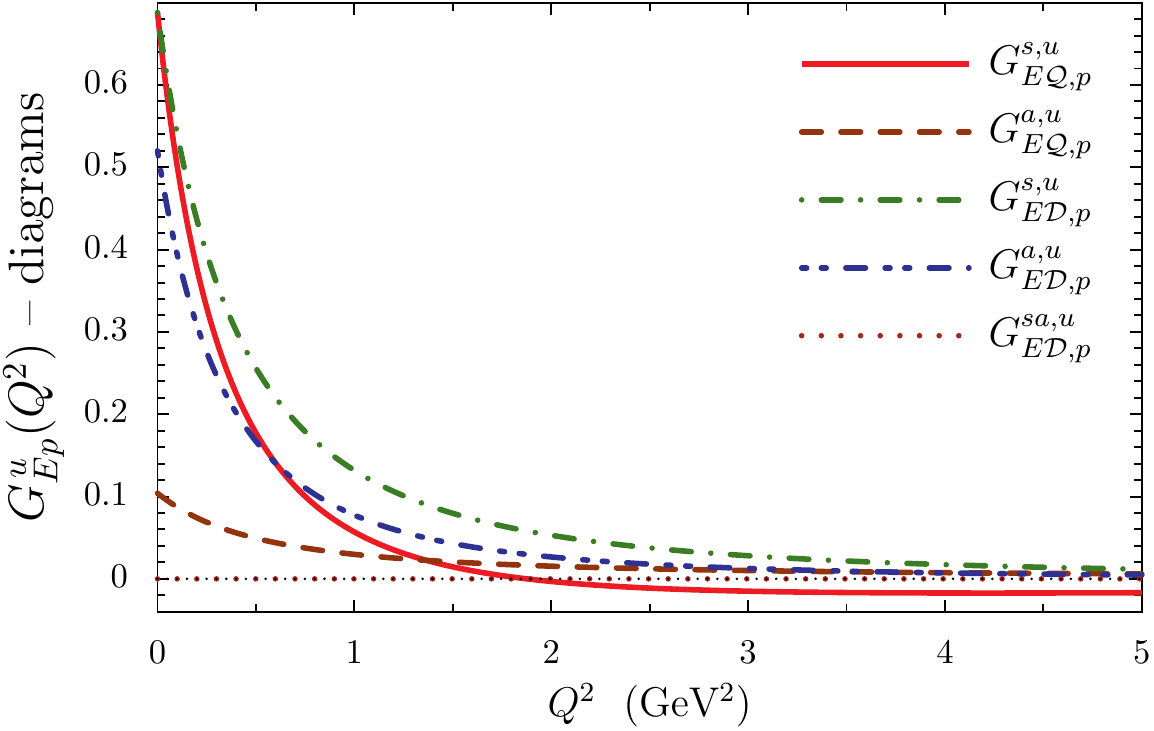}}\\
\subfloat{\centering\includegraphics[width=\columnwidth,clip=true,angle=0]{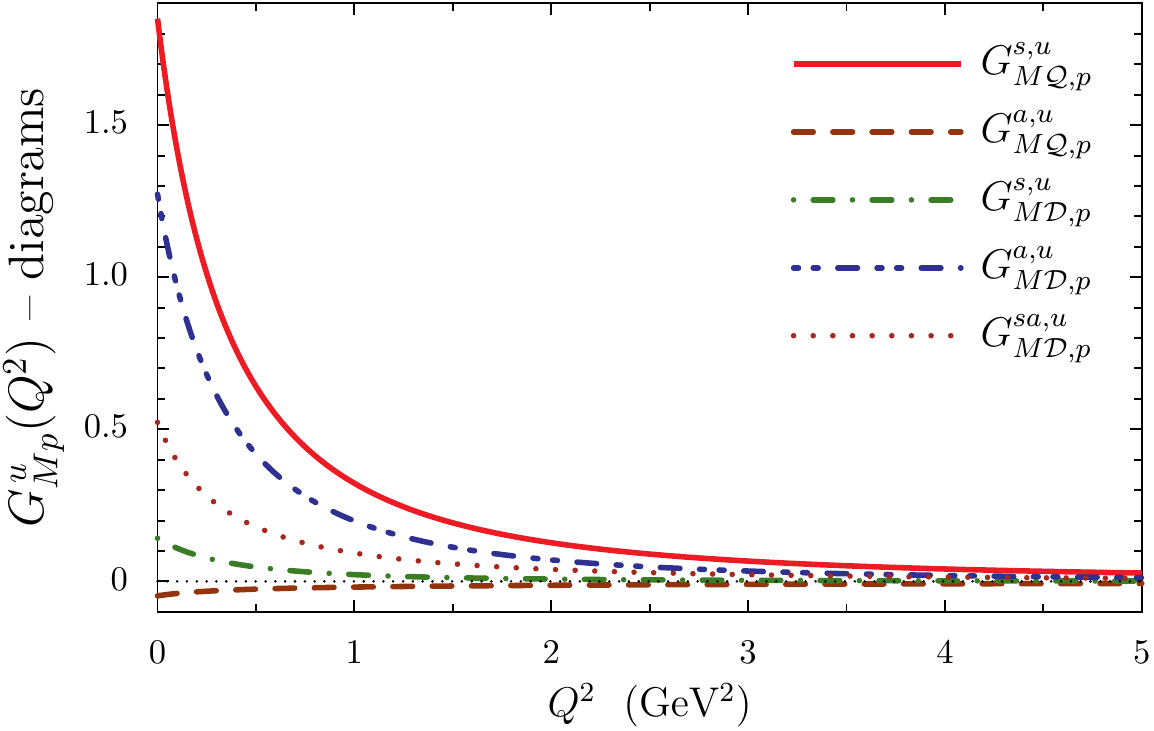}}
\caption{(Colour online) Total contributions to the proton $u$-sector Sachs form factors from each Feynman diagram 
in Fig.~\ref{fig:nucleon_em_current_feynman_diagrams}. These results include both the vector and tensor
coupling contributions and the sum gives the total $u$-sector Sachs proton form factors.}
\label{fig:sachs_u_sector_diagrams}
\end{figure}

\begin{figure}[tbp]
\subfloat{\centering\includegraphics[width=\columnwidth,clip=true,angle=0]{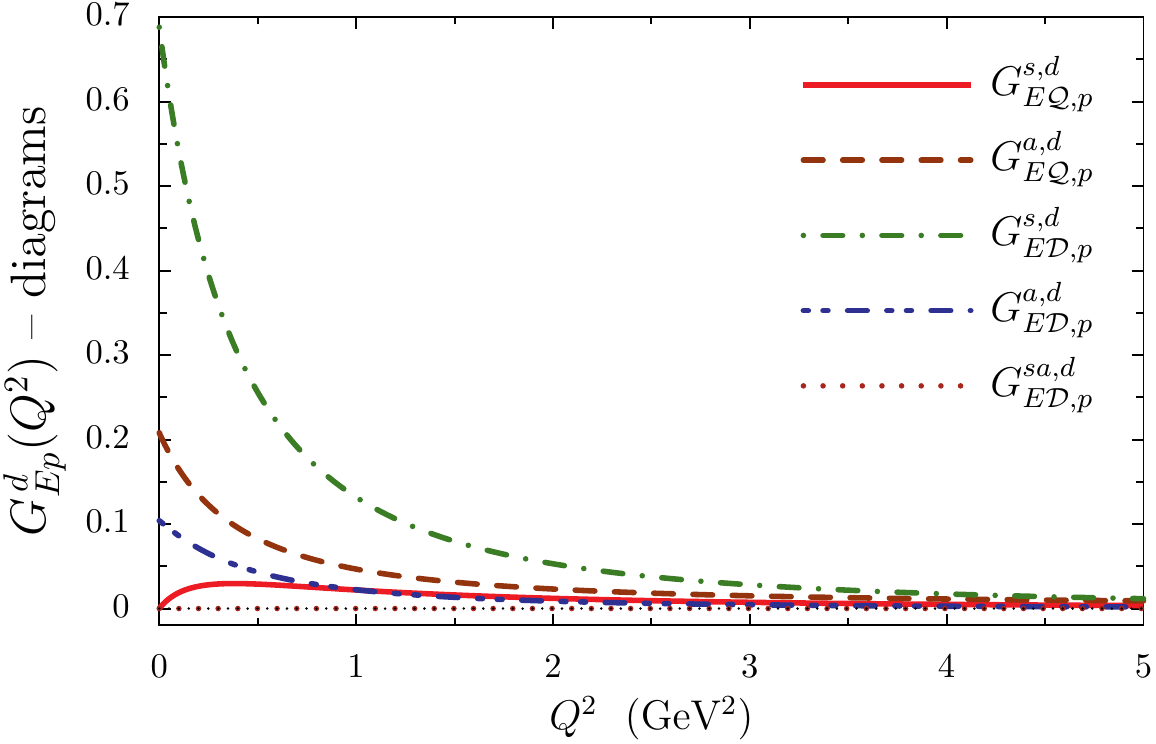}}\\
\subfloat{\centering\includegraphics[width=\columnwidth,clip=true,angle=0]{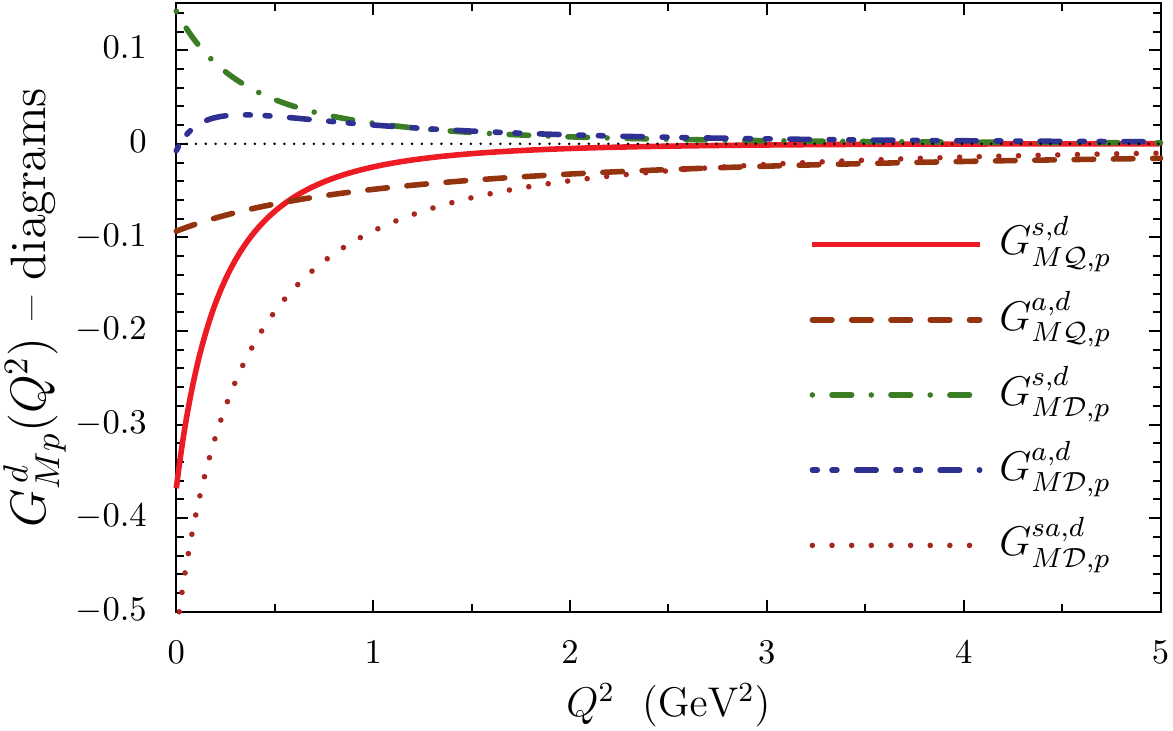}}
\caption{(Colour online) Total contributions to the neutron $u$-sector Sachs form factors from each Feynman diagram 
in Fig.~\ref{fig:nucleon_em_current_feynman_diagrams}. These results include both the vector and tensor
coupling contributions and the sum gives the total $u$-sector Sachs neutron form factors.}
\label{fig:sachs_d_sector_diagrams}
\end{figure}

Our results for the proton and neutron Sachs form factors are presented in Figs.~\ref{fig:proton_sachs}
and \ref{fig:neutron_sachs}, where in each case we have used the three variants of dressed quark form factor
discussed in Sect.~\ref{sec:QPV}. Empirical results from Ref.~\cite{Kelly:2004hm} are shown as the dotted
line. After including the vertex dressing from the BSE and also pion loop effects, the agreement with 
experiment is good. The proton electric form factor is slightly too soft and, as already discussed, possess
a zero at $Q^2 \simeq 3.7\,$GeV$^2$. For the proton magnetic form factor there is excellent agreement 
with the empirical results of Ref.~\cite{Kelly:2004hm}. Our result for the neutron electric form factor 
drops too slowly for $Q^2 \gtrsim 1\,$GeV$^2$ and the magnetic form factor lacks a little strength for
$Q^2 \lesssim 1\,$GeV$^2$. However, overall the agreement with data is very good.

Results for the the proton quark sector Sachs form factors are given in Figs.~\ref{fig:sachs_u_sector}
for the $u$-quark sector and Figs.~\ref{fig:sachs_d_sector} for the $d$-quark sector. Empirical 
results are obtained from Ref.~\cite{Kelly:2004hm} using the identities
\begin{align}
G^u_{ip} &= 2\,G_{ip} + G_{in}, &
G^d_{ip} &= G_{ip} + 2\,G_{in}, 
\end{align}
where $i=(E,\,M)$. Overall the agreement with experiment is very good. Interestingly however, is that the pion loop effects have little influence on the quark sector Sachs form factors with the notable exception of $G_{Mp}^d(Q^2)$. Here the pion cloud results in a 10-fold increase in the magnitude of the $d$ quark sector magnetic moment, from $\mu_d^{\text{(bse)}} = -0.075\,\mu_N$ to $\mu_d = -0.85\,\mu_N$. There are a number of effects that all add to give this very large pion loop correction, most importantly however, is the large $d$ quark sector anomalous magnetic moment of a dressed up quark (Eq.~\eqref{eq:quark_sector_amm_dressed_quarks}) generated by the pion cloud. This results in a large contribution to $G_{Mp}^d(Q^2)$ from the quark diagram where the photon couples to a dressed up quark, with a scalar diquark as spectator. Another notable result is that the zero in $G_{Ep}(Q^2)$ resides solely in the $u$ quark sector, as illustrated by the upper panels of Figs.~\ref{fig:sachs_u_sector} and \ref{fig:sachs_d_sector}.
 
To understand these results better in we give the total contribution from each diagram, including both
the BSE vertex dressing and pion loop effects in Figs.~\ref{fig:sachs_u_sector_diagrams} for the $u$
quark sector and Figs.~\ref{fig:sachs_d_sector_diagrams} for the $d$ quark sector. For the upper panel 
in Figs.~\ref{fig:sachs_u_sector_diagrams} it is clear that the zero in $G_{Ep}^u(Q^2)$ and therefore
$G_{Ep}(Q^2)$ has its origin solely in $G_{E\mathcal{Q},p}^{s,u}$, the $u$ quark sector contribution from 
quark diagrams with a spectator scalar diquark. We also find the interesting result that the diagrams
associated with transitions between scalar and axialvector diquarks are identically zero for the Sachs
electric form factors but do contribute sizeably to the Sachs magnetic form factors. 

\section{Analysis of Pion Cloud effects \label{sec:pioncloud}}

\begin{figure}[t]
\subfloat{\centering\includegraphics[width=\columnwidth,clip=true,angle=0]{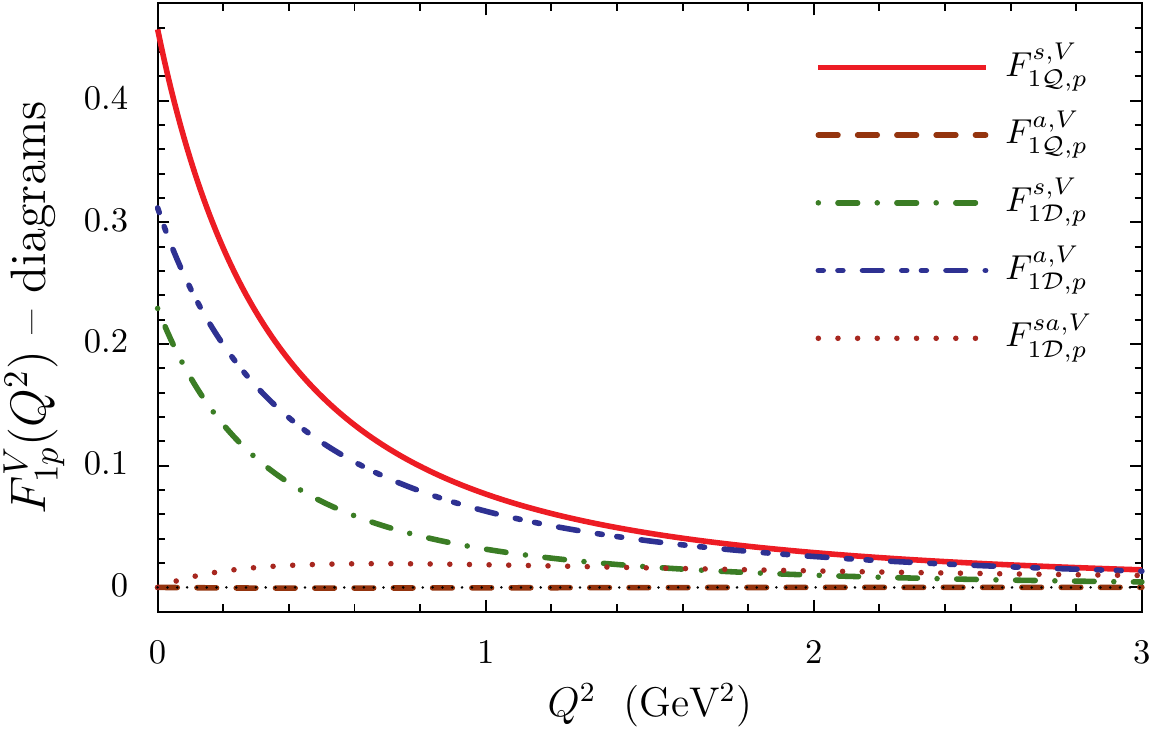}}\\
\subfloat{\centering\includegraphics[width=\columnwidth,clip=true,angle=0]{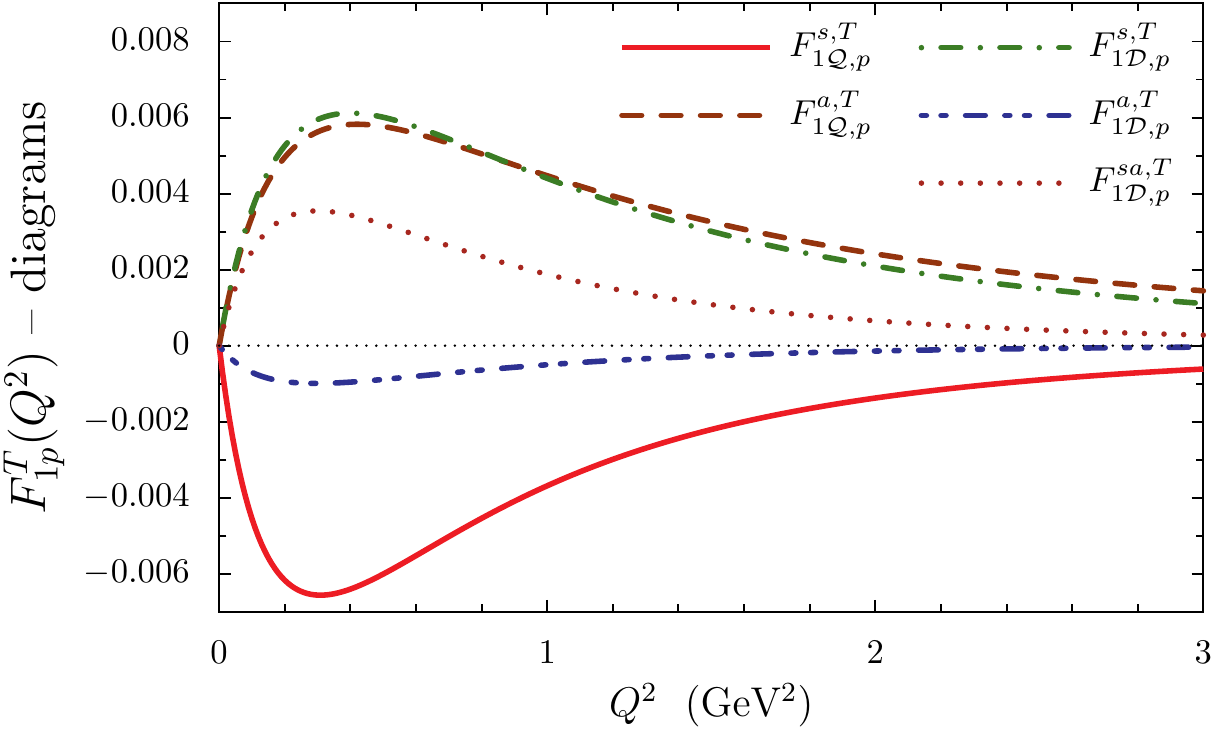}}
\caption{(Colour online) Total contributions to the proton 
Dirac form factor from each Feynman diagram 
in Fig.~\ref{fig:nucleon_em_current_feynman_diagrams} for 
a vector coupling (\textit{upper panel}) and
tensor coupling (\textit{lower panel}) to the dressed quarks. 
The sum of these 10 contributions  gives the 
total proton Dirac form factor illustrated 
in Fig.~\ref{fig:proton_form_factors} (\textit{solid} curve).}
\label{fig:proton_diagrams_F1}
\end{figure}

\begin{figure}[tbp]
\subfloat{\centering\includegraphics[width=\columnwidth,clip=true,angle=0]{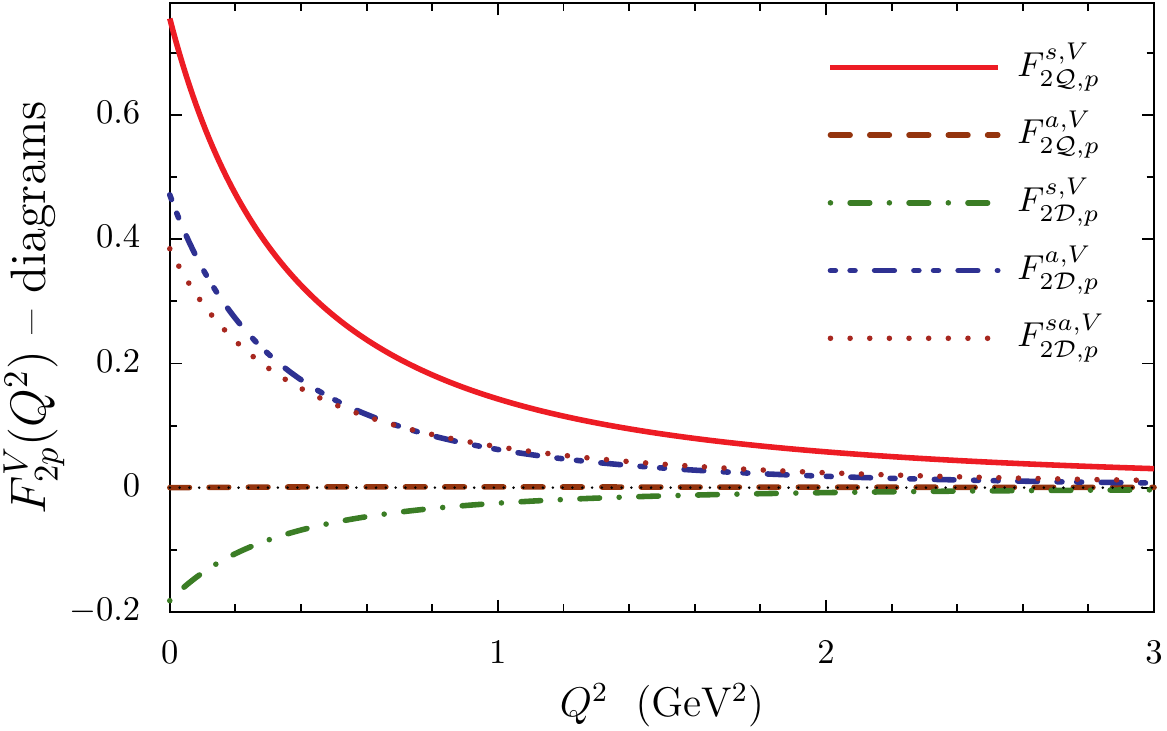}}\\
\subfloat{\centering\includegraphics[width=\columnwidth,clip=true,angle=0]{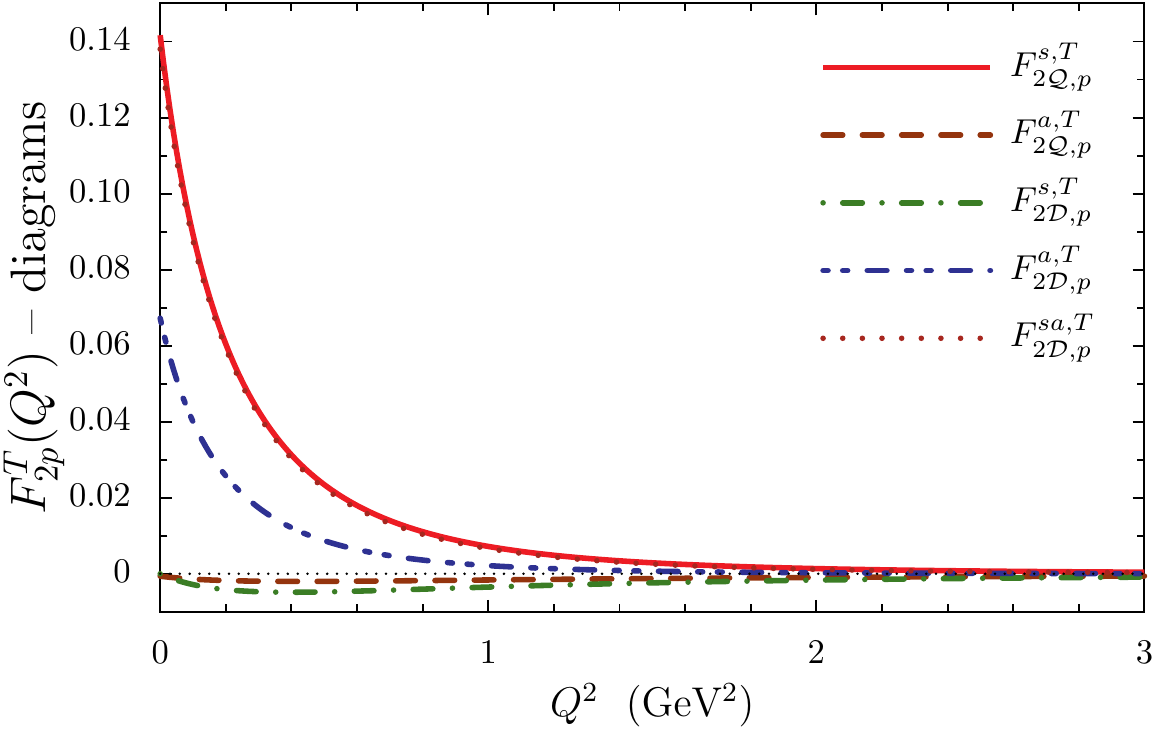}}
\caption{(Colour online) Total contributions to the proton Pauli form factor from 
each Feynman diagram 
in Fig.~\ref{fig:nucleon_em_current_feynman_diagrams} for a 
vector coupling (\textit{upper panel}) and
tensor coupling (\textit{lower panel}) to the dressed quarks. 
The sum of these 10 contributions gives 
the total proton Pauli form 
factor illustrated in Fig.~\ref{fig:proton_form_factors} (\textit{solid} curve).}
\label{fig:proton_diagrams_F2}
\end{figure}

A deeper understanding of our results is gained by decomposing the form factors in 
Figs.~\ref{fig:proton_form_factors} 
and \ref{fig:neutron_form_factors} into the various total 
contributions arising from each of the six diagrams 
represented in Fig.~\ref{fig:nucleon_em_current_feynman_diagrams},
for both the vector and tensor coupling to the dressed quarks.
In this case the nucleon form factors are expressed as
\begin{align}
F_{ip}(Q^2) = F^V_{ip}(Q^2) + F^T_{ip}(Q^2), \\
F_{in}(Q^2) = F^V_{in}(Q^2) + F^T_{in}(Q^2),
\end{align}
where $i = (1,\,2)$.
In terms of the six diagrams contained 
in Fig.~\ref{fig:nucleon_em_current_feynman_diagrams} we have
\begin{align}
\label{eq:total_diagrams_vector}
F^V_{ip} &= F^{s,V}_{i\calQ,p} + F^{a,V}_{i\calQ,p} + F^{s,V}_{i\calD,p} + F^{a,V}_{i\calD,p} + F^{sa,V}_{i\calD,p}, \\
\label{eq:total_diagrams_tensor}
F^T_{ip} &= F^{s,T}_{i\calQ,p} + F^{a,T}_{i\calQ,p} + F^{s,T}_{i\calD,p} + F^{a,T}_{i\calD,p} + F^{sa,T}_{i\calD,p},
\end{align}
and neutron expressions are obtained by $p\to n$.
All terms in Eqs.~\eqref{eq:total_diagrams_vector} 
and \eqref{eq:total_diagrams_tensor} include the 
isospin coefficients and the dressed quark form factors.

\begin{table*}[tp]
\addtolength{\tabcolsep}{9.1pt}
\addtolength{\extrarowheight}{2.2pt}
\begin{tabular}{l|cccccccccc|cccc}
\hline\hline
          & $F^{s,V}_{1\calQ}$ & $F^{a,V}_{1\calQ}$ & $F^{s,V}_{1\calD}$ & $F^{a,V}_{1\calD}$ & $F^{sa,V}_{1\calD}$ &
            $F^{s,T}_{1\calQ}$ & $F^{a,T}_{1\calQ}$ & $F^{s,T}_{1\calD}$ & $F^{a,T}_{1\calD}$ & $F^{sa,T}_{1\calD}$ & total\\[0.6ex]
\hline
$F_{1p}$   & \ph{-}0.46 & 0          & \ph{-}0.23 & \ph{-}0.31 &  0    &       0    & 0    & 0   &        0    &  0    & \ph{-}1\\
$F_{2p}$   & \ph{-}0.76 & 0          &      -0.18 & \ph{-}0.47 &  0.38 & \ph{-}0.14 & 0.0  & 0.0 &  \ph{-}0.07 &  0.14 & \ph{-}1.78\\[0.6ex]
$F_{1n}$   &      -0.23 & \ph{-}0.10 & \ph{-}0.23 &      -0.10 &  0    &       0    & 0    & 0   &        0    &  0    & \ph{-}0\\
$F_{2n}$   &      -0.38 &      -0.15 &      -0.18 &      -0.16 & -0.38 &      -0.25 & 0.0  & 0.0 &       -0.17 & -0.14 & -1.81\\
\hline\hline
\end{tabular}
\caption{Total contributions to the nucleon Dirac and Pauli form factors from 
the various diagrams at $Q^2 = 0$. The vector 
contributions are obtained from the appropriate body form factors 
at $Q^2 = 0$ multiplied by isospin factors and quark 
charges. Therefore, the vector results do not change with 
the various approximations for the dressed quark form factors.
The tensor contributions are only non-zero if 
the dressed quarks have an anomalous magnetic moment, and in this
framework this occurs solely from pion loop effects. 
Rows with an entry of ``0'' are identically zero because of charge
conservation.}
\label{tab:nucleon_diagrams_Q20}
\end{table*}

Figs.~\ref{fig:proton_diagrams_F1} and \ref{fig:proton_diagrams_F2} illustrate
the results for the total contributions from each diagram 
in Fig.~\ref{fig:nucleon_em_current_feynman_diagrams} 
to the proton form factors, where the contribution from the vector and tensor 
quark-photon couplings are shown separately. 
Results for the various contributions at $Q^2 = 0$ 
are also listed in rows 1 and 2 of Table~\ref{tab:nucleon_diagrams_Q20}. 
These results demonstrate the 
dominance of the scalar diquark component in the proton wave function. 
For example, diagrams which contain
only a scalar diquark carry 69\% of the proton charge 
and 40\% of its anomalous magnetic moment,
while diagrams that contain only axialvector diquarks 
carry 31\% of the proton charge and 
30\% of its anomalous magnetic moment. 
The remainder of the anomalous magnetic moment, 30\%, is carried by
diagrams with both scalar and axialvector diquarks. 
The contributions that arise from the tensor coupling
to the dressed quarks, 
illustrated in the lower panels of Figs.~\ref{fig:proton_diagrams_F1} 
and \ref{fig:proton_diagrams_F2}, diminish rapidly with 
increasing $Q^2$ because they are suppressed by 
the dressed quark Pauli form factors, 
illustrated in Fig.~\ref{fig:constituentquarkformfactors}, which 
vanish monotonically for increasing $Q^2$.

The coupling of the photon to the diquarks is not only necessary for charge 
conservation, but is critical for a good description of the experimental data. 
Here the axialvector diquarks play an important role. For a proton, the photon 
is twice as likely to couple to a $\{uu\}$ type axialvector diquark than one 
of type $\{ud\}$. Furthermore, axialvector diquarks of type $\{uu\}$ have a 
charge of 
$4/3$ and a magnetic moment of $3.27\,\mu_N$ 
(see Tab.~\ref{tab:rho_axial_diquark_moments}) and hence they provide the second 
largest contribution to the proton charge and anomalous magnetic moment. 
The largest contribution in each case comes from the quark diagram with a 
spectator scalar diquark. With the exception of the diquark diagram with an 
axialvector diquark spectator, the Feynman diagrams of Fig.~\ref{fig:nucleon_em_current_feynman_diagrams}, for the vector coupling, can only contribute to $F_{2p}(Q^2)$ if the quarks have non-zero orbital angular momentum. The sizeable contributions from all other diagrams (see Fig.~\ref{fig:proton_diagrams_F2}) indicate that there is significant quark orbital angular momentum in the proton wave function~\cite{Myhrer:2007cf,Thomas:2008ga}.

\begin{figure}[tbp]
\subfloat{\centering\includegraphics[width=\columnwidth,clip=true,angle=0]{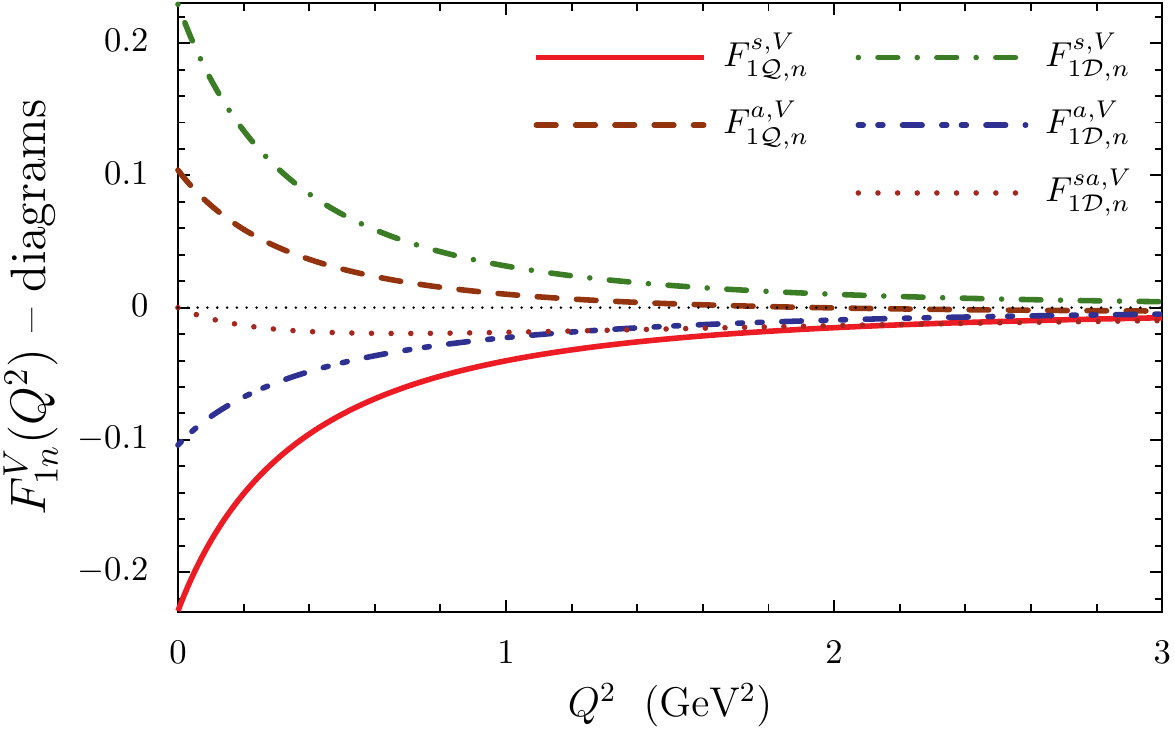}}\\
\subfloat{\centering\includegraphics[width=\columnwidth,clip=true,angle=0]{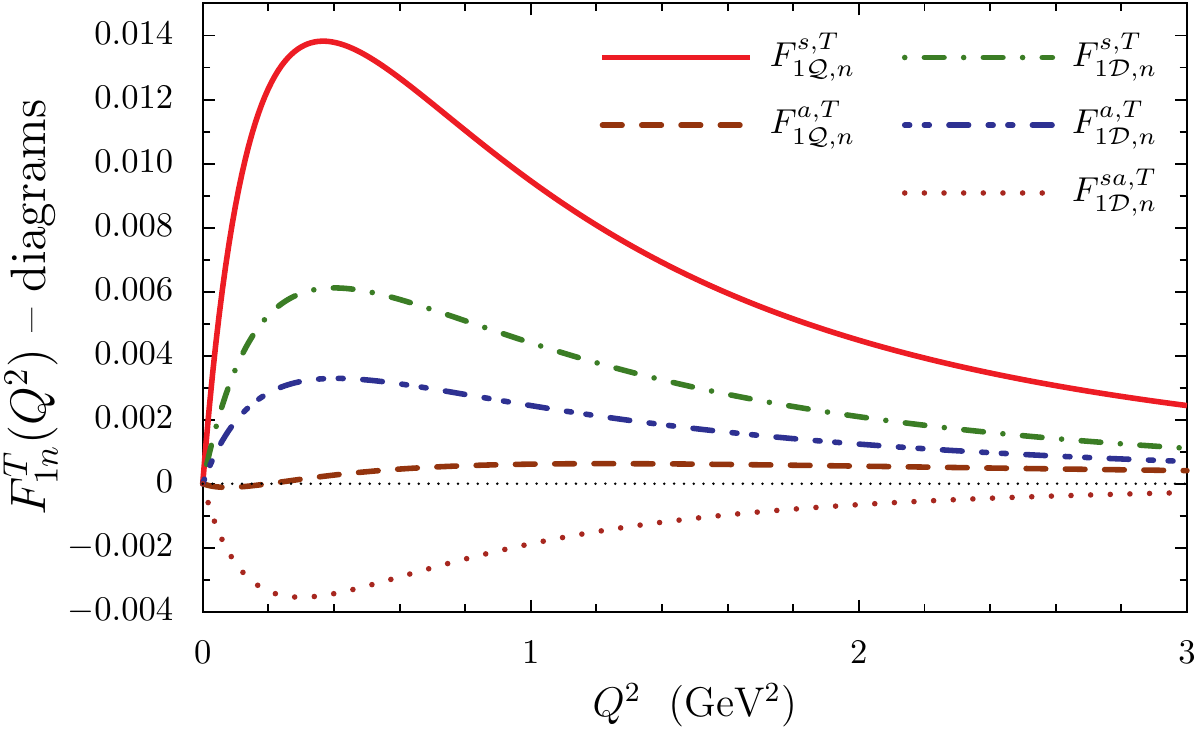}}
\caption{(Colour online) Total contributions to the neutron Dirac form factor from each Feynman diagram in Fig.~\ref{fig:nucleon_em_current_feynman_diagrams} for a vector quark coupling (\textit{upper panel}) and tensor quark coupling (\textit{lower panel}). The sum of these 10 contributions gives the total neutron Dirac form factor illustrated in Fig.~\ref{fig:neutron_form_factors} (\textit{solid} curve).}
\label{fig:neutron_diagrams_F1}
\end{figure}

\begin{figure}[tbp]
\subfloat{\centering\includegraphics[width=\columnwidth,clip=true,angle=0]{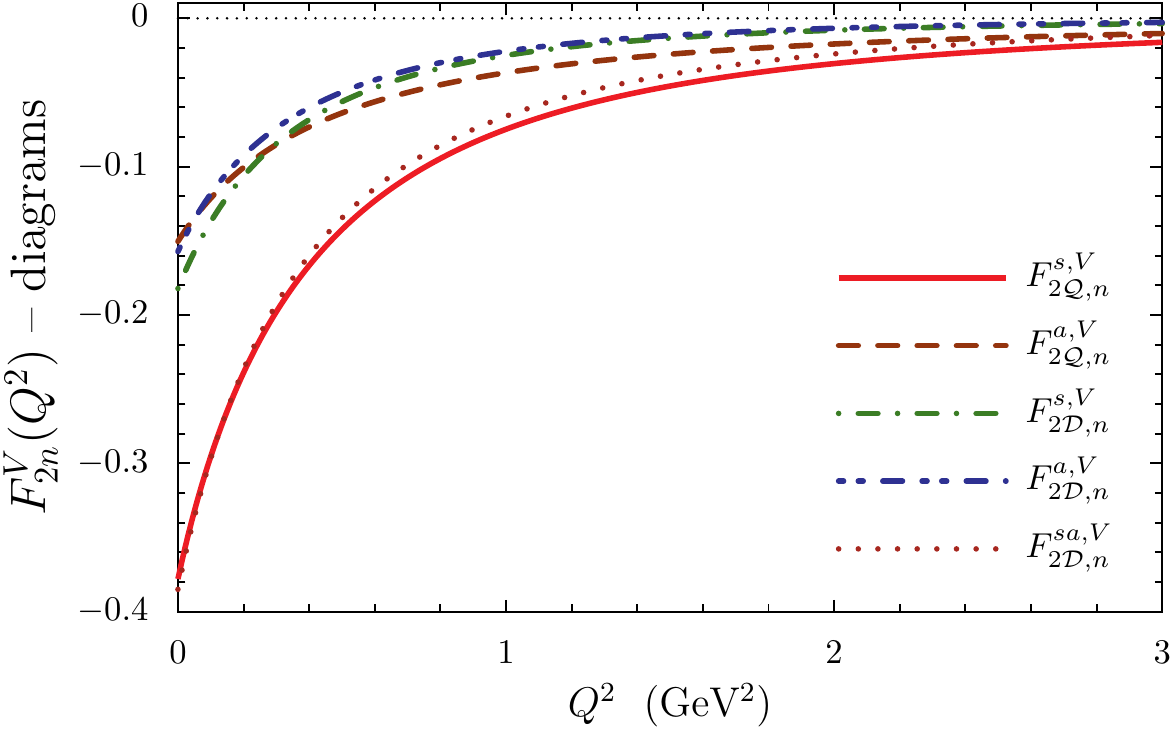}}\\
\subfloat{\centering\includegraphics[width=\columnwidth,clip=true,angle=0]{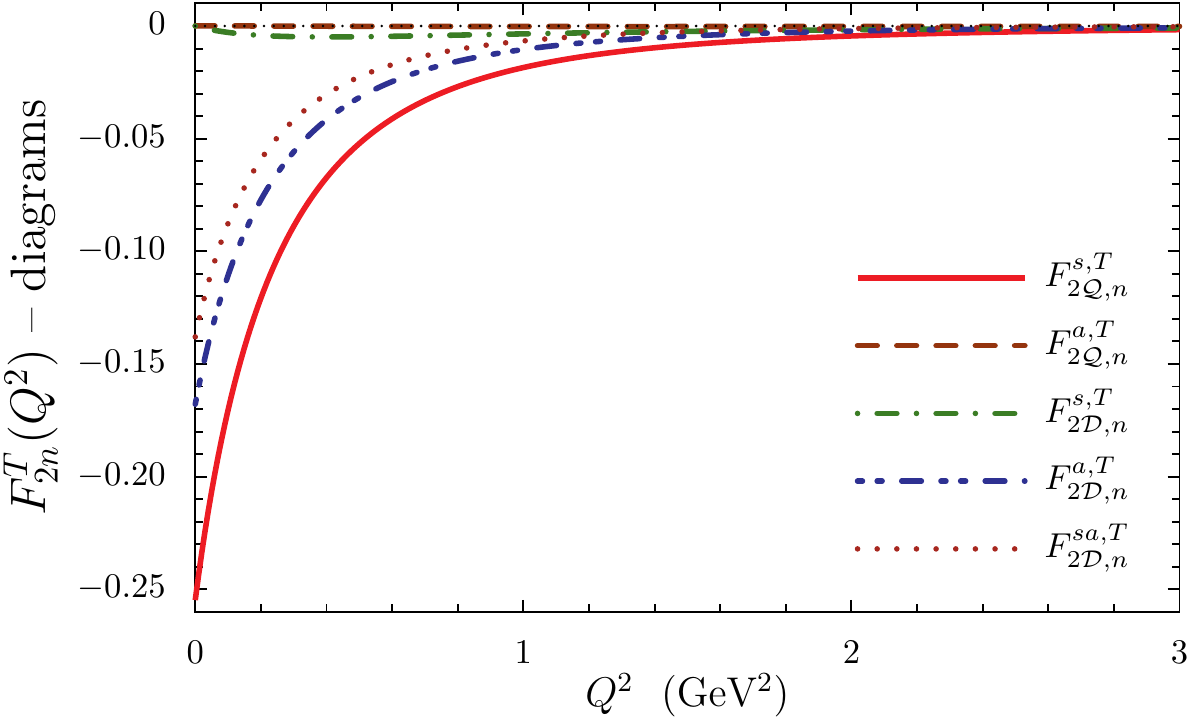}}
\caption{(Colour online) Total contributions to the neutron 
Pauli form factor from each Feynman diagram 
in Fig.~\ref{fig:nucleon_em_current_feynman_diagrams} for 
a vector quark coupling (\textit{upper panel}) and
tensor quark coupling (\textit{lower panel}). 
The sum of these 10 contributions gives the total neutron Dirac form 
factor illustrated in Fig.~\ref{fig:neutron_form_factors} (\textit{solid} curve).}
\label{fig:neutron_diagrams_F2}
\end{figure}

Figures~\ref{fig:neutron_diagrams_F1} and \ref{fig:neutron_diagrams_F2} 
and rows 3 and 4 of 
Table~\ref{tab:nucleon_diagrams_Q20} give results for 
the neutron form factors, broken down into the total 
vector and tensor coupling contributions from 
the Feynman diagrams of Fig.~\ref{fig:nucleon_em_current_feynman_diagrams}.
For the Dirac form factor the diagram pairs with 
the same quark--diquark structure 
cancel each other to give a 
charge of zero, while diagrams that contain only scalar diquarks 
carry 45\% of the neutron
anomalous magnetic moment and diagrams containing 
only axialvector diquarks carry 27\%. The remaining
28\% is carried by diagrams with both scalar and axialvector diquarks. 
The tensor coupling contributions
to $F_{1n}(Q^2)$ are much more significant compared to the proton case, 
an indication of the particular sensitivity
of the neutron Dirac form factor to  
pion cloud effects. axialvector diquarks
play a reduced role in neutron structure, compared to the proton, 
because the $\{dd\}$ type axialvector diquark
has half the charge and an anomalous magnetic moment 65\% 
the size of the $\{uu\}$ type diquark found in the
proton.

\section*{Acknowledgements}
ICC thanks Craig Roberts for many helpful discussions. This work was supported
by: the Department of Energy, Office of Nuclear Physics, 
contract no. DE-AC02-06CH11357;
the Australian Research Council through the 
ARC Centre of Excellence in Particle Physics at the Terascale 
and an ARC Australian Laureate Fellowship FL0992247 (AWT);
and the Grant in Aid for Scientific Research (Kakenhi) of the Japanese Ministry of
Education, Sports, Science and Technology, Project No. 20168769.


\begin{thebibliography}{99}
%
\bibitem{Jones:1999rz} 
  M.~K.~Jones {\it et al.}  [Jefferson Lab Hall A Collaboration],
  Phys.\ Rev.\ Lett.\  {\bf 84}, 1398 (2000)
  [nucl-ex/9910005].
%
\bibitem{Gayou:2001qd} 
  O.~Gayou {\it et al.}  [Jefferson Lab Hall A Collaboration],
  Phys.\ Rev.\ Lett.\  {\bf 88}, 092301 (2002)
  [nucl-ex/0111010].


\bibitem{Gayou:2001qt} 
  O.~Gayou, K.~Wijesooriya, A.~Afanasev, M.~Amarian, K.~Aniol, S.~Becher, K.~Benslama and L.~Bimbot {\it et al.},
  Phys.\ Rev.\ C {\bf 64}, 038202 (2001).



\bibitem{Puckett:2010ac} 
  A.~J.~R.~Puckett, E.~J.~Brash, M.~K.~Jones, W.~Luo, M.~Meziane, L.~Pentchev, C.~F.~Perdrisat and V.~Punjabi {\it et al.},
  Phys.\ Rev.\ Lett.\  {\bf 104}, 242301 (2010)
  [arXiv:1005.3419 [nucl-ex]].



\bibitem{Puckett:2011xg} 
  A.~J.~R.~Puckett, E.~J.~Brash, O.~Gayou, M.~K.~Jones, L.~Pentchev, C.~F.~Perdrisat, V.~Punjabi and K.~A.~Aniol {\it et al.},
  Phys.\ Rev.\ C {\bf 85}, 045203 (2012)
  [arXiv:1102.5737 [nucl-ex]].



\bibitem{Brodsky:1973kr} 
  S.~J.~Brodsky and G.~R.~Farrar,
  Phys.\ Rev.\ Lett.\  {\bf 31}, 1153 (1973).



\bibitem{Brodsky:1974vy} 
  S.~J.~Brodsky and G.~R.~Farrar,
  Phys.\ Rev.\ D {\bf 11}, 1309 (1975).



\bibitem{Ralston:2003mt} 
  J.~P.~Ralston and P.~Jain,
  Phys.\ Rev.\ D {\bf 69}, 053008 (2004)
  [hep-ph/0302043].



\bibitem{Pohl:2010zza} 
  R.~Pohl, A.~Antognini, F.~Nez, F.~D.~Amaro, F.~Biraben, J.~M.~R.~Cardoso, D.~S.~Covita and A.~Dax {\it et al.},
  Nature {\bf 466}, 213 (2010).



\bibitem{Antognini:1900ns} 
  A.~Antognini, F.~Nez, K.~Schuhmann, F.~D.~Amaro, FrancoisBiraben, J.~M.~R.~Cardoso, D.~S.~Covita and A.~Dax {\it et al.},
  Science {\bf 339}, 417 (2013).



\bibitem{Cloet:2010qa} 
  I.~C.~Clo\"et and G.~A.~Miller,
  Phys.\ Rev.\ C {\bf 83}, 012201 (2011)
  [arXiv:1008.4345 [hep-ph]].



\bibitem{Miller:2011yw} 
  G.~A.~Miller, A.~W.~Thomas, J.~D.~Carroll and J.~Rafelski,
  Phys.\ Rev.\ A {\bf 84}, 020101 (2011)
  [arXiv:1101.4073 [physics.atom-ph]].



\bibitem{Carroll:2011rv} 
  J.~D.~Carroll, A.~W.~Thomas, J.~Rafelski and G.~A.~Miller,
  Phys.\ Rev.\ A {\bf 84}, 012506 (2011)
  [arXiv:1104.2971 [physics.atom-ph]].



\bibitem{Gribov:1972rt} 
  V.~N.~Gribov and L.~N.~Lipatov,
  Sov.\ J.\ Nucl.\ Phys.\  {\bf 15}, 675 (1972)
  [Yad.\ Fiz.\  {\bf 15}, 1218 (1972)].



\bibitem{Altarelli:1977zs} 
  G.~Altarelli and G.~Parisi,
  Nucl.\ Phys.\ B {\bf 126}, 298 (1977).



\bibitem{Dokshitzer:1977sg} 
  Y.~L.~Dokshitzer,
  Sov.\ Phys.\ JETP {\bf 46}, 641 (1977)
  [Zh.\ Eksp.\ Teor.\ Fiz.\  {\bf 73}, 1216 (1977)].



\bibitem{Nambu:1961tp} 
  Y.~Nambu and G.~Jona-Lasinio,
  Phys.\ Rev.\  {\bf 122}, 345 (1961).



\bibitem{Nambu:1961fr} 
  Y.~Nambu and G.~Jona-Lasinio,
  Phys.\ Rev.\  {\bf 124}, 246 (1961).



\bibitem{Vogl:1991qt} 
  U.~Vogl and W.~Weise,
  Prog.\ Part.\ Nucl.\ Phys.\  {\bf 27}, 195 (1991).



\bibitem{Hatsuda:1994pi} 
  T.~Hatsuda and T.~Kunihiro,
  Phys.\ Rept.\  {\bf 247}, 221 (1994)
  [hep-ph/9401310].



\bibitem{Klevansky:1992qe} 
  S.~P.~Klevansky,
  Rev.\ Mod.\ Phys.\  {\bf 64}, 649 (1992).



\bibitem{Thomas:1981vc} 
  A.~W.~Thomas, S.~Theberge and G.~A.~Miller,
  Phys.\ Rev.\ D {\bf 24}, 216 (1981).



\bibitem{Theberge:1982xs} 
  S.~Theberge and A.~W.~Thomas,
  Nucl.\ Phys.\ A {\bf 393}, 252 (1983).



\bibitem{Thomas:1982kv} 
  A.~W.~Thomas,
  Adv.\ Nucl.\ Phys.\  {\bf 13}, 1 (1984).



\bibitem{Ishii:1993np} 
  N.~Ishii, W.~Bentz and K.~Yazaki,
  Phys.\ Lett.\ B {\bf 301}, 165 (1993).



\bibitem{Ishii:1993rt} 
  N.~Ishii, W.~Bentz and K.~Yazaki,
  Phys.\ Lett.\ B {\bf 318}, 26 (1993).



\bibitem{Ishii:1995bu} 
  N.~Ishii, W.~Bentz and K.~Yazaki,
  Nucl.\ Phys.\ A {\bf 587}, 617 (1995).



\bibitem{Mineo:2003vc} 
  H.~Mineo, W.~Bentz, N.~Ishii, A.~W.~Thomas and K.~Yazaki,
  Nucl.\ Phys.\ A {\bf 735}, 482 (2004)
  [nucl-th/0312097].



\bibitem{Cloet:2005rt} 
  I.~C.~Clo\"et, W.~Bentz and A.~W.~Thomas,
  Phys.\ Rev.\ Lett.\  {\bf 95}, 052302 (2005)
  [nucl-th/0504019].



\bibitem{Cloet:2005pp} 
  I.~C.~Clo\"et, W.~Bentz and A.~W.~Thomas,
  Phys.\ Lett.\ B {\bf 621}, 246 (2005)
  [hep-ph/0504229].



\bibitem{Cloet:2006bq} 
  I.~C.~Clo\"et, W.~Bentz and A.~W.~Thomas,
  Phys.\ Lett.\ B {\bf 642}, 210 (2006)
  [nucl-th/0605061].



\bibitem{Cloet:2007em} 
  I.~C.~Clo\"et, W.~Bentz and A.~W.~Thomas,
  Phys.\ Lett.\ B {\bf 659}, 214 (2008)
  [arXiv:0708.3246 [hep-ph]].



\bibitem{Ito:2009zc} 
  T.~Ito, W.~Bentz, I.~C.~Clo\"et, A.~W.~Thomas and K.~Yazaki,
  Phys.\ Rev.\ D {\bf 80}, 074008 (2009)
  [arXiv:0906.5362 [nucl-th]].



\bibitem{Matevosyan:2010hh} 
  H.~H.~Matevosyan, A.~W.~Thomas and W.~Bentz,
  Phys.\ Rev.\ D {\bf 83}, 074003 (2011)
  [arXiv:1011.1052 [hep-ph]].



\bibitem{Matevosyan:2011vj} 
  H.~H.~Matevosyan, W.~Bentz, I.~C.~Clo\"et and A.~W.~Thomas,
  Phys.\ Rev.\ D {\bf 85}, 014021 (2012)
  [arXiv:1111.1740 [hep-ph]].



\bibitem{Matevosyan:2012ga} 
  H.~H.~Matevosyan, A.~W.~Thomas and W.~Bentz,
  Phys.\ Rev.\ D {\bf 86}, 034025 (2012)
  [arXiv:1205.5813 [hep-ph]].



\bibitem{Bentz:2001vc} 
  W.~Bentz and A.~W.~Thomas,
  Nucl.\ Phys.\ A {\bf 696}, 138 (2001)
  [nucl-th/0105022].



\bibitem{Buballa:2003qv} 
  M.~Buballa,
  Phys.\ Rept.\  {\bf 407}, 205 (2005)
  [hep-ph/0402234].



\bibitem{Ebert:1996vx} 
  D.~Ebert, T.~Feldmann and H.~Reinhardt,
  Phys.\ Lett.\ B {\bf 388}, 154 (1996)
  [hep-ph/9608223].



\bibitem{Hellstern:1997nv} 
  G.~Hellstern, R.~Alkofer and H.~Reinhardt,
  Nucl.\ Phys.\ A {\bf 625}, 697 (1997)
  [hep-ph/9706551].



\bibitem{Maris:2003vk} 
  P.~Maris and C.~D.~Roberts,
  Int.\ J.\ Mod.\ Phys.\ E {\bf 12}, 297 (2003)
  [nucl-th/0301049].



\bibitem{Roberts:2011cf} 
  H.~L.~L.~Roberts, L.~Chang, I.~C.~Clo\"et and C.~D.~Roberts,
  Few Body Syst.\  {\bf 51}, 1 (2011)
  [arXiv:1101.4244 [nucl-th]].



\bibitem{Close:1988br} 
  F.~E.~Close and A.~W.~Thomas,
  Phys.\ Lett.\ B {\bf 212}, 227 (1988).



\bibitem{Schreiber:1991qx} 
  A.~W.~Schreiber, P.~J.~Mulders, A.~I.~Signal and A.~W.~Thomas,
  Phys.\ Rev.\ D {\bf 45}, 3069 (1992).



\bibitem{Hess:1998sd} 
  M.~Hess, F.~Karsch, E.~Laermann and I.~Wetzorke,
  Phys.\ Rev.\ D {\bf 58}, 111502 (1998)
  [hep-lat/9804023].



\bibitem{Bender:1996bb} 
  A.~Bender, C.~D.~Roberts and L.~Von Smekal,
  Phys.\ Lett.\ B {\bf 380}, 7 (1996)
  [nucl-th/9602012].



\bibitem{Buck:1992wz} 
  A.~Buck, R.~Alkofer and H.~Reinhardt,
  Phys.\ Lett.\ B {\bf 286}, 29 (1992).



\bibitem{Mineo:1999eq} 
  H.~Mineo, W.~Bentz and K.~Yazaki,
  Phys.\ Rev.\ C {\bf 60}, 065201 (1999)
  [nucl-th/9907043].



\bibitem{Mineo:2002bg} 
  H.~Mineo, W.~Bentz, N.~Ishii and K.~Yazaki,
  Nucl.\ Phys.\ A {\bf 703}, 785 (2002)
  [nucl-th/0201082].



\bibitem{Ernst:1960zza} 
  F.~J.~Ernst, R.~G.~Sachs and K.~C.~Wali,
  Phys.\ Rev.\  {\bf 119}, 1105 (1960).



\bibitem{Young:2006jc} 
  R.~D.~Young, J.~Roche, R.~D.~Carlini and A.~W.~Thomas,
  Phys.\ Rev.\ Lett.\  {\bf 97}, 102002 (2006)
  [nucl-ex/0604010].



\bibitem{Aniol:2005zf} 
  K.~A.~Aniol {\it et al.}  [HAPPEX Collaboration],
  Phys.\ Rev.\ Lett.\  {\bf 96}, 022003 (2006)
  [nucl-ex/0506010].



\bibitem{Armstrong:2005hs} 
  D.~S.~Armstrong {\it et al.}  [G0 Collaboration],
  Phys.\ Rev.\ Lett.\  {\bf 95}, 092001 (2005)
  [nucl-ex/0506021].



\bibitem{Baunack:2009gy} 
  S.~Baunack, K.~Aulenbacher, D.~Balaguer Rios, L.~Capozza, J.~Diefenbach, B.~Glaser, D.~von Harrach and Y.~Imai {\it et al.},
  Phys.\ Rev.\ Lett.\  {\bf 102}, 151803 (2009)
  [arXiv:0903.2733 [nucl-ex]].



\bibitem{Cates:2011pz} 
  G.~D.~Cates, C.~W.~de Jager, S.~Riordan and B.~Wojtsekhowski,
  Phys.\ Rev.\ Lett.\  {\bf 106}, 252003 (2011)
  [arXiv:1103.1808 [nucl-ex]].



\bibitem{Maris:1999bh} 
  P.~Maris and P.~C.~Tandy,
  Phys.\ Rev.\ C {\bf 61}, 045202 (2000)
  [nucl-th/9910033].



\bibitem{Amendolia:1986wj} 
  S.~R.~Amendolia {\it et al.}  [NA7 Collaboration],
  Nucl.\ Phys.\ B {\bf 277}, 168 (1986).



\bibitem{Amendolia:1984nz} 
  S.~R.~Amendolia, B.~Badelek, G.~Batignani, G.~A.~Beck, F.~Bedeschi, E.~H.~Bellamy, E.~Bertolucci and D.~Bettoni {\it et al.},
  Phys.\ Lett.\ B {\bf 146}, 116 (1984).



\bibitem{Horn:2006tm} 
  T.~Horn {\it et al.}  [Jefferson Lab F(pi)-2 Collaboration],
  Phys.\ Rev.\ Lett.\  {\bf 97}, 192001 (2006)
  [nucl-ex/0607005].



\bibitem{Tadevosyan:2007yd} 
  V.~Tadevosyan {\it et al.}  [Jefferson Lab F(pi) Collaboration],
  Phys.\ Rev.\ C {\bf 75}, 055205 (2007)
  [nucl-ex/0607007].



\bibitem{Blok:2008jy} 
  H.~P.~Blok {\it et al.}  [Jefferson Lab Collaboration],
  Phys.\ Rev.\ C {\bf 78}, 045202 (2008)
  [arXiv:0809.3161 [nucl-ex]].



\bibitem{Lepage:1979zb} 
  G.~P.~Lepage and S.~J.~Brodsky,
  Phys.\ Lett.\ B {\bf 87}, 359 (1979).



\bibitem{Farrar:1979aw} 
  G.~R.~Farrar and D.~R.~Jackson,
  Phys.\ Rev.\ Lett.\  {\bf 43}, 246 (1979).



\bibitem{Gluck:1998xa} 
  M.~Gluck, E.~Reya and A.~Vogt,
  Eur.\ Phys.\ J.\ C {\bf 5}, 461 (1998)
  [hep-ph/9806404].



\bibitem{Chang:2013nia} 
  L.~Chang, I.~C.~Clo\"et, C.~D.~Roberts, S.~M.~Schmidt and P.~C.~Tandy,
  Phys.\ Rev.\ Lett.\  {\bf 111}, 141802 (2013)
  [arXiv:1307.0026 [nucl-th]].



\bibitem{Dally:1982zk} 
  E.~B.~Dally, J.~M.~Hauptman, J.~Kubic, D.~H.~Stork, A.~B.~Watson, Z.~Guzik, T.~S.~Nigmanov and V.~D.~Ryabtsov {\it et al.},
  Phys.\ Rev.\ Lett.\  {\bf 48}, 375 (1982).



\bibitem{Frankfurt:1977vc} 
  L.~L.~Frankfurt and M.~I.~Strikman,
  Nucl.\ Phys.\ B {\bf 148}, 107 (1979).



\bibitem{Brodsky:1992px} 
  S.~J.~Brodsky and J.~R.~Hiller,
  Phys.\ Rev.\ D {\bf 46}, 2141 (1992).



\bibitem{Kelly:2004hm} 
  J.~J.~Kelly,
  Phys.\ Rev.\ C {\bf 70}, 068202 (2004).



\bibitem{Beringer:1900zz} 
  J.~Beringer {\it et al.}  [Particle Data Group Collaboration],
  Phys.\ Rev.\ D {\bf 86}, 010001 (2012).



\bibitem{Zhan:2011ji} 
  X.~Zhan, K.~Allada, D.~S.~Armstrong, J.~Arrington, W.~Bertozzi, W.~Boeglin, J.~-P.~Chen and K.~Chirapatpimol {\it et al.},
  Phys.\ Lett.\ B {\bf 705}, 59 (2011)
  [arXiv:1102.0318 [nucl-ex]].



\bibitem{Bernauer:2010wm} 
  J.~C.~Bernauer {\it et al.}  [A1 Collaboration],
  Phys.\ Rev.\ Lett.\  {\bf 105}, 242001 (2010)
  [arXiv:1007.5076 [nucl-ex]].



\bibitem{Sick:2012zz} 
  I.~Sick,
  Prog.\ Part.\ Nucl.\ Phys.\  {\bf 67}, 473 (2012).



\bibitem{Cloet:2008re} 
  I.~C.~Clo\"et, G.~Eichmann, B.~El-Bennich, T.~Klahn and C.~D.~Roberts,
  Few Body Syst.\  {\bf 46}, 1 (2009)
  [arXiv:0812.0416 [nucl-th]].

\bibitem{Cloet:2013gva} 
  I.~C.~Clo\"et, C.~D.~Roberts and A.~W.~Thomas,
  Phys.\ Rev.\ Lett.\  {\bf 111}, 101803 (2013)
  [arXiv:1304.0855 [nucl-th]].

\bibitem{Riordan:2010id} 
  S.~Riordan, S.~Abrahamyan, B.~Craver, A.~Kelleher, A.~Kolarkar, J.~Miller, G.~D.~Cates and N.~Liyanage {\it et al.},
  Phys.\ Rev.\ Lett.\  {\bf 105}, 262302 (2010)
  [arXiv:1008.1738 [nucl-ex]].



\bibitem{Diehl:2003ny} 
  M.~Diehl,
  Phys.\ Rept.\  {\bf 388}, 41 (2003)
  [hep-ph/0307382].



\bibitem{Belitsky:2005qn} 
  A.~V.~Belitsky and A.~V.~Radyushkin,
  Phys.\ Rept.\  {\bf 418}, 1 (2005)
  [hep-ph/0504030].



\bibitem{Itzykson:1980rh} 
  C.~Itzykson and J.~B.~Zuber,
  New York, Usa: Mcgraw-hill (1980) 705 P.(International Series In Pure and Applied Physics)



\bibitem{Myhrer:2007cf} 
  F.~Myhrer and A.~W.~Thomas,
  Phys.\ Lett.\ B {\bf 663}, 302 (2008)
  [arXiv:0709.4067 [hep-ph]].



\bibitem{Thomas:2008ga} 
  A.~W.~Thomas,
  Phys.\ Rev.\ Lett.\  {\bf 101}, 102003 (2008)
  [arXiv:0803.2775 [hep-ph]].
%
\end{thebibliography}

\end{document}